\documentclass[a4paper,fleqn]{cas-sc}

\usepackage{graphicx}
\usepackage{xcolor}
\usepackage{graphicx,subfig,epsfig}
\usepackage{amssymb,amsmath,amsbsy,bm,bbm}
\usepackage{algorithm,algpseudocode,theorem}
\usepackage{multirow}
\usepackage{tabularx}
\usepackage{mathtools, cuted}
\usepackage[numbers,sort&compress]{natbib}

\def\tsc#1{\csdef{#1}{\textsc{\lowercase{#1}}\xspace}}
\tsc{WGM}
\tsc{QE}

\begin{document}
\def\floatpagepagefraction{1}
\def\textpagefraction{.001}

\shorttitle{Designing experimental conditions for Lotka-Volterra model to infer interaction types}    

\shortauthors{H. Cho, A. Lewis, K. Storey, and H. Byrne}  

\title [mode = title]{Designing experimental conditions to use the Lotka-Volterra model to infer tumor cell line interaction types}  



%

\author[1,4]{Heyrim Cho}

\author[2]{Allison L. Lewis}

\author[2]{Kathleen M. Storey*}

\author[3]{Helen M. Byrne}







\address[1]{organization={Department of Mathematics},
            addressline={University of California, Riverside}, 
            state={CA},
            country={US}
            }
            

\address[2]{organization={Department of Mathematics},
            addressline={Lafayette College}, 
            city={Easton},
            state={PA},
            country={US}
            }
            
\address[3]{organization={Department of Mathematics},
            addressline={University of Oxford}, 
            city={Oxford},
            country={UK}}

\address[4]{organization={Interdisciplinary Center for Quantitative Modeling in Biology, University of California Riverside},
            country={US}
            }
            
            




\cortext[3]{Corresponding author}



\begin{abstract}
The Lotka-Volterra model is widely used to model interactions between two species. Here, we generate synthetic data mimicking competitive, mutualistic and antagonistic interactions between two tumor cell lines, and then use the Lotka-Volterra model to infer the interaction type. Structural identifiability of the Lotka-Volterra model is confirmed, and practical identifiability is assessed for three experimental designs: (a) use of a single data set, with a mixture of both cell lines observed over time, (b) a sequential design where growth rates and carrying capacities are estimated using data from experiments in which each cell line is grown in isolation, and then interaction parameters are estimated from an experiment involving a mixture of both cell lines, and (c) a parallel experimental design where all model parameters are fitted to data from two mixtures (containing both cell lines but with different initial ratios) simultaneously. Each design is tested on data generated from the Lotka-Volterra model with noise added, to determine efficacy in an ideal sense. In addition to assessing each design for practical identifiability, we investigate how the predictive power of the model---i.e., its ability to fit data for initial ratios other than those to which it was calibrated---is affected by the choice of experimental design. The parallel calibration procedure is found to be optimal and is further tested on \textit{in silico} data generated from a spatially-resolved cellular automaton model, which accounts for oxygen consumption and allows for variation in the intensity level of the interaction between the two cell lines. We use this study to highlight the care that must be taken when interpreting parameter estimates for the spatially-averaged Lotka-Volterra model when it is calibrated against data produced by the spatially-resolved cellular automaton model, since baseline competition for space and resources in the CA model may contribute to a discrepancy between the type of interaction used to generate the CA data and the type of interaction inferred by the LV model. 
\end{abstract}




\maketitle 

\section{Introduction}\label{sec:intro}

In many cancers, cellular heterogeneity plays a significant role in resistance to treatment and tumor recurrence \cite{McGranahan2017,Tabassum2015}. Thus, it is crucial to gain a better understanding of how a tumor's growth dynamics and its response to treatment influence---and are influenced by---cellular heterogeneity and the interactions between different cell populations, in order to make more accurate predictions about treatment responses. Mathematical modeling is an important tool that can be used to gain insight into the mechanisms that drive tumor heterogeneity and to make patient-specific predictions. Ordinary differential equation models are especially useful when fitting to sparse data, and the efficacy of such models has been compared in multiple studies \cite{Byrne2010, Altrock2015, ChoSpringer, Laleh2022, benzekry2014,gerlee2013,Simpson}. However, to ensure practical utility of such models for making predictions that are accurate and contain limited uncertainty, unique parameter identifiability---in the sense that the same model output cannot be produced by two different sets of parameter values---must be achieved. A thorough discussion of the importance of parameter identifiability in systems biology models is provided in \cite{Wieland}. Further, Simpson et al. (2022) performed an identifiability analysis on a series of population growth models, of the type often used to simulate tumor growth \cite{Simpson}. Here, we investigate the parameter identifiability of the Lotka-Volterra model, a two-compartment ordinary differential equation model, in order to assess its ability to distinguish different types of interactions between two tumor cell lines. 

The Lotka-Volterra (LV) model is widely used to model interactions between two populations \cite{Lotka1925, Volterra1926}. Its structure is sufficiently flexible to describe competitive, mutualistic, and antagonistic interactions.  In particular, the competitive LV model has been widely used to study competition between different subpopulations of cancer cells, the host tissue, and other cell populations \cite{gatenby1996reaction,cho2020impact,diego2013,cunningham2018,isea2015}. However, recent studies suggest that interactions between different cancer populations may be more complex than competition, and may contribute to tumor invasion and treatment response \cite{Paczkowski2021,chapman2014,wu2010interaction,caswell2017,noble2021,susswein2022}. 
Since tumor cell lines can interact in multiple ways, with consequential implications for tumor progression and recurrence, we seek to determine whether it is possible to infer interaction type when fitting the LV model to data from two interacting tumor cell lines. In Paczkowski et al. (2021), a preliminary investigation was performed by fitting the LV model to dynamic time course data from tumor spheroids cultured from prostate cancer cell lines \cite{Paczkowski2021}. Previously, a range of techniques has been used to estimate parameter values in the LV model from experimental data  \cite{Sim2012,KLOPPERS2013817,Liepe,MARASCO201649}. We extend this work by first considering the structural identifiability of the LV model given perfect, noise-free data about the volumes of two cell lines. We then assess its practical identifiability using synthetic data generated from the LV model with noise added, and from a spatially-resolved cellular automaton (CA) model. Our CA model incorporates various types of cellular interaction and adds to the growing body of literature in which the growth of tumor spheroids is simulated using different types of agent-based models (e.g., on-lattice models in \cite{lowengrub2009nonlinear,macklin2010multiscale,van2015simulating,Poleszczuk2016} and off-lattice models in \cite{hyun2013improved,venugopalan2014multicellular,kim2014mathematical,ghaffarizadeh2018physicell}).

As part of the practical identifiability investigation, we compare the ability of different experimental design structures to fit the LV model and to infer the type of interaction between the two cell lines given dynamic tumor data on the volumes (and/or proportions) of the two cell types. We consider three experimental design structures: (a) individual calibration, where we fit to a single data set which contains dynamic volume information for both cell lines; (b) sequential calibration of the cell-line specific intrinsic growth rates and carrying capacities using dynamic data from each cell line, and subsequent calibration of the interaction parameters using time course data from an experiment in which the two cell lines are co-cultured; and (c) parallel calibration of all model parameters using data from two mixture experiments, which differ in terms of their initial conditions. 

We use synthetic data from the LV model to identify the optimal experimental design scheme, to eliminate issues relating to model discrepancy. We then calibrate the model to synthetic data generated from the CA model. The CA model accounts for spatial heterogeneity, oxygen consumption, and allows the intensity of the imposed cell-cell interactions to be varied. This enables us to study the robustness of the calibration procedure on a data set that better approximates \textit{in vitro} experimental data for multicellular tumor spheroids. 

The results from our structural and practical identifiability analyses provide insight into the type of experimental design and data needed to obtain identifiable parameter estimates and infer the interaction type between two cell lines. Our investigations suggest that where volumes data about both cell lines is available for multiple sets of initial conditions, the LV model can be fitted to the data and used to make predictions that may inform treatment decisions in the clinic. At the same time, care is needed when interpreting parameter estimates of the (spatially-averaged) LV model when it is fit to spatially-resolved data.

In Section \ref{sec:Methods}, we describe the LV model and the methods that we use to assess its structural and practical identifiability. In Section \ref{sec:results} we demonstrate the structural identifiability of the LV model, and assess its practical identifiability under the three proposed calibration procedures using synthetic data generated by the LV model with noise added. We then test the parallel calibration procedure on data generated from the CA model. We also discuss the limitations of using a spatially-averaged model, such as the LV model, to make inferences about spatially-resolved data. In Section \ref{sec:Disc}, we summarize our findings and describe possible directions for future work. Further details on the CA model are described in Appendix A, and the sequential calibration results of fitting the LV model to the CA data are presented in Appendix B.

\section{Methods}\label{sec:Methods}

\subsection{Lotka-Volterra Model and Synthetic Data} \label{subsec:LVmodel}

We use the Lotka-Volterra (LV) model \cite{Lotka1925, Volterra1926} to describe the growth of a heterogeneous tumor spheroid comprised of two cancer cell types, which we term Type-$S$ and Type-$R$. 
In isolation, each population is assumed to 
undergo logistic growth, with growth rates $r_S$ and $r_R$ and carrying capacities $K_S$ and $K_R$ respectively. Interactions between the two populations are determined by the signs and magnitudes of the parameters $\gamma_S$ and $\gamma_R$. Written in terms of these parameters, the governing equations read:
\begin{eqnarray}
\frac{dS}{dt} &=& r_SS\left(1-\frac{S}{K_S}-\frac{\gamma_RR}{K_S}\right), \label{eqn:dSdt} \\ 
\frac{dR}{dt} &=& r_R R\left(1-\frac{R}{K_R}-\frac{\gamma_SS}{K_R}\right), \label{eqn:dRdt}
\end{eqnarray}
where $S(t)$ and $R(t)$ represent the volume (mm$^3$) of Type-$S$ and Type-$R$ cancer cells at time $t$, respectively. We close Equations (\ref{eqn:dSdt})-(\ref{eqn:dRdt}) by assuming that the initial volumes of the Type-$S$ and Type-$R$ cells are known, so that:
\begin{equation}
S(0) = S_0 \quad \mbox{and} \quad R(0) = R_0. 
\end{equation}
The parameter set consists of $\theta=\{r_S, r_R, K_S, K_R, \gamma_S, \gamma_R\}$, where $r_S$, $r_R$, $K_S$, and $K_R$ are assumed to be non-negative, while $\gamma_S$ and $\gamma_R$ may be positive, negative, or zero. The signs of $\gamma_R$ and $\gamma_S$ define the type of interaction between the two cell lines: it is competitive if $\gamma_R$ and $\gamma_S$ are positive, mutualistic if they are both negative, and antagonistic if they are of opposite signs (see Table \ref{tab:LVsign} for a summary).

\begin{table}[!htb]
    \centering
    \begin{tabular}{|c|c|c|c|} \hline 
         & $\gamma_S > 0$ & $\gamma_S = 0 $& $\gamma_S < 0$  \\ \hline 
      $\gamma_R > 0$    & Competitive &  & R antagonizes S  \\ \hline 
      $\gamma_R = 0$    &  & Neutral & \\ \hline 
      $\gamma_R < 0$    & S antagonizes R &  & Mutual \\ \hline 
    \end{tabular}
    \caption{A summary of how the signs of the parameters $\gamma_S$ and $\gamma_R$ in Equations (\ref{eqn:dSdt})-(\ref{eqn:dRdt}) determine the type of interaction between the Type-$S$ and Type-$R$ cells.}
    \label{tab:LVsign}
\end{table}

In Section \ref{sec:practicaltoLV}, we assess the practical identifiability of the LV model parameters by fitting it to data from the LV model. We generate 29 data sets: two pure experiments (i.e., homogeneous tumor data comprising only Type-$S$ or Type-$R$ cells), and nine mixture experiments with initial ratios 1:9, 2:8,..., 9:1 of Type-$S$:Type-$R$ cells, for competitive, mutual, and antagonistic interaction types (assuming, without loss of generality, that Type-$R$ cells antagonize Type-$S$ cells).  Following \cite{Paczkowski2021}, we fix $r_S = r_R = 0.36$ and $K_S = K_R = 0.85$, and vary $\gamma_S$ and $\gamma_R$ to simulate the different interaction types; Table \ref{tab:LVparams} summarizes the parameter values used to generate the data sets.

\begin{table}[!bth]
    \centering
    \begin{tabular}{|c|c|c|c|c|c|c|}
    \hline 
        \textbf{Interaction Type} & $r_S$ & $r_R$ & $K_S$ & $K_R$ & $\gamma_S$ & $\gamma_R$  \\
        \hline 
         Competitive  & 0.36 & 0.36 & 0.85 & 0.85 & 0.5 & 0.5 \\
         \hline 
         Mutual &  0.36 & 0.36 & 0.85 & 0.85 & -0.5 & -0.5 \\
         \hline 
         Antagonistic &  0.36 & 0.36 & 0.85 & 0.85 & -0.5 & 0.5 \\
         \hline 
    \end{tabular}
    \caption{Parameter values used to generate synthetic data from the Lotka-Volterra model.}
    \label{tab:LVparams}
\end{table}

\begin{figure}[!bth]
\begin{centering}  
\rotatebox{90}{$\qquad\qquad$ \rotatebox{-90}{(a)}}
\includegraphics[width=4in]{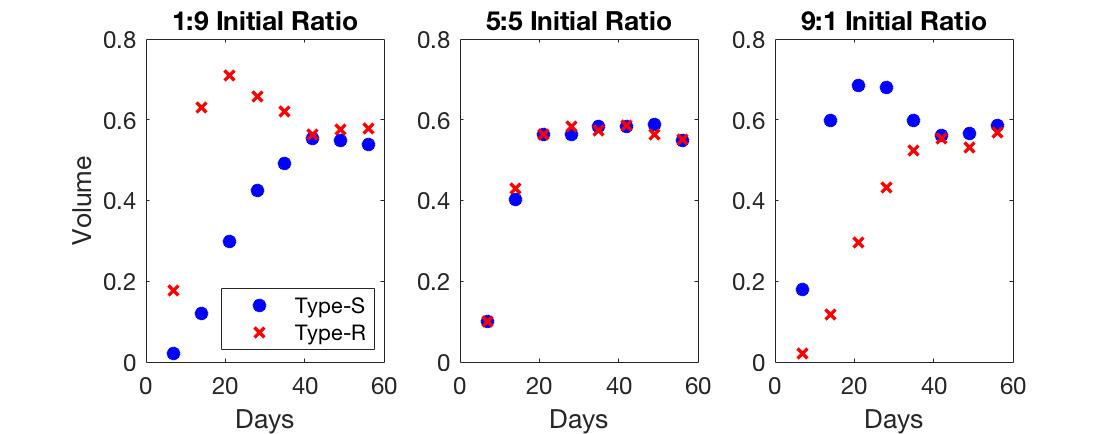} \\ 
\rotatebox{90}{$\qquad\qquad$ \rotatebox{-90}{(b)}}
\includegraphics[width=4in]{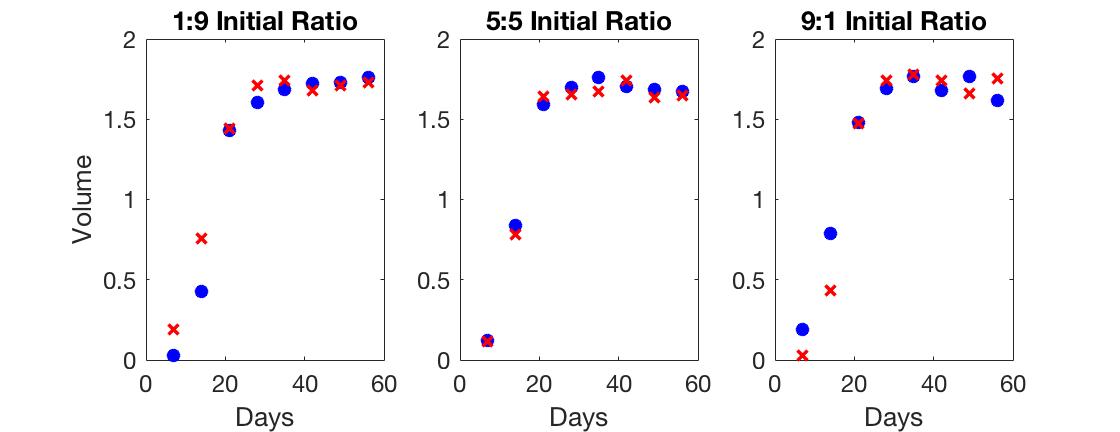} \\ 
\rotatebox{90}{$\qquad\qquad$ \rotatebox{-90}{(c)}}
\includegraphics[width=4in]{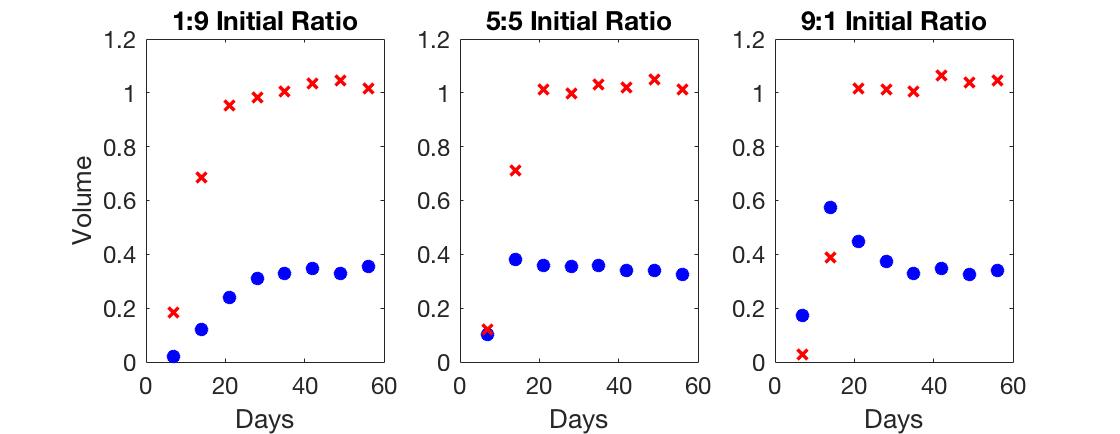}\\ 
\end{centering}
\caption{Synthetic data from the Lotka-Volterra model, generated with the parameter values listed in Table \ref{tab:LVparams} and
initial Type-$S$:Type-$R$ ratios of 1:9, 5:5, and 9:1, for (a) competitive, (b) mutual, and (c) antagonistic interaction types. }
\label{fig:LVdata}
\end{figure}

All simulated tumors are initialized with a total tumor volume of 0.02 mm$^3$; thus, the nine sets of initial conditions are given by $(S_0^j, R_0^j) = (0.002j, 0.02-0.002j)$ for $j = 1,2 ..., 9$. Volume measurements for both cell lines are generated at $t=7, 14, 21, \dots, 56$ days. 
For a given initial ratio, and at each data point---denoted $(S_i,R_i)$ for $i=1,\dots,8$---we add 5\% noise: i.e., $S_{\text{noise}}(t_i) = S(t_i)( 1  + 0.05 \xi_{S,i})$ and $R_{\text{noise}}(t_i) = R(t_i)(1+0.05 \xi_{R,i})$, where $\xi_S$ and $\xi_R$ are uniform random variables on $[-1,1]$. Sample synthetic data for the 1:9, 5:5, and 9:1 initial ratios for the three interaction types are shown in Figure \ref{fig:LVdata}. Since the intrinsic growth rates and carrying capacities for both cell lines are identical, the simulations for the competitive and mutual cases predict long-time coexistence of the two cell lines at equal volumes. In the antagonistic cases, the dominance of the Type-$R$ cells results in a higher predicted volume for the Type-$R$ population, regardless of the initial conditions.

\subsection{Cellular Automaton Model and Synthetic Data}
\label{subsec:CAmodel}

As a test of the ability of the LV model to fit \textit{in vitro} tumor spheroid data, we extend a cellular automaton (CA) model from  \cite{Paczkowski2021,ChoSpringer,ChoJCM}. Cells are arranged on a two-dimensional lattice, representing a two-dimensional cross-section of a three-dimensional tumor spheroid \textit{in vitro}. Each cell can be classified as proliferating, quiescent, or necrotic. Cell state is determined by the local oxygen concentration, which is modeled using a reaction-diffusion equation. As in the LV model, we consider two types of tumor cells, Type-$S$ and Type-$R$, which consume oxygen, proliferate, and die at rates which depend on the local $O_2$ concentration. As in the LV data generation discussed in Section \ref{subsec:LVmodel}, we further assume that the two cell types
are identical except for how they interact with each other, 
although this could later be modified to represent differences in other characteristics, e.g.~radiosensitive versus radioresistant.

In the CA model, we consider four interaction types: neutral, competitive, mutualistic, and antagonistic (where, without loss of generality, the Type-$R$ population antagonizes the Type-$S$ population, and the Type-$S$ population promotes the Type-$R$ population). The interaction type affects both the oxygen consumption and proliferation rates. For example, in the antagonistic case, Type-$S$ cells surrounded by a large number of Type-$R$ cells consume less oxygen, and as a result, divide at a slower rate; likewise, Type-$R$ cells increase their rates of oxygen consumption and division when surrounded by Type-$S$ cells. 

The model includes a parameter, $I$, which determines the intensity of the cellular interactions, with $I=1$ for neutral interactions (all cell cycle and oxygen consumption rates fixed at their baseline values), and $I>1$ for stronger interactions (the cell cycle and oxygen consumption rates altered to reflect the interaction type). With the exception of the sensitivity analysis presented in Figure \ref{fig:comparesign_intlevel}, we fix $I=4$ for competitive, mutual, and antagonistic interactions. Further details about the CA model are included in \ref{sec:app:CA}.

We generate \textit{in silico} data from the CA model for the four interaction types, and different values of the initial proportion of Type-$S$ cells, as for the LV data. In all cases, we initialize the cells in a circular region with a radius of 9 cells, corresponding to a total tumor volume of approximately 0.02 mm$^3$. The cell types at each initial site are randomly chosen, according to the specified initial conditions. For each choice of initial conditions and parameter values, we simulate 10 model replicates and record the average volumes of Type-$S$ cells and Type-$R$ cells at $t= 7, 14, 21, \dots, 56$, to reduce the impact of stochastic outliers on the calibration process.

\begin{figure}[!htb]
    \centering
    \includegraphics[width=\textwidth]{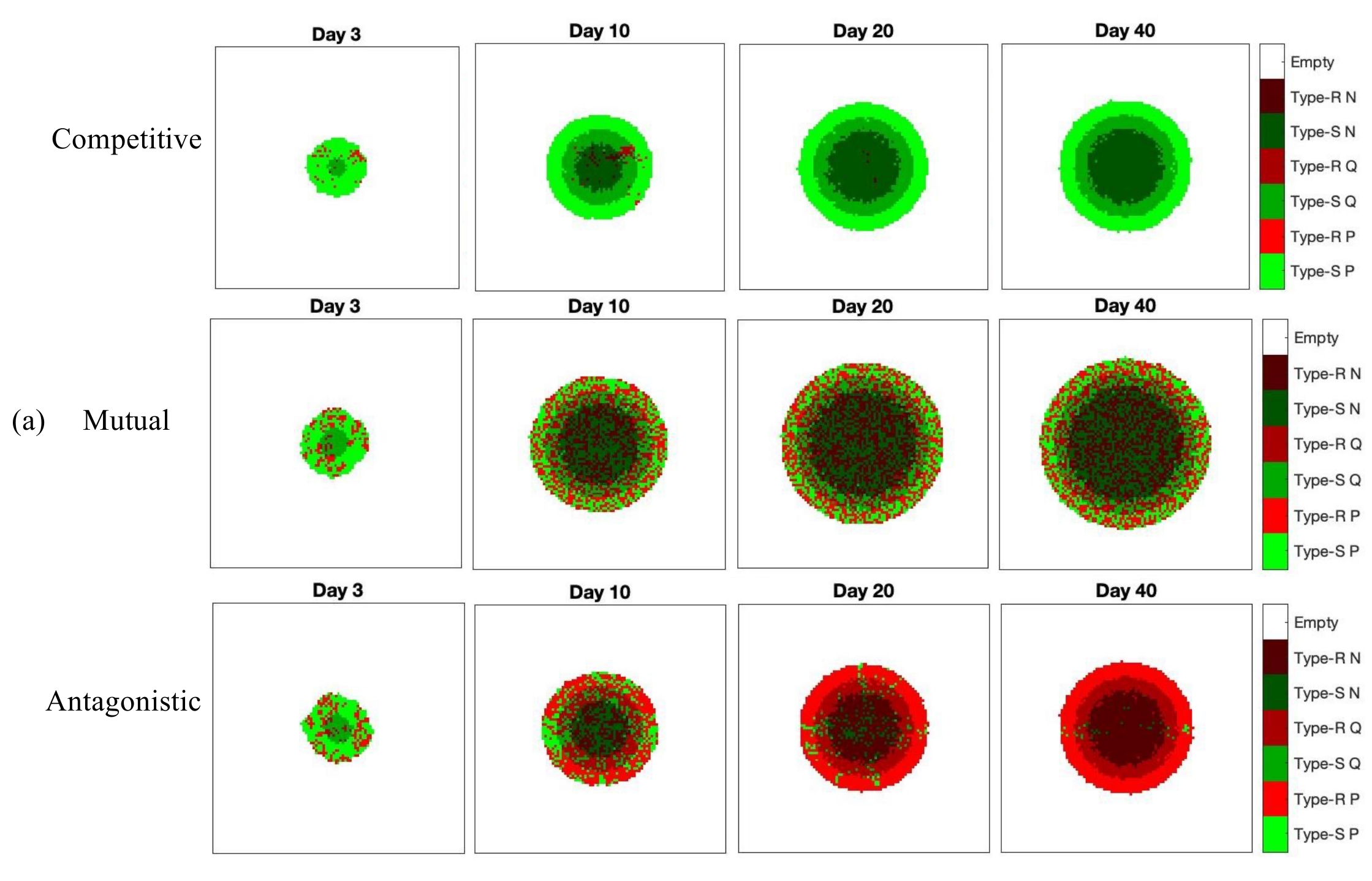}
    \includegraphics[width=\textwidth]{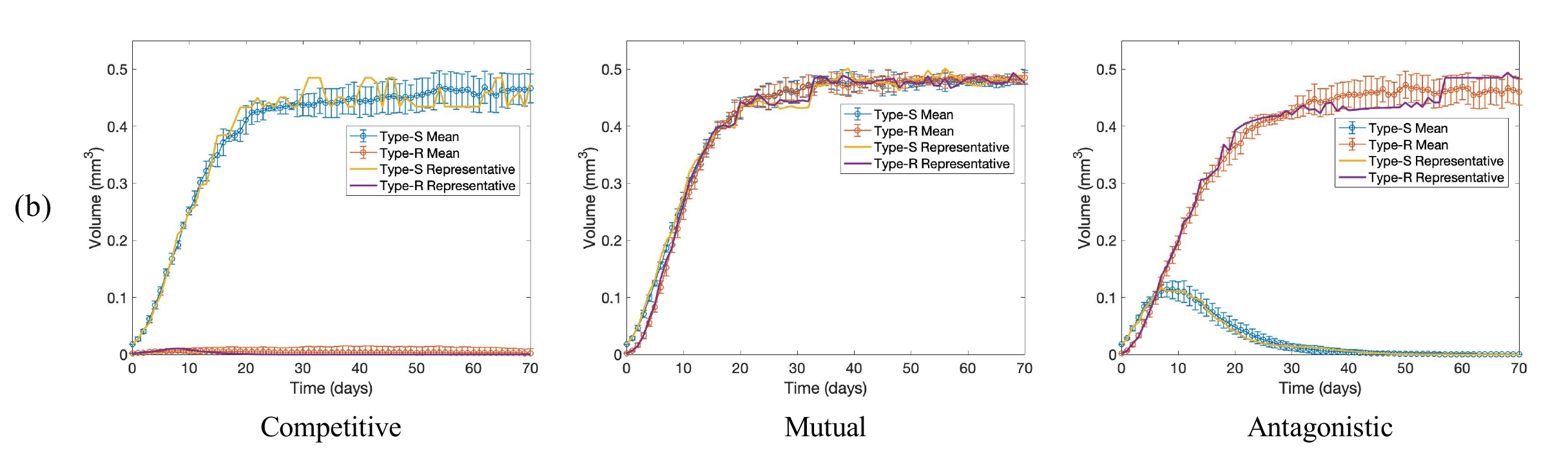}
    \caption{CA simulation outputs, showing (a) two-dimensional cross sections of spheroids at Days 3, 10, 20, and 40, for competitive, mutual, and antagonistic interaction types, and (b) mean trajectories, with error bars indicating the standard deviation, from 10 replicates of the CA. The yellow and purple curves correspond to the specific simulations shown in part (a). In all cases, the simulations were initialized with a 9:1 ratio of Type-$S$ to Type-$R$ cells, with parameters as listed in Table \ref{table:CA_pars}.}
    \label{fig:CA_spheroids}
\end{figure}

Figure \ref{fig:CA_spheroids}(a) shows how the spatial distributions of Type-$S$ and Type-$R$ cells change over time for typical simulations, with an initial ratio of Type-$S$:Type-$R$ cells of 9:1. This figure compares the distribution of cells for the competitive, mutual, and antagonistic interaction types. Figure \ref{fig:CA_spheroids}(b) shows how the total numbers of Type-$S$ cells and Type-$R$ cells change over time for the particular simulations shown in part (a) and the averaged replicates. We observe that in the competitive case, the Type-$S$ cells quickly become dominant, due to the large proportion of Type-$S$ cells at $t=0$. In the mutual case, the Type-$R$ population catches up with the Type-$S$ population size by day 10, and both cell types are able to co-exist throughout the tumor. In the antagonistic case, the Type-$R$ cells have a selective advantage, so they overtake the Type-$S$ cells by day 10 and proceed to dominate the spheroid. In contrast to the data generated by the LV model in Section \ref{subsec:LVmodel}, asymptotic coexistence of the two cell lines is predicted only for the mutual case.  The parameter values used to generate the simulated tumor spheroids are given in Table \ref{table:CA_pars}.

\subsection{Model Calibration} \label{subsec:calibration}

Throughout the investigation, we use Bayesian inference to perform parameter estimation.  In more detail, we use a Metropolis Hastings algorithm to generate Markov parameter chains whose stationary distributions represent the posterior distributions of the parameters. These posterior densities can be used to estimate optimal values of the parameters when the LV model is fitted to the data, denoted by $\hat \theta = \{\hat r_S, \hat r_R, \hat K_S, \hat K_R, \hat \gamma_S, \hat \gamma_R\}$, and reflect the uncertainty in the estimates. We employ non-informative uniform prior distributions of $\mathcal{U}(0,1)$ for $\{r_S, r_R, K_S, K_R\}$, and $\mathcal{U}(-3,3)$ for the interaction parameters $\gamma_S$ and $\gamma_R$. The lower bounds for $r_S$, $r_R$, $K_S$, and $K_R$ are chosen to ensure positivity of these parameter estimates, while the ranges for $\gamma_S$ and $\gamma_R$ are chosen to allow inference of all interaction types (mutual, competitive, and antagonistic). The optimal parameter estimates are identified by maximizing the likelihood function; if we assume that errors are independent and normally distributed as $\varepsilon \sim \mathcal{N}(0,\sigma^2)$, then the likelihood function is of the form $$\mathcal{L}(y|\theta) = \frac{1}{(2\pi \sigma^2)^{n/2}} \exp{\left(\frac{1}{2\sigma^2}\sum\limits_{i=1}^n [y_i-f_i(\theta)]^2\right)},$$ where $y_i$ represents the $i$th observed data point for $i=1,\dots,n$, and $f_i(\theta)$ represents the quantity of interest for parameter set $\theta$ at time $t_i$. With this formulation, maximizing the likelihood function is equivalent to minimizing the sum-of-squares error between the observed data and the model predictions. The Markov chain of accepted candidates is constructed using the proposal function $J(\theta^* | \theta^{k-1}) = \mathcal{N}(\theta^{k-1},C)$, where $C$ represents the estimated covariance matrix for the parameter set $\theta$. That is, each parameter candidate is sampled from a multivariate normal distribution centered at the previous accepted candidate. If the new candidate, $\theta^*$, improves the likelihood function (i.e., yields a smaller sum-of-squares error than the previous candidate), we set $\theta^k = \theta^*$, accepting this candidate into our Markov chain. Else, we accept the candidate with probability $1-\alpha$, where $\alpha$ is defined as the ratio of the likelihoods using the current candidate as the numerator and the previous candidate as the denominator---see \cite{Smith} for details---and reject otherwise. The occasional acceptance of a parameter candidate whose likelihood is inferior to those already sampled promotes well-mixing of the Markov parameter chains and full exploration of the admissible parameter space, permitting quicker convergence to the stationary distribution and ensuring that the chain does not stall in a local optimum. Upon convergence, these Markov chains can be used to visualize the parameter posterior densities.  Full details regarding the implementation of the Metropolis Hastings algorithm are provided at \cite{Smith, Haario}. For all calibration procedures, we use the MCMC toolbox for MATLAB with the Delayed Rejection Adaptive Metropolis variation of the Metropolis Hastings algorithm \cite{Laine}.

\subsection{Identifiability Analysis}
\label{subsec:identifiability}

\paragraph{Structural Identifiability} Before undertaking parameter estimation, it is important to establish whether the model is \textit{a priori}  structurally identifiable, i.e., is not over-parametrized.  At the input level, we say that a parameter is not structurally identifiable if perturbing the parameter can still result in the same model output through compensation via changes in other parameters \cite{Wieland, Chis, Smith}; else, it is structurally identifiable. A model is only structurally identifiable if all of its parameters are structurally identifiable; essentially, we require a 1-to-1 mapping between the parameter set and model output.  Several algorithms for assessing structural identifiability have been proposed, including methods that employ generating series, power series expansions, differential algebra, and  differential geometry. For further details, we refer the interested reader to \cite{Villaverde, Bates, Chis2, Raue}.

Previous studies have used algebraic methods and Lie derivatives to investigate the structural identifiability of the LV model \cite{Remien, Greene}.  In this investigation,  structural identifiability of the LV model (\ref{eqn:dSdt})-(\ref{eqn:dRdt}) will be assessed using the Taylor series expansion method as outlined in \cite{Chappell, Pohjanpalo}.

We denote by $y(t;\theta)$ the time-dependent observable quantity (or quantities) evaluated at parameter set $\theta$. Assuming that $y(t;\theta)$ is analytic in a neighborhood of the initial conditions $t=t_0$, we can express $y(t;\theta)$ and each of its time derivatives in terms of the model parameters and initial conditions using the Taylor series expansion 
\begin{eqnarray*}
y(t;\theta) = y(t_0; \theta)+y^{(1)}(t_0;\theta)\ t+y^{(2)}(t_0;\theta)\ \frac{t^2}{2!}+\dots+y^{(i)}(t_0;\theta)\ \frac{t^i}{i!}+\dots,
\end{eqnarray*}
where $y^{(i)}(t_0;\theta)$ denotes the $i$th time derivative of the observable quantity evaluated at the initial condition $t_0$ with parameter set $\theta$. 

By considering the coefficients of the Taylor series expansion, we can restate the problem of structural identifiability as a system of algebraic equations relating our unknown parameter set $\theta$ to the observable quantity $y(t;\theta)$ and its time derivatives at $t=t_0$, where the number of equations to be considered depends upon the number of parameters to be estimated. If these equations admit a unique solution for $\theta$ in terms of $y(t;\theta)$ and its derivatives at $t=t_0$, we declare the model to be globally structurally identifiable. If no such solution exists, we conclude that the model is either non-identifiable, or locally structurally identifiable, in the sense that the identifiability of the model parameters depends on the region of parameter space in which it is evaluated. 

Establishing structural identifiability of nonlinear systems is a challenging task, whose complexity increases with the dimension of the parameter space. The GenSSI (Generating Series for testing Structural Identifiability) MATLAB toolbox enables non-experts to perform such analyses \cite{Chis}. It employs the generating series approach and produces identifiability tableaux, which can be used to determine how to handle parameters classified as non-identifiable \cite{Balsa-Canto}.  When assessing structural identifiability of the LV model in Section \ref{sec:structID}, we will confirm the results of our Taylor series identifiability analysis using the GenSSI toolbox. 

\paragraph{Practical Identifiability} The focus of practical, or \textit{a posteriori}, identifiability is to determine whether model parameters can be inferred from potentially noisy data via model calibration. Potential issues with practical identifiability include: difficulty inferring parameter values due to measurement errors, model discrepancy (when a model is unable to accurately represent the underlying system), and an experimental design that is unable to naturally ``excite"---i.e., activate---certain parameters. Each issue may preclude \textit{a posteriori} parameter identification, even for structurally identifiable models. These issues depend on the quality, design, and availability of the data, rather than the model structure. 

For our purposes, we say that a parameter is practically identifiable if the posterior distribution obtained from the Metropolis Hastings algorithm is unimodal, exhibiting a clear and unique optimum. Otherwise, the parameter is deemed practically non-identifiable, in the sense that multiple values of the input may yield the same model output. In this way, non-identifiability of a parameter may manifest as a posterior distribution that is multimodal---suggesting multiple local maxima---or as a distribution that is relatively unchanged from its non-informative prior distribution, indicating that the parameter is uninformed by the data. A model is pronounced practically identifiable given a particular data set only if all model parameters are practically identifiable.

\subsection{Experimental Design Options}
\label{subsec:ex_designs}

For our practical identifiability investigation, we compare three methods to calibrate the LV model to data. First, we consider whether the parameters in the LV model may be uniquely identified using data from a single experiment---containing time course volumetric data about both cell lines---with the initial ratio of the two cell lines specified, as described in Section \ref{subsec:LVmodel}.  We term this method ``individual calibration". 

Our second method is termed ``sequential calibration."  Here, we first estimate $(r_S, K_S)$ and $(r_R, K_R)$, the intrinsic growth rates and carrying capacities for each cell line, using data describing the dynamics of one cell line grown in isolation. Subsequently, these parameters are fixed at their estimated values $\{\hat r_S, \hat r_R, \hat K_S, \hat K_R\}$, and the interaction parameters $\gamma_S$ and $\gamma_R$ are inferred from a third data set containing a mixture of the two cell lines. Thus, this method requires three data sets---two relating to homogeneous growth of each cell line and one to heterogeneous growth of a mixture. 

Our final calibration procedure is referred to as ``parallel calibration," as all six parameters are estimated simultaneously using two mixture data sets which differ only in terms of their initial conditions.  For example, we might attempt to infer the complete parameter set $\{r_S, r_R, K_S, K_R, \gamma_S, \gamma_R\}$ by fitting the LV model to growth data initialized with ratios of 1:9 and 9:1 for the Type-$S$ and Type-$R$ cells.  

\begin{figure}
    \centering
    \includegraphics[width=\textwidth]{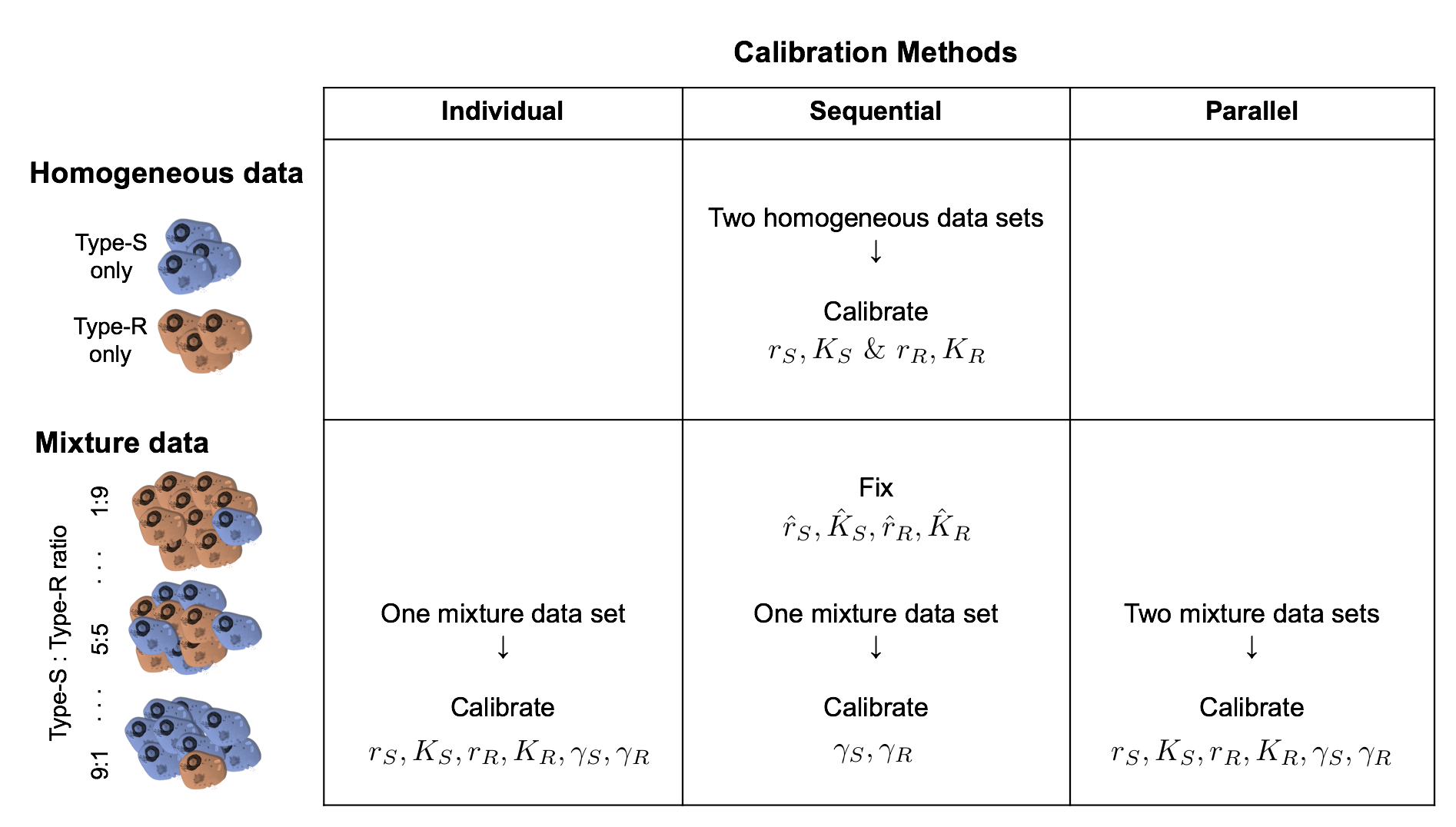}
    \caption{Summary of the three calibration procedures outlined in Section \ref{subsec:ex_designs}.}
    \label{fig:flowchart}
\end{figure}

A summary of the three calibration methods is presented in Figure \ref{fig:flowchart}. All methods are assessed in Section \ref{sec:practicaltoLV}, using synthetic data generated from the LV model.  In Section \ref{sec:CAresults}, we conduct further tests, applying the parallel procedure to data generated from the CA model.  

\subsection{Assessment Metrics}
\label{subsec:metrics}

We use three metrics---termed $E_1$, $E_2$, and $E_3$---to assess the interaction inference, accuracy, and predictive power of the resulting model fits. Metric $E_1$ is a label which indicates the interaction type inferred by the LV model, based upon the signs of the interaction parameters:
\begin{eqnarray}
E_1 = \left\{ \begin{array}{cc} C, & \text{ if  } \ \hat \gamma_S > 0, \ \hat \gamma_R >0 \\ M, & \text{ if  } \ \hat \gamma_S <0, \ \hat \gamma_R <0 \\ A, & \text{ if  }\  \hat \gamma_S <0, \ \hat \gamma_R >0. \end{array} \right. \label{eqn:E1}
\end{eqnarray}
where $C$, $M$, and $A$ indicate ``competitive", ``mutual", and ``antagonistic" relationships, respectively. We note that the interaction type inferred by the LV model may not match the type of interaction used to generate the data. 

Metric $E_2$ measures the accuracy and predictive power of the inferred model fit. That is, in addition to measuring the error of the model fit to the training data set used for calibration, we quantify the accuracy with which the model reproduces the dynamics of other heterogeneous tumors which differ from the training data set in terms of their initial conditions, using testing data sets generated from the other eight initial ratio combinations. We denote the observed data by $d_i^j = (s_i^j, r_i^j),$ where $s_i^j$ and $r_i^j$ represent the volume measurements of Type-$S$ and Type-$R$ cells, respectively, with initial ratio $j:(10-j)$, for $j=1,\dots,9$ at time $t_i$ for $i=1,\dots,N$. We denote the solution to the LV model at time $t_i$, parameterized with parameters $\theta$ and initial conditions $[S(0), R(0)]$, by $y(t_i; \theta, [S(0),R(0)]) = (S(t_i),R(t_i))$. Metric $E_2$ is then defined to be the log of the sum-of-squares error between the training and testing data sets and the predictions of the LV model for all nine initial conditions. Thus, we have 
\begin{eqnarray}
E_2 = \ln \left( \sum_{i=1}^{N} \sum_{j=1}^9 ||d_i^j - y(t_i; \hat{\theta}, [S_0^j, R_0^j]  ) ||_2^2 \right), \label{eqn:E2}
\end{eqnarray}
where $\hat \theta$ represents the inferred parameter set resulting from the calibration of the LV model to the training data set.

Our third metric measures the ability of the LV model and proposed calibration procedure to recover the true values of the parameters used to generate the data. As such, it is used only in Section \ref{sec:practicaltoLV} where we use the LV model to generate the training and testing data. If the true parameter values are denoted by $\tilde \theta = \{\tilde r_S, \tilde r_R, \tilde K_S, \tilde K_R, \tilde \gamma_S, \tilde \gamma_R\}$ and the estimated parameters by $\hat \theta = [\hat r_S, \hat r_R, \hat K_S, \hat K_R, \hat \gamma_S, \hat \gamma_R]$, then metric $E_3$ is defined as 
\begin{eqnarray}
E_3 =\|\tilde \theta - \hat \theta\|_1. \label{eqn:E3}
\end{eqnarray} Accordingly, small values of $E_3$ indicate that the recovered parameter estimates are close to the true parameter values.

\section{Results} \label{sec:results}

\subsection{Structural Identifiability} \label{sec:structID}

We illustrate that the LV model is structurally identifiable.
Using the Taylor series expansion method discussed in Section \ref{subsec:identifiability}, we evaluate $y(t; \theta) = \{S,R\}$ and its successive time derivatives in terms of the model parameters and initial conditions at time $t = 0$, assuming that the observables $S(t)$ and $R(t)$ are analytic in a neighborhood of $t=0^+$. Recall, these derivatives represent the coefficients of the Taylor series expansion (up to a constant multiple) of the observable. If we introduce the following notation,

$$a_0 = S(0^+), \ \ a_1 = S'(0^+), \ \ a_2 = S''(0^+), \ \ a_3 = S'''(0^+),$$
$$b_0 = R(0^+), \ \ b_1 = R'(0^+), \ \ b_2 = R''(0^+), \ \ b_3 = R'''(0^+),$$ 
\noindent then it is straightforward to show that Equations (\ref{eqn:dSdt})-(\ref{eqn:dRdt}) supply the following identities:
\begin{eqnarray}
\frac{a_1}{a_0} &=& r_S - a_0\left(\frac{r_S}{K_S}\right)-b_0\left(\frac{r_S\gamma_R}{K_S}\right), \label{eqn:newdSdt} \\ \ \nonumber \\
\frac{b_1}{b_0} &=& r_R - b_0\left(\frac{r_R}{K_R}\right)-a_0\left(\frac{r_R\gamma_S}{K_R}\right). \label{eqn:newdRdt} \\ \ \nonumber 
\end{eqnarray}

We view Equations (\ref{eqn:newdSdt})-(\ref{eqn:newdRdt}) as two algebraic equations that relate the unknown model parameters to measurable quantities $a_i$ and $b_i$ ($i=0,1,2,\dots$).  Since we seek to identify six parameters, at least four more equations are needed.  We obtain these additional equations by differentiating with respect to time and substituting for $a_i$ and $b_i$.  In this way, we obtain the following equations:

\begin{eqnarray}
\frac{a_2}{a_1} &=& r_S - 2a_0\left(\frac{r_S}{K_S}\right) - \left(\frac{a_0b_1}{a_1}+b_0\right)\left(\frac{r_S\gamma_R}{K_S}\right) \label{eqn:derdSdt} \\ \ \nonumber \\
\frac{b_2}{b_1} &=& r_R - 2b_0\left(\frac{r_R}{K_R}\right) - \left(a_0+\frac{a_1b_0}{b_1}\right)\left(\frac{r_R\gamma_S}{K_R}\right) \label{eqn:derdRdt} \\ \ \nonumber \\
\frac{a_3}{a_2} &=& r_S - 2\left(a_0+\frac{a_1^2}{a_2}\right)\left(\frac{r_S}{K_S}\right) -\left(\frac{a_0b_2}{a_2}+\frac{2a_1b_1}{a_2}+b_0\right)\left(\frac{r_S\gamma_R}{K_S}\right) \label{eqn:secderdSdt} \\ \ \nonumber \\
\frac{b_3}{b_2} &=& r_R - 2\left(b_0+\frac{b_1^2}{b_2}\right)\left(\frac{r_R}{K_R}\right) -\left(a_0+\frac{2a_1b_1}{b_2}+\frac{a_2b_0}{b_2}\right)\left(\frac{r_R\gamma_S}{K_R}\right) \label{eqn:secderdRdt} \\ \ \nonumber 
\end{eqnarray}

\noindent If we make the substitution 
\begin{eqnarray}
p_1 = r_S, \ \ \ \ p_2 = r_R, \ \ \ \ p_3 = \frac{r_S}{K_S}, \ \ \ \ p_4 = \frac{r_R}{K_R}, \ \ \ \ p_5 = \frac{r_s\gamma_R}{K_S}, \ \ \ \ p_6 = \frac{r_R\gamma_S}{K_R}, \label{eqn:substitution}
\end{eqnarray}  

\noindent then it is straightforward to show that Equations (\ref{eqn:newdSdt})-(\ref{eqn:secderdRdt}) admit a unique solution for the six parameters $\{p_1, p_2, p_3, p_4, p_5, p_6\}$ provided that all initial conditions are known:
\begin{eqnarray*}
p_1 &=& -\frac{a_0^3a_2b_2-a_0^3a_3b_1-2a_0^2a_1^2b_2+4a_0^2a_1a_2b_1+a_0^2a_1a_3b_0-a_0^2a_2^2b_0-2a_0a_1^3b_1-2a_0a_1^2a_2b_0+2a_1^4b_0}{a_0^3(a_1b_2-a_2b_1)}, \\ \ \\
p_2 &=& \frac{a_0b_0^2b_1b_3-a_0b_0^2b_2^2-2a_0b_0b_1^2b_2+2a_0b_1^4-a_1b_0^3b_3+4a_1b_0^2b_1b_2-2a_1b_0b_1^3+a_2b_0^3b_2-2a_2b_0^2b_1^2}{b_0^3(a_1b_2-a_2b_1)},\\ \ \\
p_3 &=& -\frac{a_0^2a_2b_2-a_0^2a_3b_1-a_0a_1^2b_2+3a_0a_1a_2b_1-2a_1^3b_1}{a_0^3(a_1b_2-a_2b_1}, \\ \ \\
p_4 &=& -\frac{a_1b_0^2b_3-3a_1b_0b_1b_2+2a_1b_1^3-a_2b_0^2b_2+a_2b_0b_1^2}{b_0^3(a_1b_2-a_2b_1}, \\ \ \\
p_5 &=& -\frac{a_0^2a_1a_3-a_0^2a_2^2-2a_0a_1^2a_2+2a_1^4}{a_0^3(a_1b_2-a_2b_1)}, \\ 
p_6 &=& \frac{b_0^2b_1b_3-b_0^2b_2^2-2b_0b_1^2b_2+2b_1^4}{b_0^3(a_1b_2-a_2b_1)}.\\
\end{eqnarray*}

\noindent The original parameter set $\{r_S, r_R, K_S, K_R, \gamma_S, \gamma_R\}$ can be recovered using Equation \eqref{eqn:substitution}. Since Equations \eqref{eqn:newdSdt}-\eqref{eqn:secderdRdt} admit a unique solution for the full parameter set, we conclude that the model is structurally identifiable given observables $y(t; \theta) = \{S,R\}$ and known initial conditions $\{a_0, a_1, a_2, a_3, b_0, b_1, b_2, b_3\}$.  

The structural identifiability analysis performed above requires knowledge of the Type-$S$ and Type-$R$ observables; it assumes that, in addition to data on total tumor volumes, the proportions of the two cell lines are also known. In practice, it may not be feasible to collect such data in a clinical setting. Thus, it is natural to ask whether the system remains structurally identifiable if the total tumor volume, $V = S+R$, is the only observable quantity.  It is possible to show that a unique parameterization exists in this case, but \textit{only} if the initial volumes of the two cell lines, $S(0)$ and $R(0)$, are known. Given these initial conditions, structural identifiability can be established by proceeding to 6th order in the Taylor series expansion for $\frac{dV}{dt} = \frac{dS}{dt}+\frac{dR}{dt}$ (results not shown).

To verify the results from the Taylor series method, we use the GenSSI toolbox for both cases: (1) assuming observables $y(t; \theta) = \{S,R\}$ with available information about $S(0)$, $R(0)$, and their derivatives, and (2) assuming only a single observable $y(t; \theta) = \{V = S+R\}$ with initial estimates of both cell lines, $S(0)$ and $R(0)$. For the first case, GenSSI confirms that the model is globally structurally identifiable.  For the second case, with a single observable, $V = S+R$, GenSSI confirms that the model is, at least, locally structurally identifiable. 

\subsection{Practical Identifiability Using Synthetic LV Data} \label{sec:practicaltoLV}

Having established that the LV model parameters are structurally identifiable, we focus now on assessing practical identifiability. We begin by attempting to fit the LV model to noisy data generated from the LV model using the three methods discussed in Section \ref{subsec:ex_designs}.

\paragraph{Individual Calibration}

\begin{figure}[!bth]
\begin{centering}
\rotatebox{90}{$\qquad\quad$ \rotatebox{-90}{(a)}}
\includegraphics[width=5in]{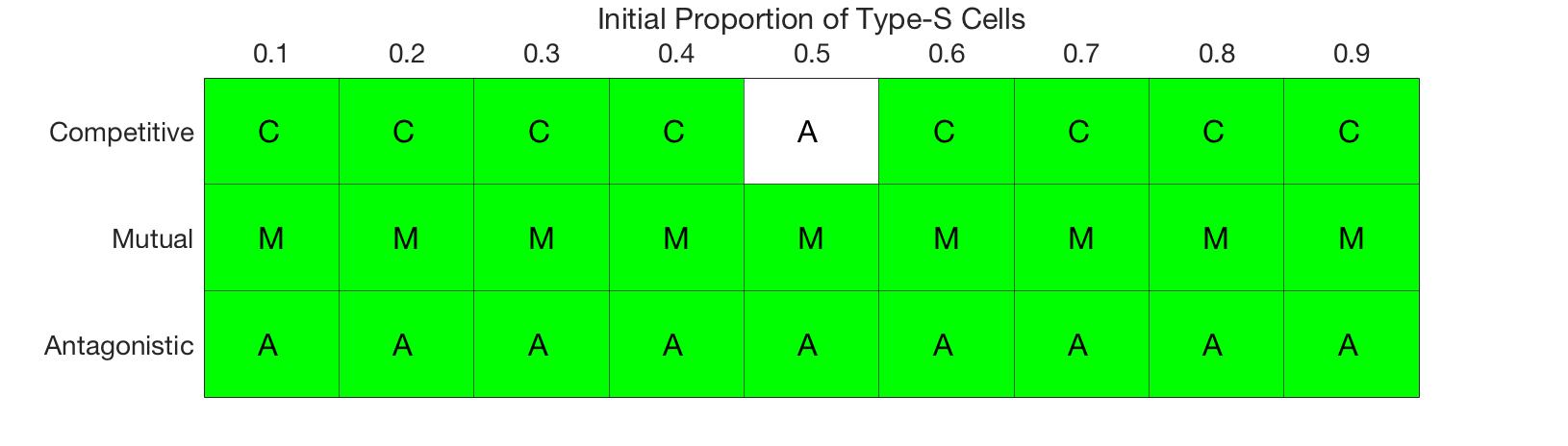} \\
\rotatebox{90}{$\qquad\quad\,\,\,$ \rotatebox{-90}{(b)}}
\includegraphics[width=5in]{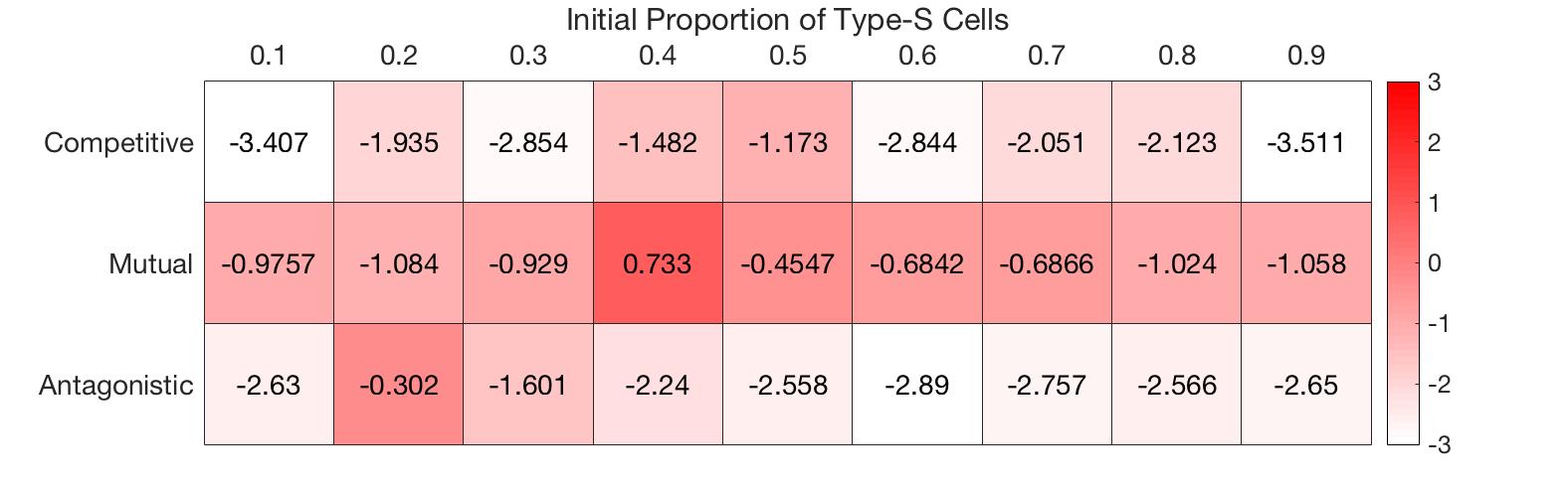} \\
\rotatebox{90}{$\qquad\quad\,\,\,$ \rotatebox{-90}{(c)}}
\includegraphics[width=5in]{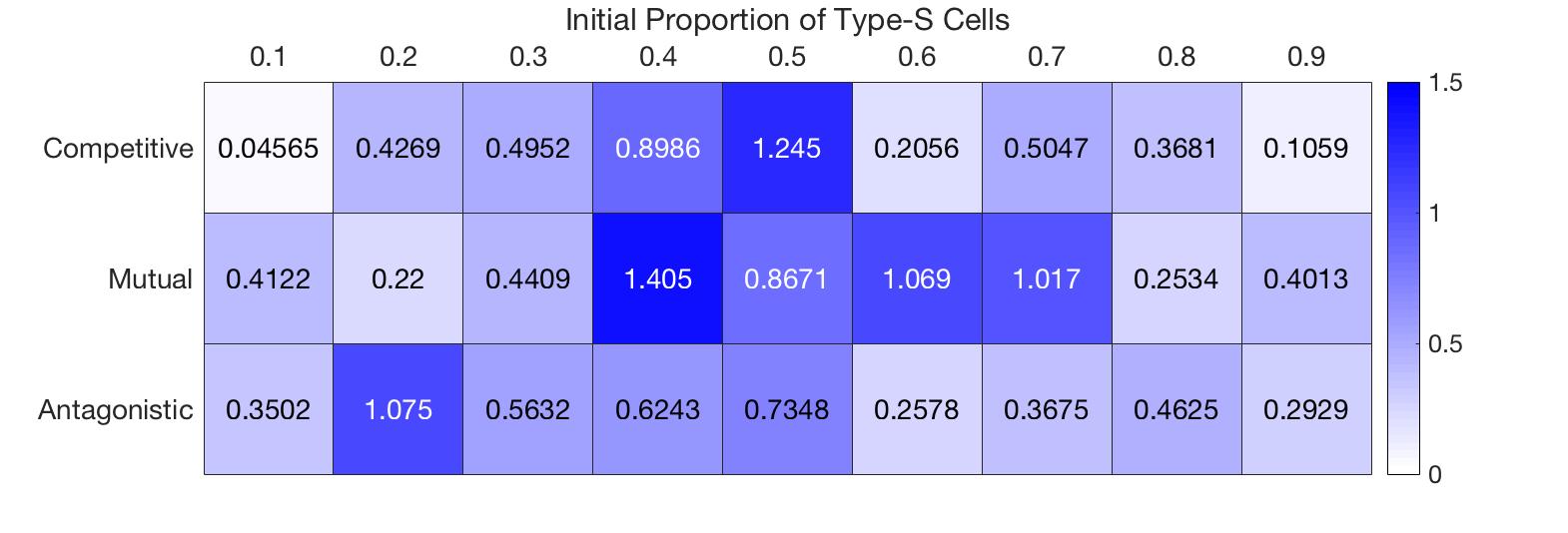}\\
\end{centering}
\caption{(Individual Calibration) (a) Metric $E_1$ in Eq.~\eqref{eqn:E1}, stating the inferred interaction type (green indicates a match between the interaction type used to generate the data and that inferred by the model), (b) metric $E_2$ in Eq.~\eqref{eqn:E2}, measuring the log of sum of squares error across all nine data sets, and (c) metric $E_3$ in Eq.~\eqref{eqn:E3}, measuring the distance from the true parameters, for the competitive, mutual, and antagonistic cases with calibration using a single data set. }
\label{fig:indcalLV_errors}
\end{figure}

\begin{figure}[!bth]
\begin{centering}
\includegraphics[width=\textwidth]{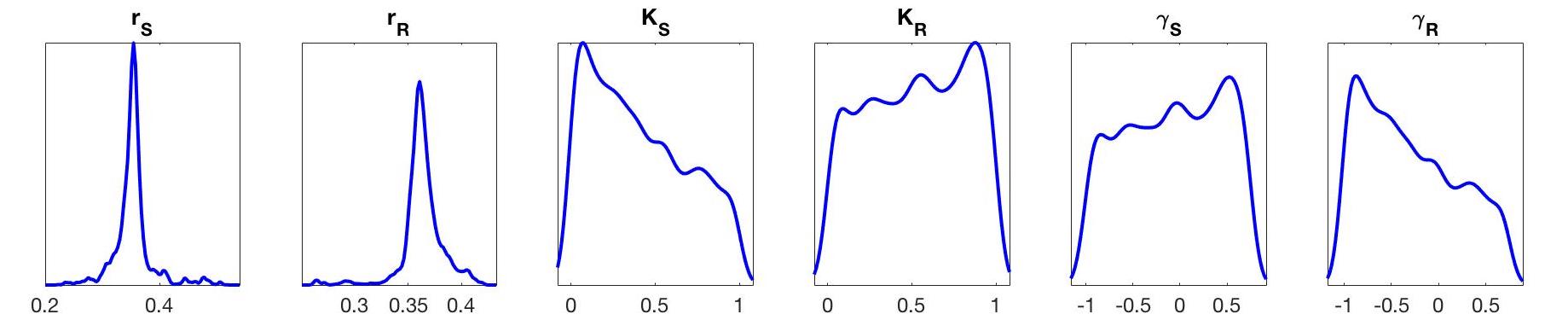} \\
\end{centering}
\caption{(Individual Calibration) Posterior parameter densities computed by Metropolis Hastings algorithm for the 5:5 initial ratio of Type-$S$ to Type-$R$ cells.}
\label{fig:ind_pt5densities}
\end{figure}

The results presented in Figure \ref{fig:indcalLV_errors}, show how the metrics $E_1$, $E_2$, and $E_3$---as defined in Section \ref{subsec:metrics}---change as the initial proportion of Type-$S$ cells varies, when we use the individual calibration procedure. Figure \ref{fig:indcalLV_errors}(a) shows that, in most cases, the model fit to the data predicts the correct interaction type, as indicated by the signs of the inferred interaction parameters. The only exception is the competitive case with an initial ratio of Type-$S$ to Type-$R$ cells of 5:5, for which an antagonistic relationship is predicted. However, the posterior parameter densities for this case reveal that the carrying capacity and interaction parameters are uninformed by the data, suggesting that the chosen parameter estimates are non-unique and that the parameters are unidentifiable 
(see Figure \ref{fig:ind_pt5densities}). 

\begin{figure}[!bth]
    \centering
    \includegraphics[width=0.47\textwidth]{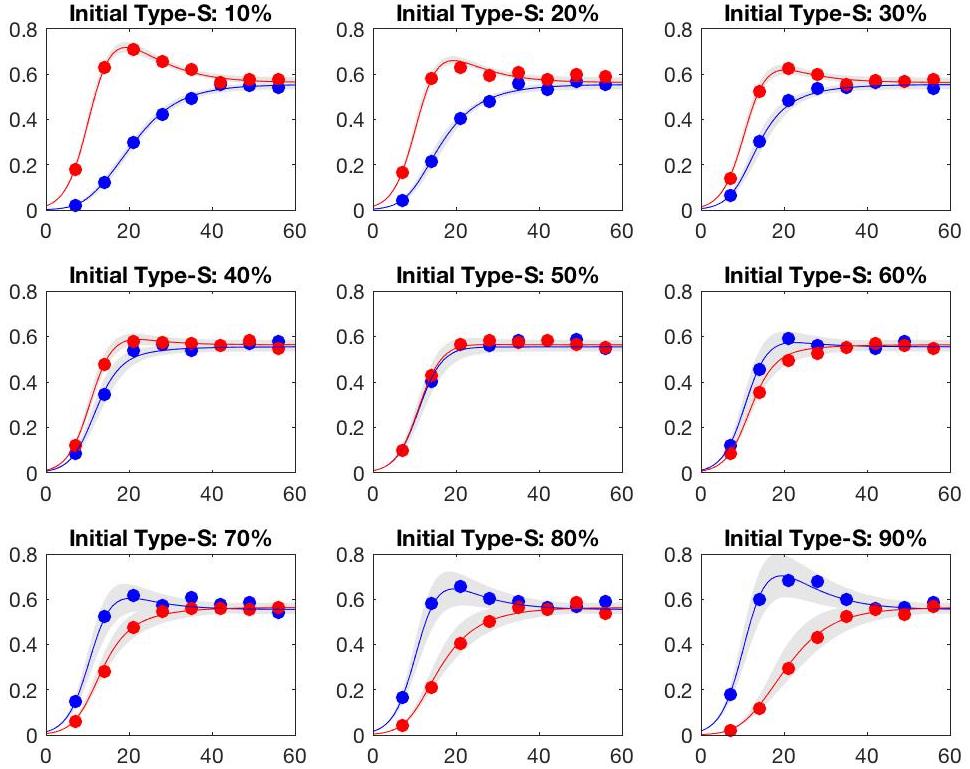} \hspace{7pt} \includegraphics[width=0.47\textwidth]{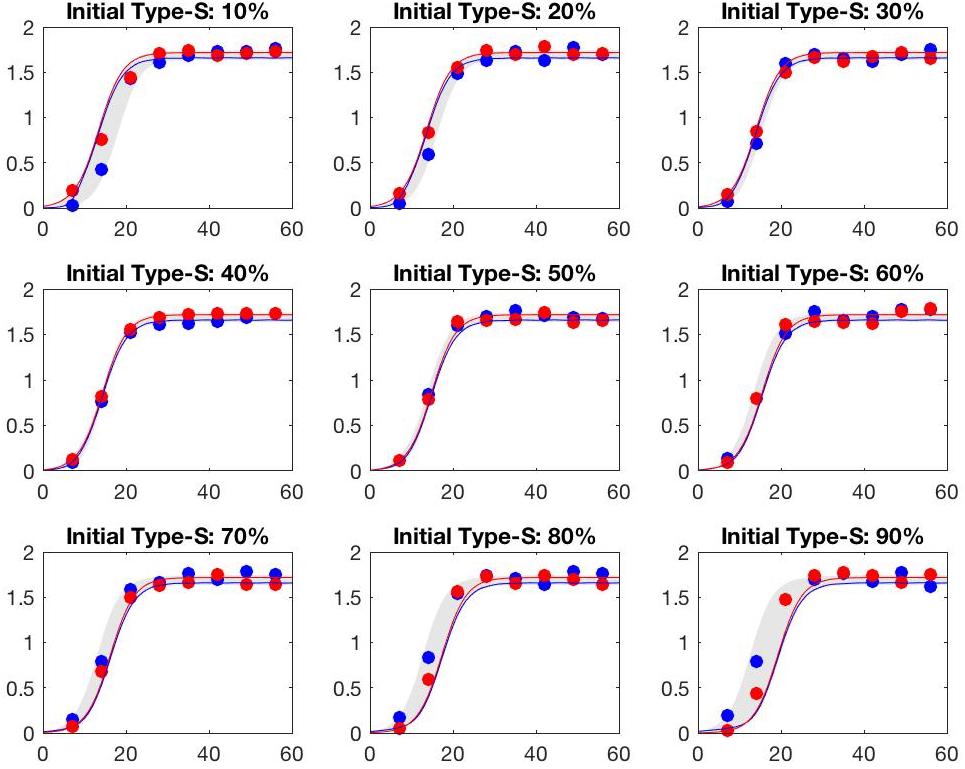} \\ 
    (a) \hspace{200pt} (b)
    \caption{(Individual Calibration) (a) Competitive case model fits using the parameter values obtained from the 1:9 experiment, and (b) Mutual case model fits using the parameter values obtained from the 4:6 experiment. }
    \label{fig:modelfits_ind}
\end{figure}

Results for the prediction error metric $E_2$ are presented in Figure \ref{fig:indcalLV_errors}(b). Errors in the model fits are generally highest for the mutual interaction data sets, although even the worst case scenario yields a reasonable visual fit---see Figure \ref{fig:modelfits_ind} for a comparison of the case with the smallest error (competitive interaction, with an initial ratio of 1:9) and the case with the largest error (mutual interaction, with an initial ratio of 4:6). Results for the parameter error metric $E_3$ are presented in Figure \ref{fig:indcalLV_errors}(c). In general, high values of $E_2$ correspond to parameter estimates that differ markedly from those used to generate the data, as indicated by large values of $E_3$. While it is unsurprising that the errors are large when the original parameter values cannot be recovered, these results indicate which experimental data sets are difficult to fit.

\begin{figure}[!bth]
\begin{centering}
\rotatebox{90}{$\qquad\quad$ \rotatebox{-90}{(a)}}
\includegraphics[width=5in]{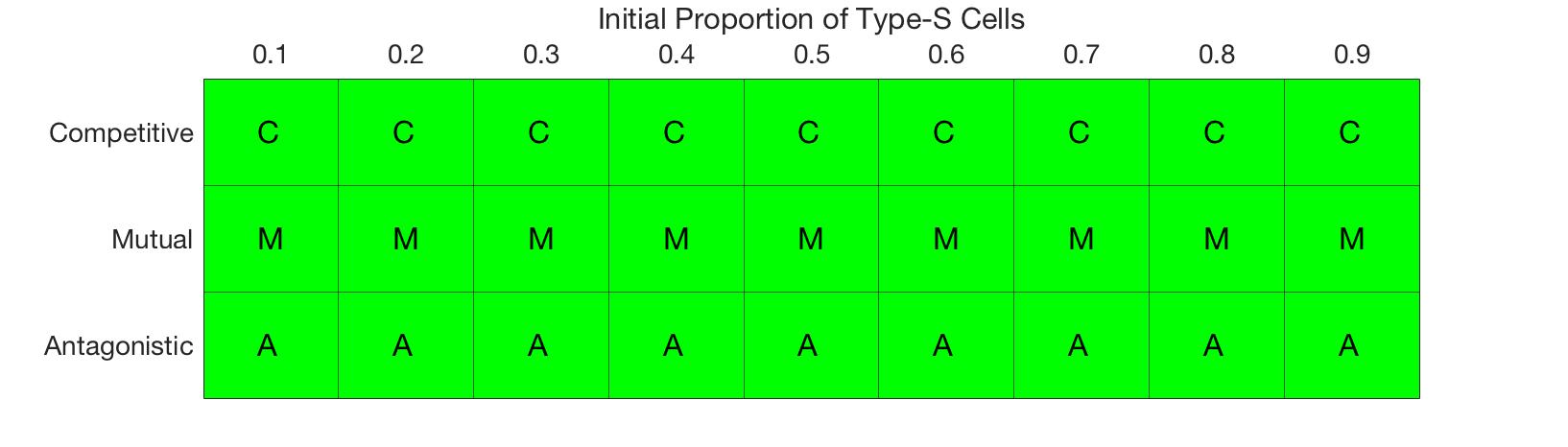} \\
\rotatebox{90}{$\qquad\quad\,\,\,$ \rotatebox{-90}{(b)}}
\includegraphics[width=5in]{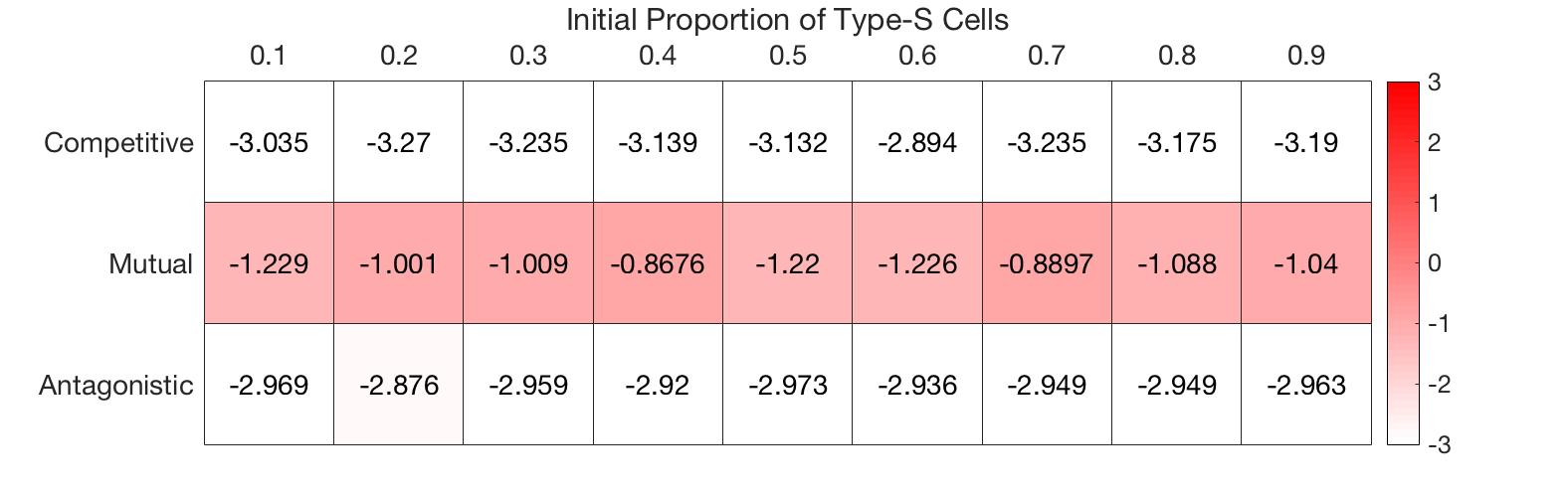} \\
\rotatebox{90}{$\qquad\quad\,\,\,$ \rotatebox{-90}{(c)}}
\includegraphics[width=5in]{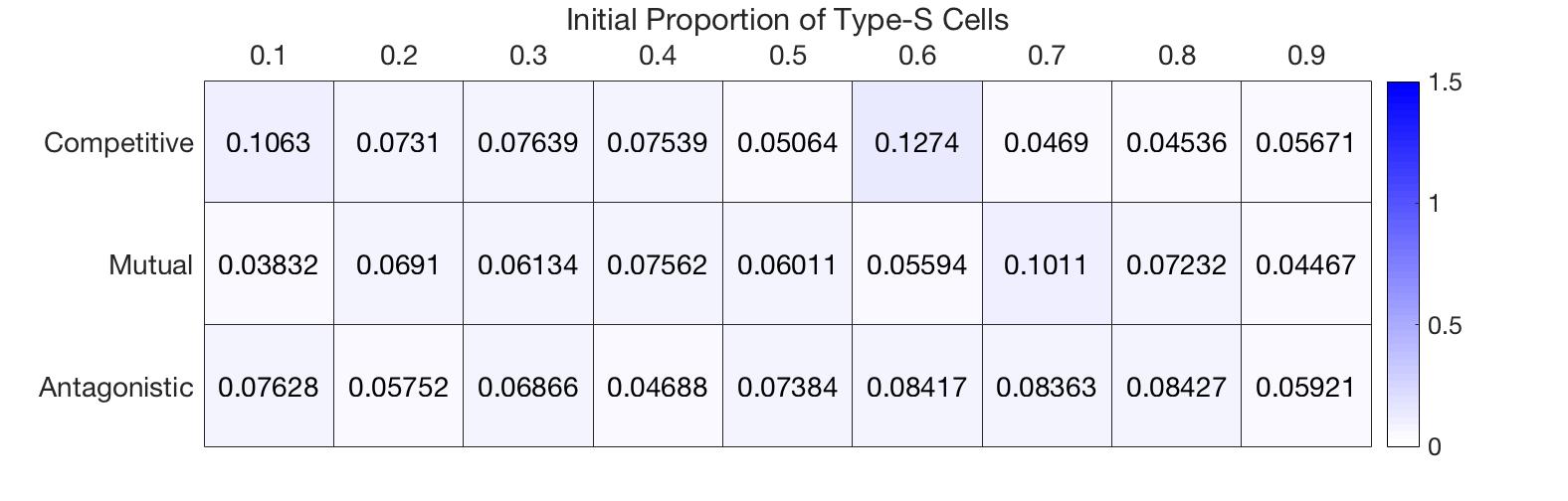}\\
\end{centering}
\caption{(Sequential Calibration) (a) Metric $E_1$ in Eq.~\eqref{eqn:E1}, stating the inferred interaction type (green indicates a match between the interaction type used to generate the data and that inferred by the model), (b) metric $E_2$ in Eq.~\eqref{eqn:E2}, measuring the log of sum of squares error across all nine data sets, and (c) metric $E_3$ in Eq.~\eqref{eqn:E3}, measuring the distance from the true parameters, for the competitive, mutual, and antagonistic cases with calibration using a single data set. }
\label{fig:sequentialLV_errors}
\end{figure}

\paragraph{Sequential Calibration}
The $E_1-E_3$ metrics for the sequential calibration procedure are reported in Figure \ref{fig:sequentialLV_errors}(a)-(c), respectively. As a reminder, these fits are generated using three data sets: a pure data set containing only Type-$S$ cells, a pure data set containing only Type-$R$ cells, and a mixture data set with a known value of the initial ratio of Type-$R$ to Type-$S$ cells (the $x$-axes in Figure \ref{fig:sequentialLV_errors} indicate the value of this initial ratio). In all cases, the interaction type is correctly inferred. We observe that while the errors in the model fits for the mutual data sets are comparable to those for the individual procedure, the errors for the competitive and antagonistic cases are markedly reduced. Further, the difference between the true and estimated parameter values are smaller for the sequential procedure than for the individual procedure across all three interaction types.

\begin{figure}[!htb]
\centerline{(a) \hspace{4.2cm} (b) \hspace{4.2cm}  (c)}
    \centering
    \begin{minipage}{0.325\textwidth}
    \includegraphics[width=\textwidth]{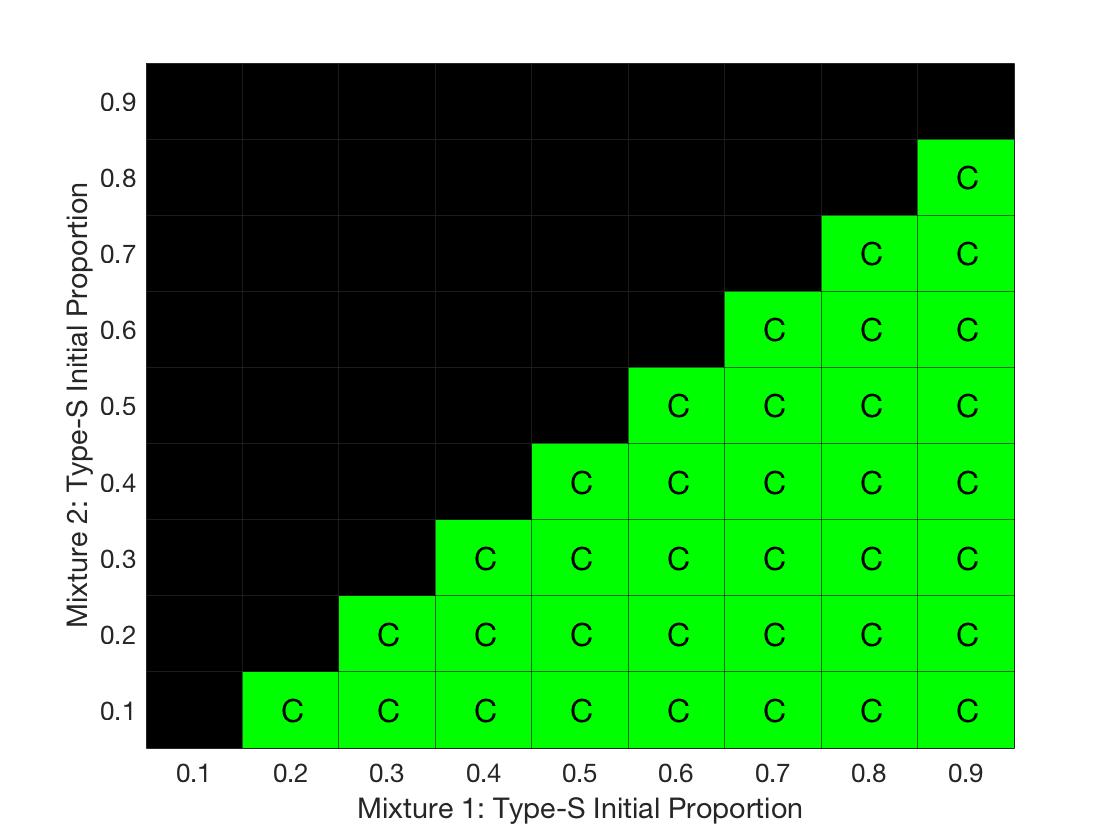}
    \end{minipage}
    \begin{minipage}{0.325\textwidth}
    \includegraphics[width=\textwidth]{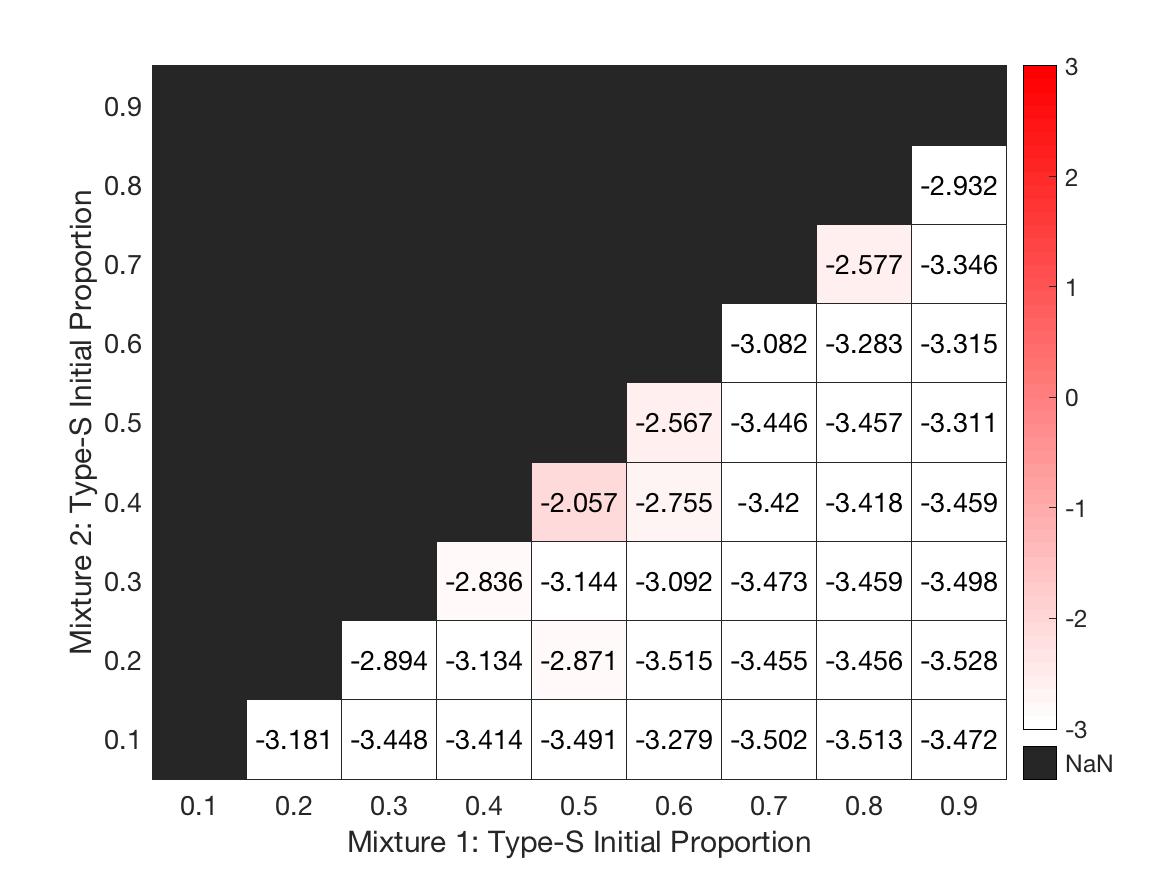}
    \end{minipage}
    \begin{minipage}{0.325\textwidth}
    \includegraphics[width=\textwidth]{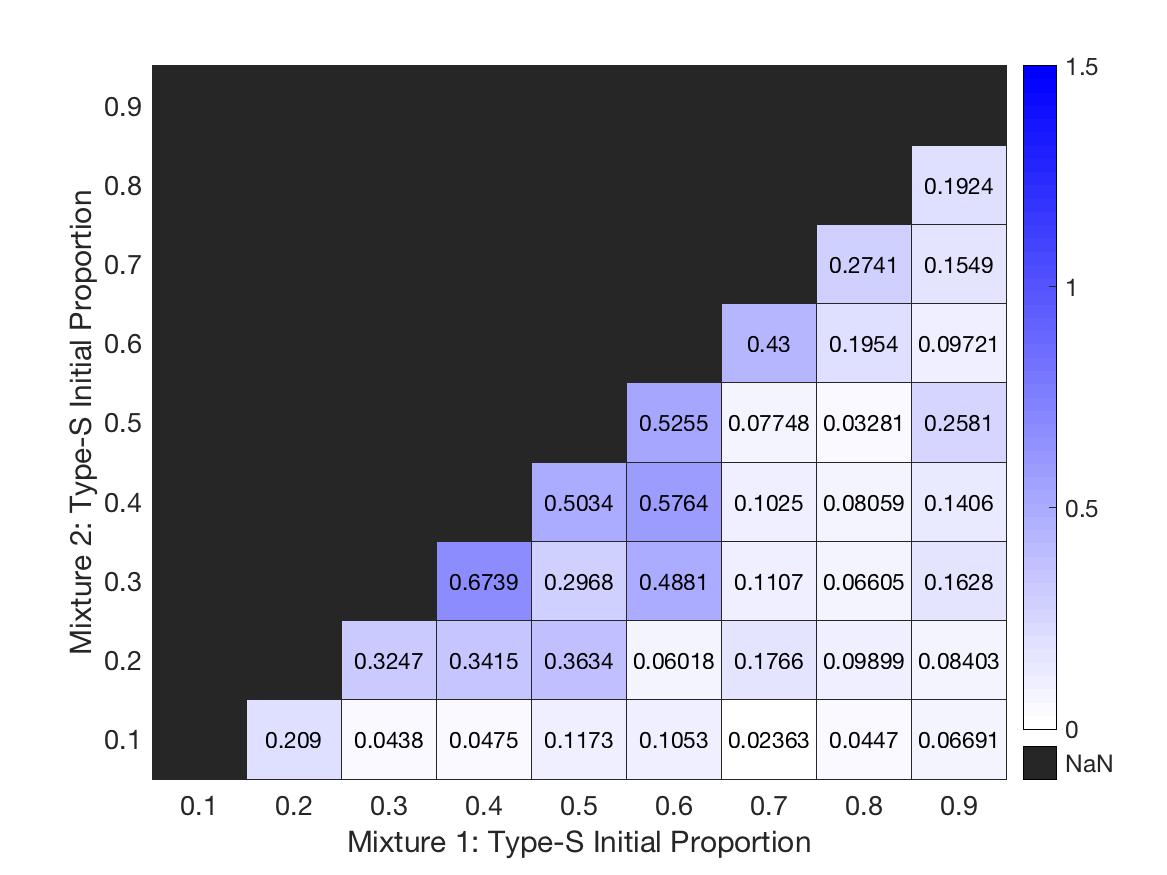}
    \end{minipage}\\ \ \\
    \begin{minipage}{0.325\textwidth}
    \includegraphics[width=\textwidth]{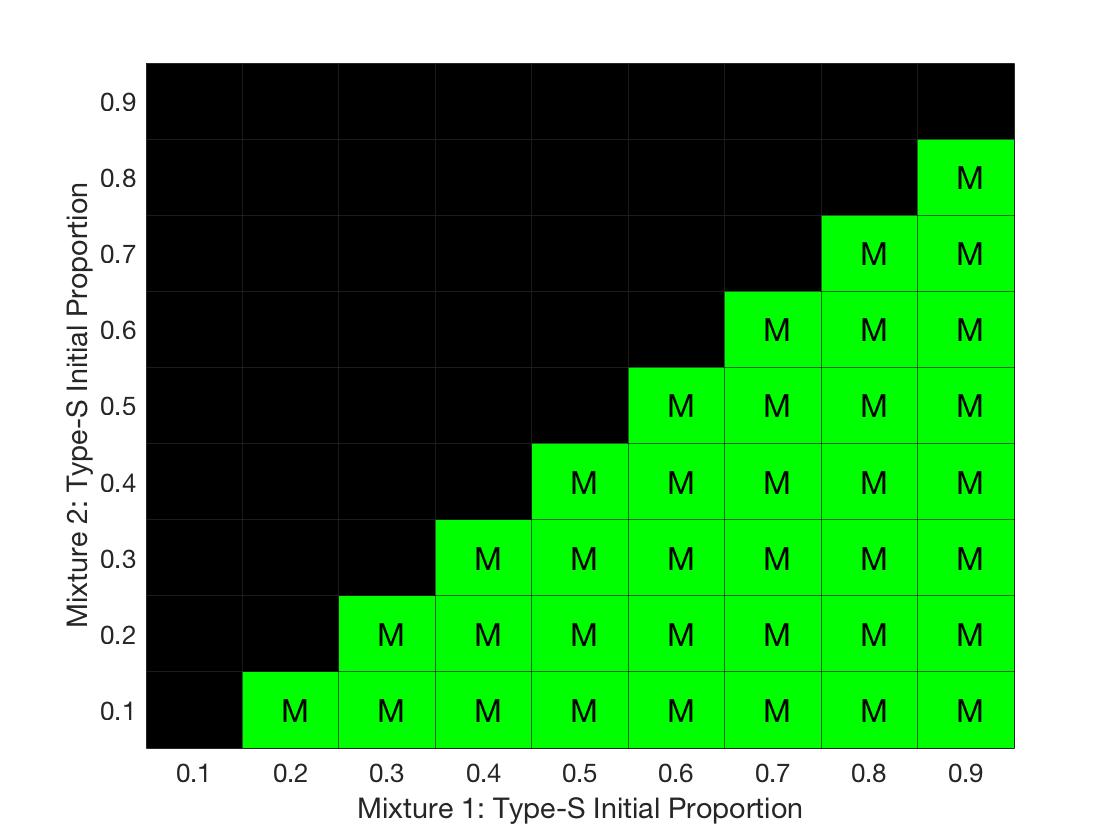}
    \end{minipage}
    \begin{minipage}{0.325\textwidth}
    \includegraphics[width=\textwidth]{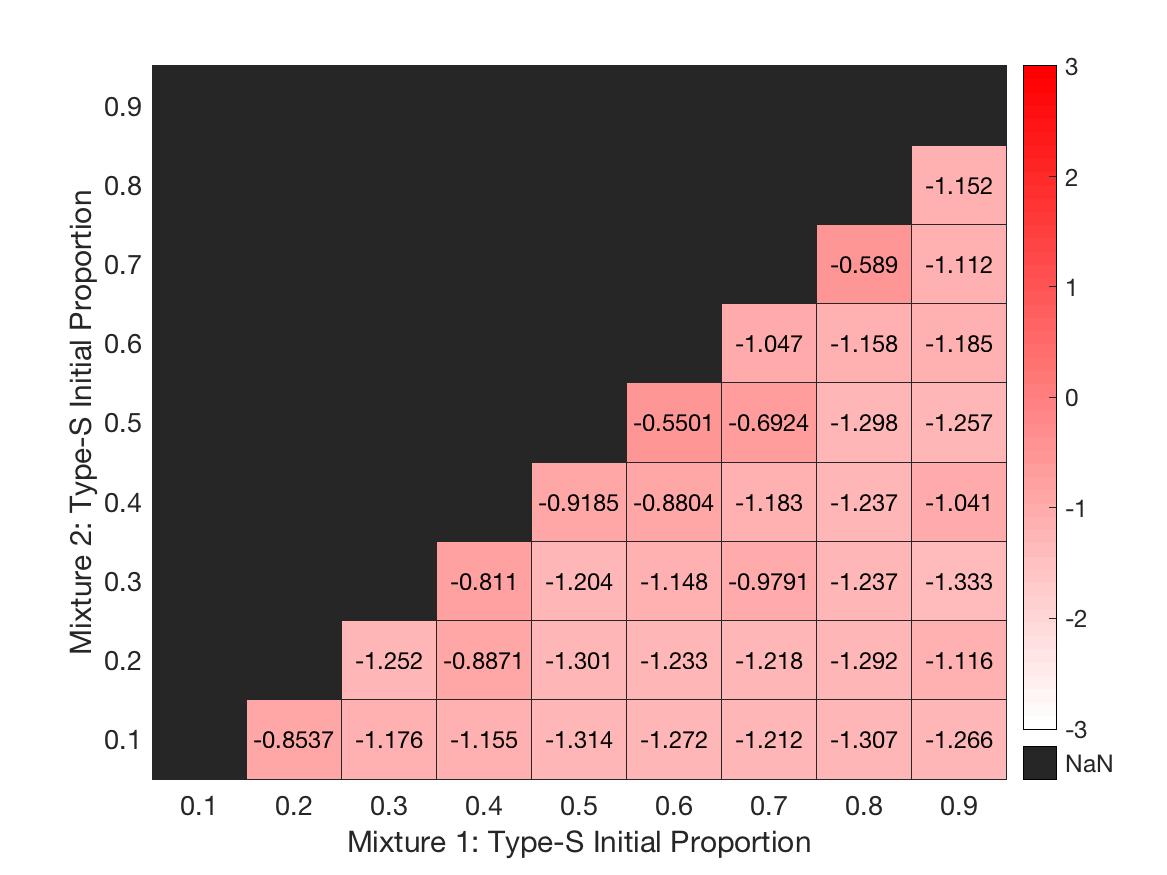}
    \end{minipage}
    \begin{minipage}{0.325\textwidth}
    \includegraphics[width=\textwidth]{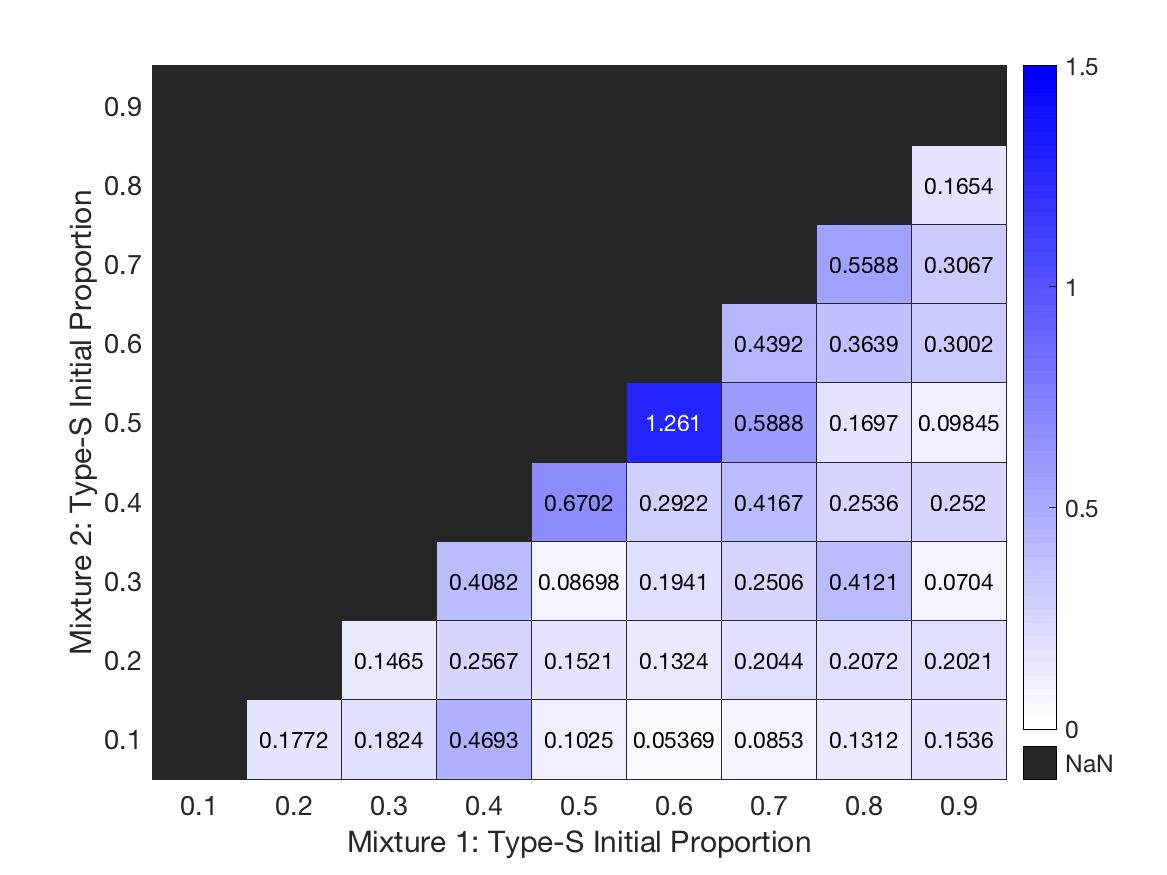}
    \end{minipage}\\ \ \\
    \begin{minipage}{0.325\textwidth}
    \includegraphics[width=\textwidth]{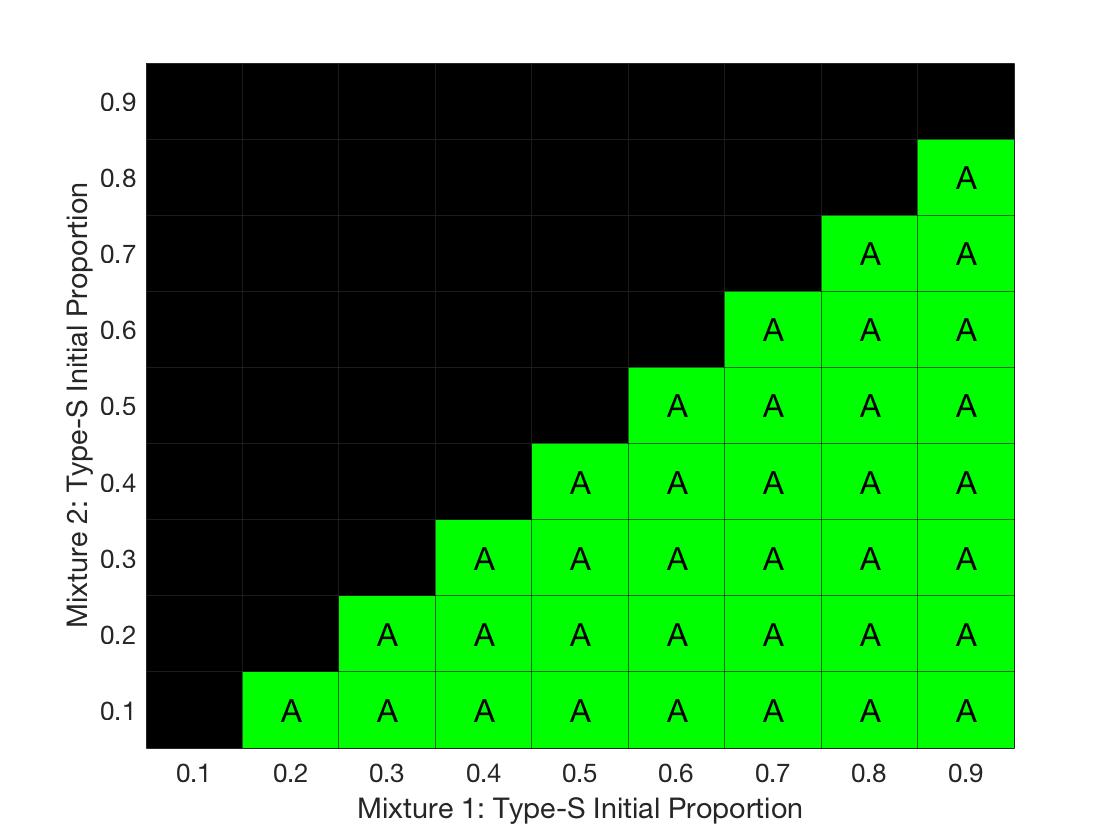}
    \end{minipage}
    \begin{minipage}{0.325\textwidth}
    \includegraphics[width=\textwidth]{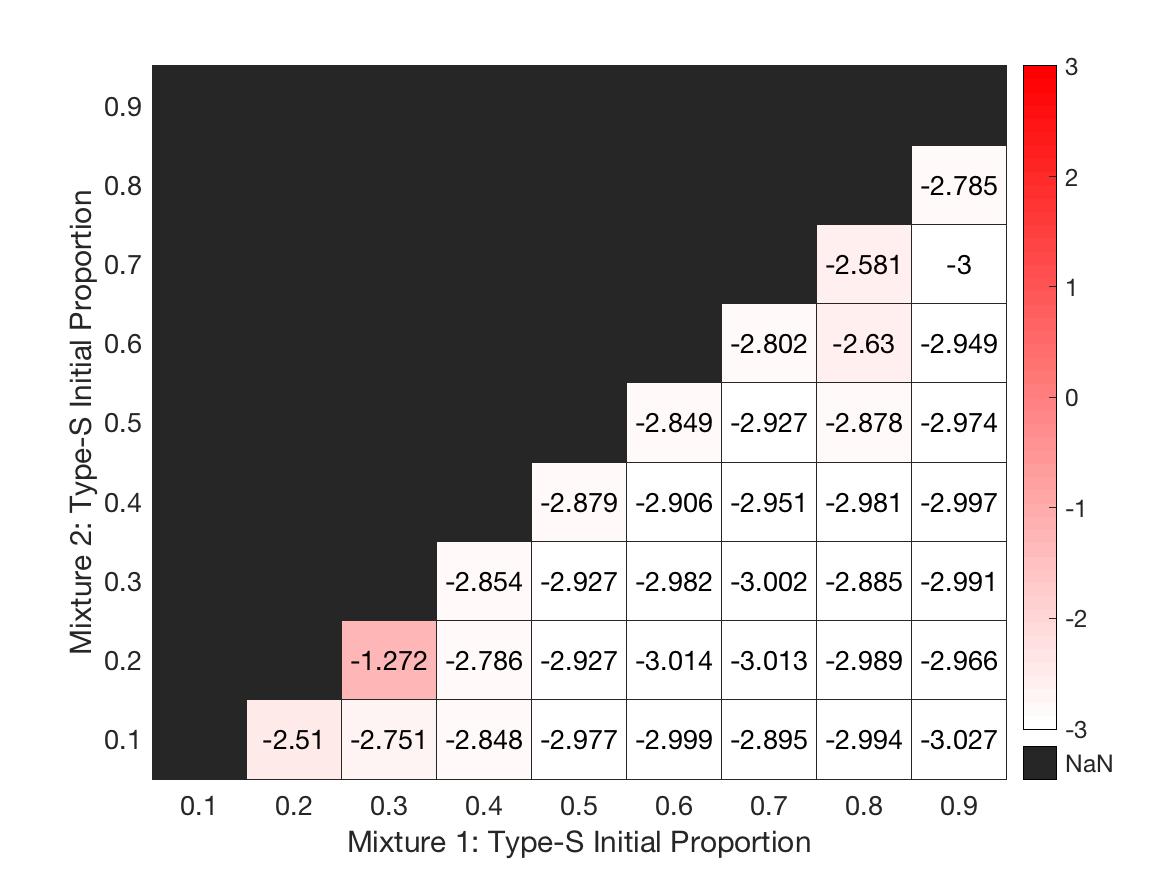}
    \end{minipage}
    \begin{minipage}{0.325\textwidth}
    \includegraphics[width=\textwidth]{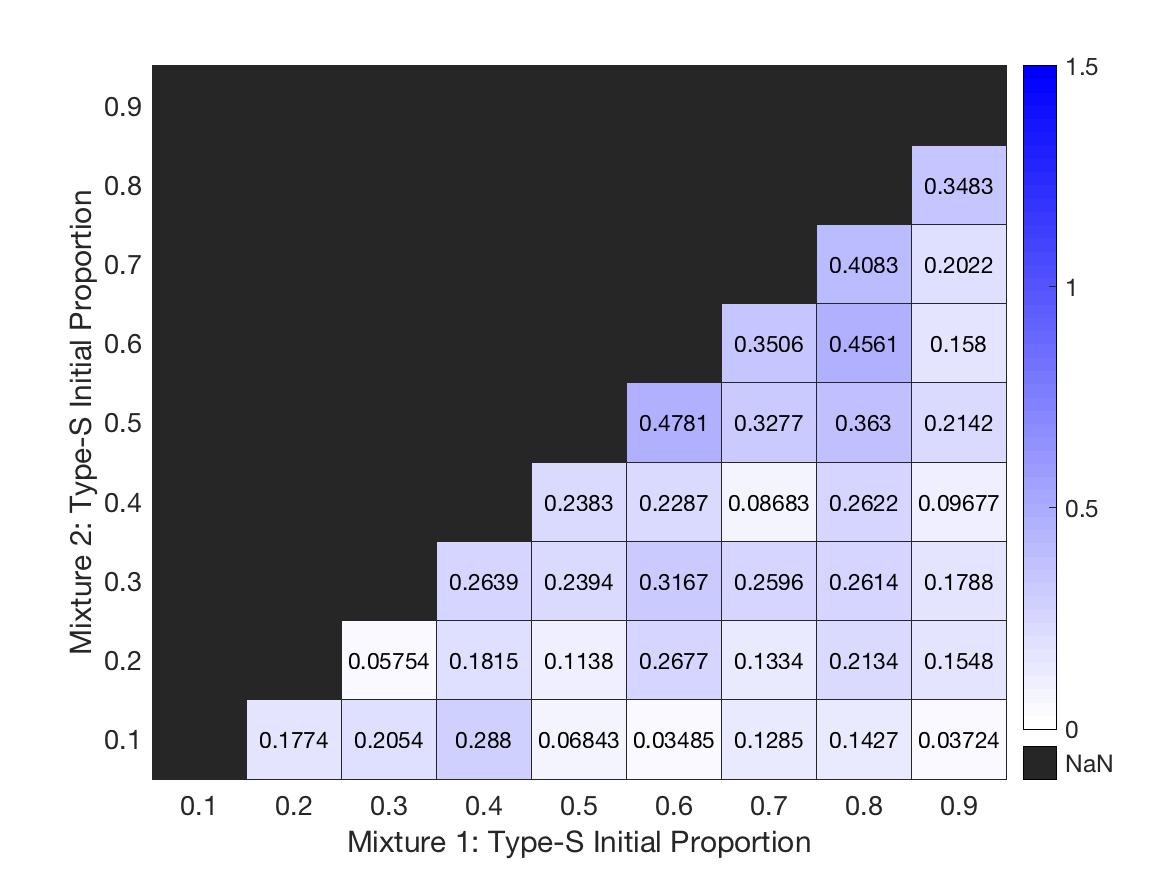}
    \end{minipage}\\ \ \\
    \caption{(Parallel Calibration) (a) Inferred interaction type (metric $E_1$ in Eq.~\eqref{eqn:E1}, left), (b) errors in model calibration across all nine data sets (metric $E_2$ in Eq.~\eqref{eqn:E2}, middle) and (c) distance from true parameter set (metric $E_3$ in Eq.~\eqref{eqn:E3}, right) for  competitive (top), mutual (middle), and antagonistic (bottom) interaction types.}
    \label{fig:errors_parallelLV}
\end{figure}

\paragraph{Parallel Calibration}
For the parallel scheme, all six parameters are estimated simultaneously using two different mixture data sets. The results for metrics $E_1-E_3$ are displayed in Figure \ref{fig:errors_parallelLV}(a)-(c) for all three interaction types. The $x$- and $y-$axes indicate the initial proportions of Type-$S$ cells employed in the two mixtures. As indicated by metric $E_1$, the inferred interaction matches the interaction type used to generate the data in all cases. In general, results for the $E_2$ and $E_3$ metrics are comparable to those from the sequential procedure, although certain mixture combinations yield better results than others.  In particular, without knowing \textit{a priori} the interaction type of the observed data, calibrating the model parameters using two experimental mixtures at the extremes---i.e., 1:9 and 9:1---would be a reasonable strategy for ensuring reasonable errors in model fits and estimated parameter values that are close to their true values.

\begin{figure}[!htb]
    \centering
    \includegraphics[width=\textwidth]{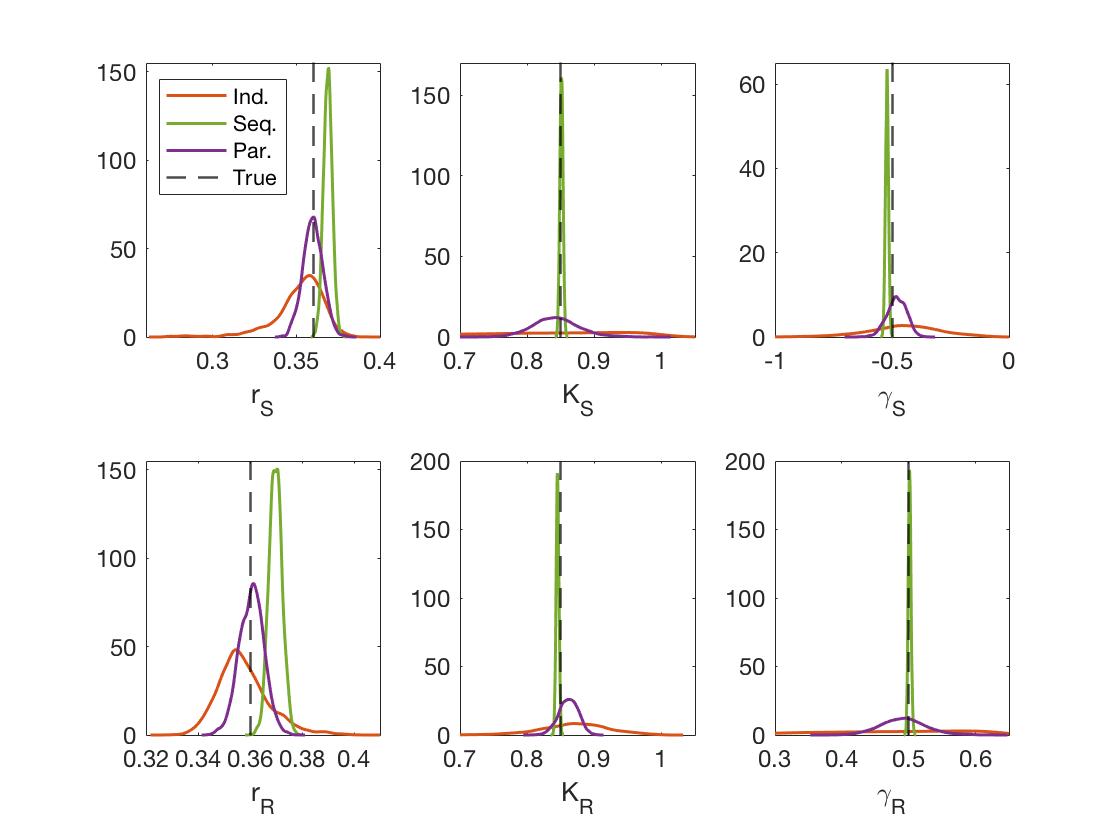}
    \caption{Comparison of posterior densities for three calibration methods (individual using 1:9, sequential using two pure data sets and 1:9, and parallel using 1:9 and 9:1 mixtures) for the antagonistic interaction type.  True values used to generate the data are depicted with a dashed black line.}
    \label{fig:compareposteriors}
\end{figure}

\paragraph{Comparison of Calibration Procedures}
To further compare the three calibration methods, we present sample parameter posterior densities in Figure \ref{fig:compareposteriors}, using the 1:9 ratio mixture for the individual calibration, the pure data sets and 1:9 ratio mixture for the sequential method, and the 1:9 and 9:1 mixtures for the parallel calibration.  We focus on the antagonistic case.  Our first observation is that the posterior densities from the individual calibration are highly uninformed by the data, particularly those for the carrying capacity and interaction parameters (further reflecting the trend noted in Figure \ref{fig:ind_pt5densities}). This is indicative of parameter non-identifiability, in the sense that parameters may not be uniquely inferred from the data. On the contrary, the sequential and parallel posterior densities are well-informed by the data, as illustrated by the narrower distributions and clear \textit{maximum a posteriori} (or MAP) point estimates. It is perhaps unsurprising that the posterior densities for the sequential procedure are the narrowest, since it employs three data sets whereas the parallel procedure uses only two. Further, the parameters are estimated two-at-a-time in the sequential procedure, reducing the total variability at any given step.

\begin{figure}[!bth]
    \centering
    \includegraphics[width=4in]{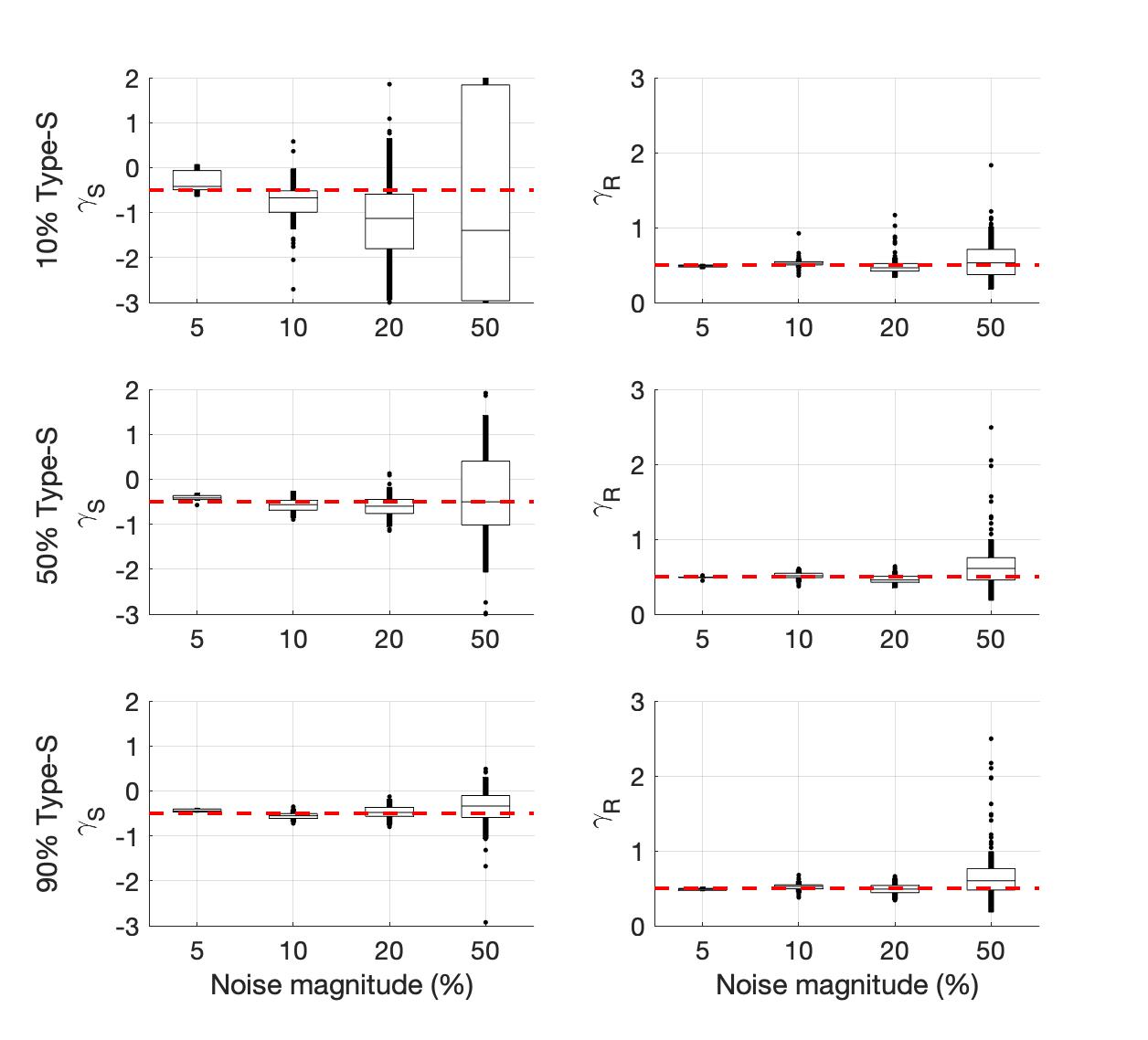} 
        \caption{(Sequential Calibration) Comparing fitted values of the interaction parameters $ \gamma_S$ and $\gamma_R$ when Type-$R$ cells antagonize Type-$S$ cells, using 100 sets each of synthetic data generated from the LV model with noise intensity levels of 5\%, 10\%, 20\%, and 50\%. The true values of $\gamma_S = -0.5$ and $\gamma_R=0.5$ are marked as red dashed lines. }
    \label{fig:odedata_param}
\end{figure}

Even so, the sequential procedure has drawbacks. While the sequential procedure ensures the practical identifiability of the LV model parameters when fitting to data generated from the LV model, the accuracy of inferred parameters may deteriorate when data is error-prone. In particular, using a homogeneous data set with large noise level can bias the intrinsic growth rates and carrying capacities for one or both cell lines; this bias must then be compensated for in the later inference of the interaction parameters. To illustrate this drawback, we fit the LV model to 100 sets of synthetic LV data for four different noise levels and the antagonistic interaction type. Figure \ref{fig:odedata_param} summarizes how the inferred values of the interaction parameters, $\hat \gamma_S$ and $\hat \gamma_R$, change as the noise level varies on the set $\{5\%, 10\%, 20\%, 50\%\}$ and the initial ratio of Type-$S$ to Type-$R$ cells varies on the set $\{$1:9, 5:5, 9:1$\}$. The accuracy of the estimated interaction parameter $\hat \gamma_S$ depends on the initial proportion of Type-$S$ cells. When the initial proportion of Type-$S$ cells is large (e.g., 90\%), the inferred interaction parameter values are more accurate and robust despite the noise, even for the largest noise level of 50\%. However, when the initial proportion of Type-$S$ cells is small (e.g., 10\%), the variance of the estimated $\hat \gamma_S$ is significantly larger and the accuracy deteriorates as the noise level increases. 
Thus, an experimental design using small initial proportions of Type-$S$ is not an appropriate design with which to accurately infer parameter values in cases where Type-$R$ cells antagonize Type-$S$ cells when performing a sequential calibration. When a large proportion of the antagonizing population (e.g. Type-$R$ in case of Type-$R$ antagonizing Type-$S$) is mixed with a small amount of the antagonized population, 
the antagonizing population tends to dominate quickly. The resulting dynamics are visually similar to those for a competitive interaction with small Type-$S$ initial proportion; thus, the inferred interaction type often manifests as competitive, with $\hat \gamma_S>0$. As the initial proportion of
Type-$S$ cells increases, the antagonizing effect of Type-$R$ is better recognized, since the Type-$R$ ultimately dominates despite its small initial proportion. This is a major drawback of sequentially calibrating the model, since the interaction type may be inconsistently inferred across different values of the initial ratio of the Type-$S$ and Type-$R$ cells if the measurement noise is large.

\begin{figure}[!htb]
    \centering
    \includegraphics[width=4in]{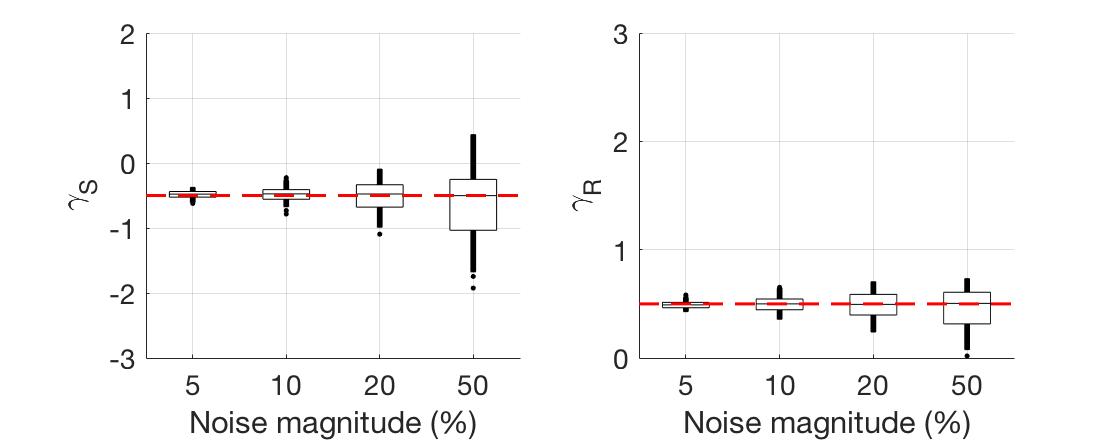}
        \caption{(Parallel Calibration) Comparing fitted values of the interaction parameters $\gamma_S$ and $\gamma_R$ when Type-$R$ cells antagonize Type-$S$ cells, using 100 sets each of synthetic data generated from the LV model with 1:9 and 9:1 mixtures and noise intensity levels of 5\%, 10\%, 20\%, and 50\%. The true values of $\gamma_S = -0.5$ and $\gamma_R=0.5$ are marked as red dashed lines. } 
    \label{fig:odedata_paramParallel}
\end{figure}

In contrast, the parallel calibration procedure is less sensitive to measurement noise, since all parameters are estimated simultaneously and two initial mixture ratios are employed. Interaction parameter estimates for the parallel model fits to 100 sets of synthetic LV data with initial ratios 1:9 and 9:1 are shown in Figure \ref{fig:odedata_paramParallel}. For the first three noise levels---5\%, 10\%, and 20\%---all parameter estimates correctly predict an antagonistic relationship. Only for the largest noise levels, 50\%, do we observe positive estimates for $\gamma_S$ in nine of the 100 cases, so that the interaction manifests as competitive according to the LV model definition. The remaining 91 interaction types are inferred as antagonistic. The increased consistency in the inferred interaction type is a benefit of using the parallel procedure. Additionally, when comparing the sequential results in Figure \ref{fig:odedata_param} to the parallel results in Figure \ref{fig:odedata_paramParallel}, we observe that the variability in the parameter estimates across the 100 fits is much smaller for the parallel calibration, since compensation for the bias in the error-prone data can be absorbed across all six parameters instead of isolated to the interaction parameters alone. This phenomenon can also be observed in Figure \ref{fig:compareposteriors}, where the point estimates associated with the parallel procedure are closer to the true parameter values than those for the sequential method ($E_3 = 0.03724$ for the parallel procedure using 1:9 and 9:1 mixtures, and $E_3 = 0.07628$ for the sequential method using the 1:9 mixture).

Thus, the choice of optimal method depends on the question of interest. If the goal is to minimize the uncertainty in the parameter estimates, a sequential experimental design may be preferable due to the narrower posterior densities obtained from the use of three data sets, although biases in the values of parameters estimated early in the sequential design may limit its efficacy depending on the chosen initial ratio. If minimizing experimental costs is important, then the parallel procedure is preferred, as it requires two data sets rather than three.  Given that obtaining clinical or \textit{in vitro} data for tumor growth measurements can be technically challenging and expensive, and given that the parallel procedure is more robust to noise and provides point estimates which are closest to the true parameter values, we focus on this method for the rest of the investigation.

\subsection{Practical Identifiability Using Synthetic Cellular Automaton Data} \label{sec:CAresults}

As the parallel procedure was chosen as the preferred calibration methodology in Section \ref{sec:practicaltoLV}, henceforth we focus on results for the parallel calibration to the CA data. Results of the sequential calibration to synthetic CA data are included in \ref{sec:app:CAseqresults} for comparison. 

\begin{figure}[!bth]
\centering 
\includegraphics[width=0.4\textwidth]{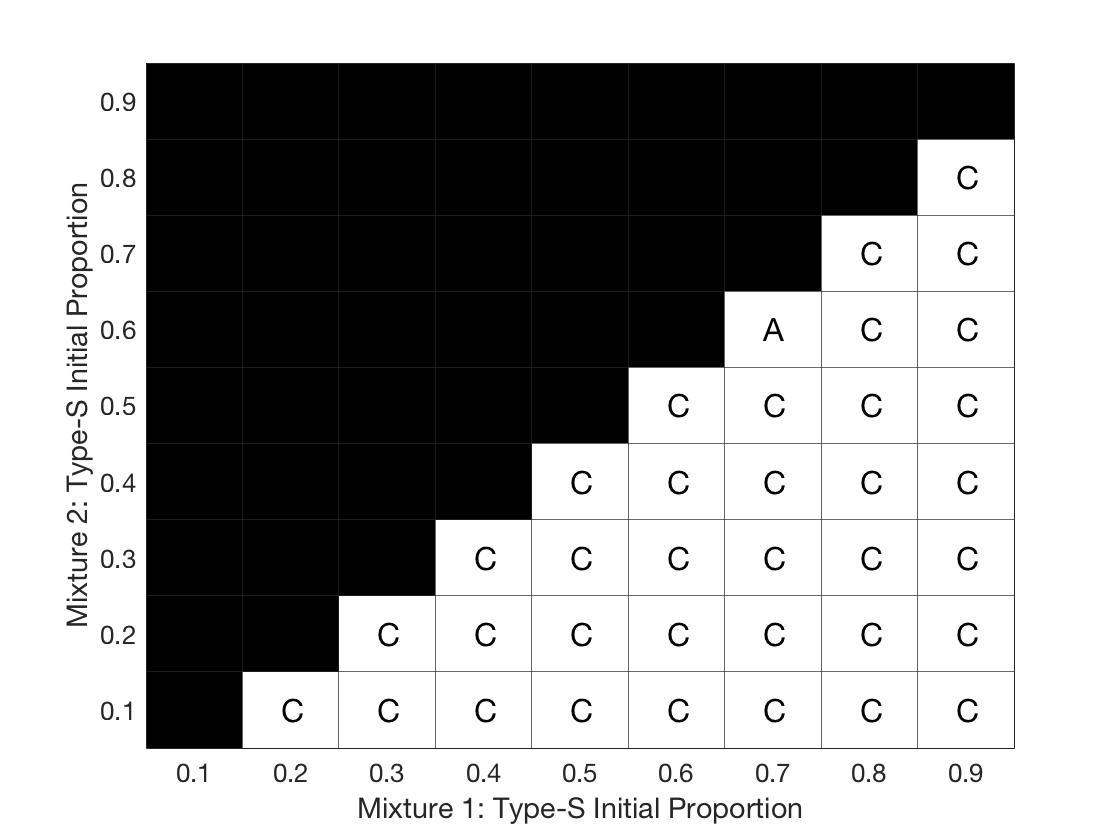}
\includegraphics[width=0.4\textwidth]{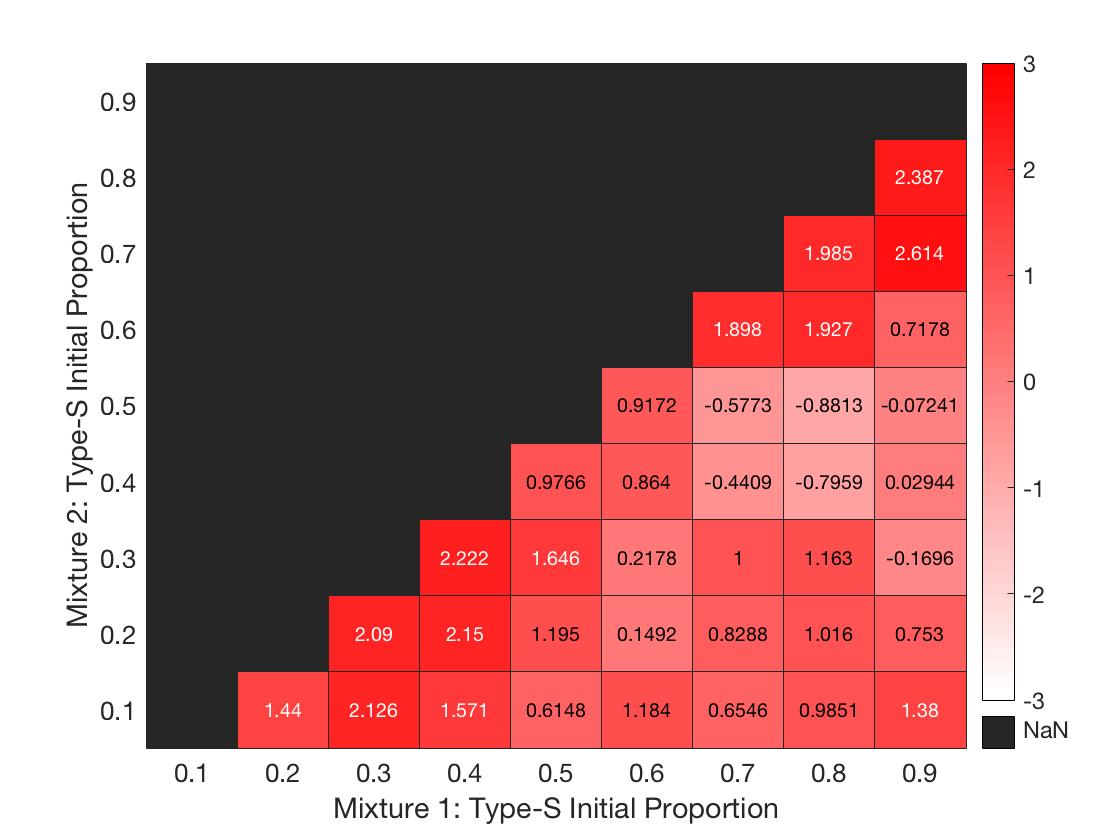}
\begin{minipage}{0.3\textwidth} \begin{center}(a)\end{center} \end{minipage}
\begin{minipage}{0.3\textwidth} \begin{center}(b)\end{center} \end{minipage}
\caption{(Parallel Calibration) (a) Metric $E_1$ in Eq.~\eqref{eqn:E1} and (b) metric $E_2$ in Eq.~\eqref{eqn:E2} for the calibration of the Lotka-Volterra model to synthetic CA data for the neutral interaction type. }
\label{fig:parallelCA_E1neutral}
\end{figure}

We begin by fitting the LV model to CA simulations generated using a neutral interaction, as described in Section \ref{subsec:CAmodel} and \ref{sec:app:CA}. The results for the $E_1$ and $E_2$ metrics are shown in Figure \ref{fig:parallelCA_E1neutral}. Notably, nearly all LV model fits infer a competitive interaction type, with the exception of one inferred antagonistic relationship. This discrepancy is a drawback of employing a spatially-averaged model (Lotka-Volterra) to fit data generated from a spatially-resolved model (cellular automaton).  In the spatially-resolved CA model, any imposed interaction mechanism is enforced atop a baseline competition for space and resources (e.g., oxygen and nutrients), which cannot be separately accounted for in the spatially-averaged LV model defined in Equations \eqref{eqn:dSdt}-\eqref{eqn:dRdt}. Thus, caution is needed when interpreting the results of the interaction inference for the remainder of this section. 

\begin{figure}[!bth]
\centering 
\includegraphics[width=0.32\textwidth]{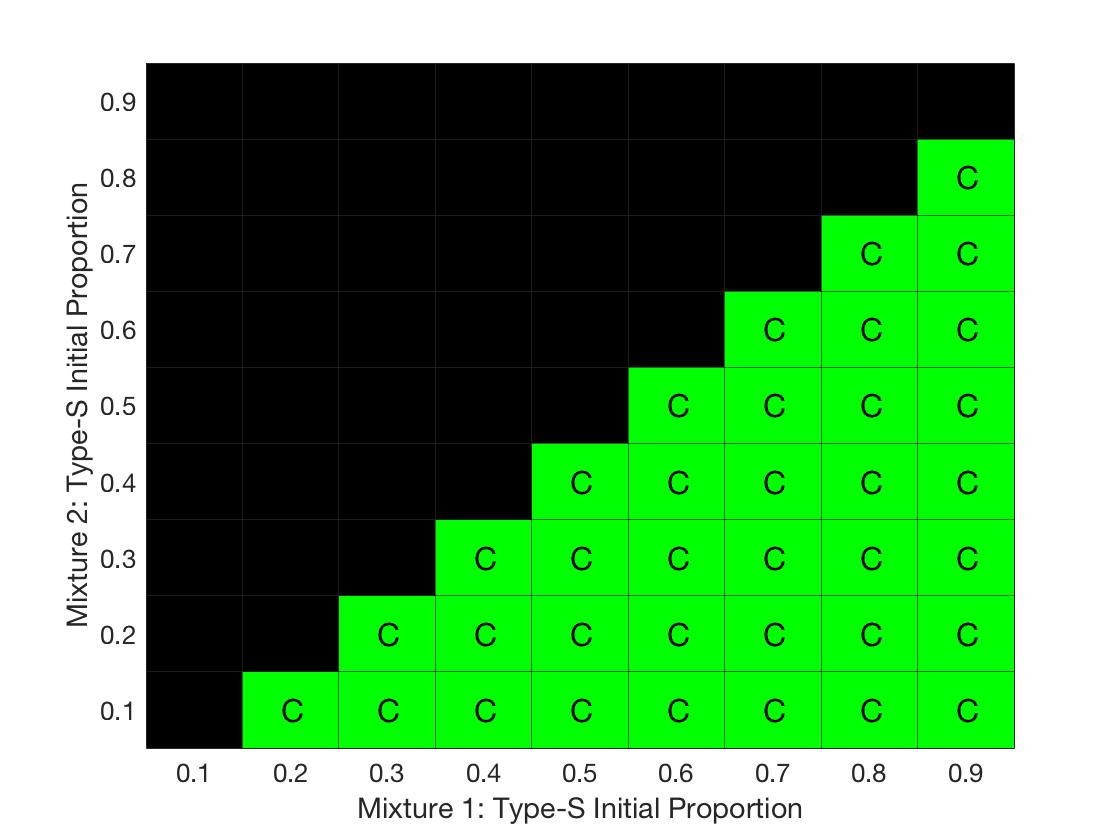}
\includegraphics[width=0.32\textwidth]{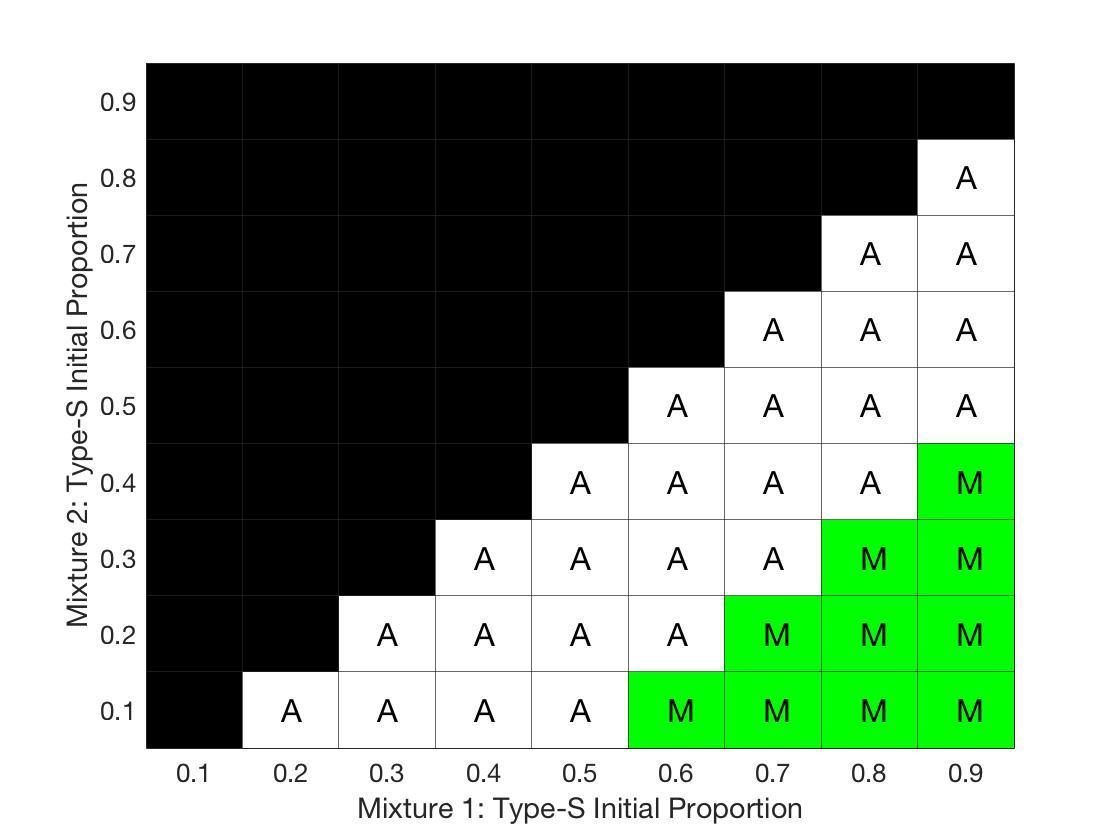}
\includegraphics[width=0.32\textwidth]{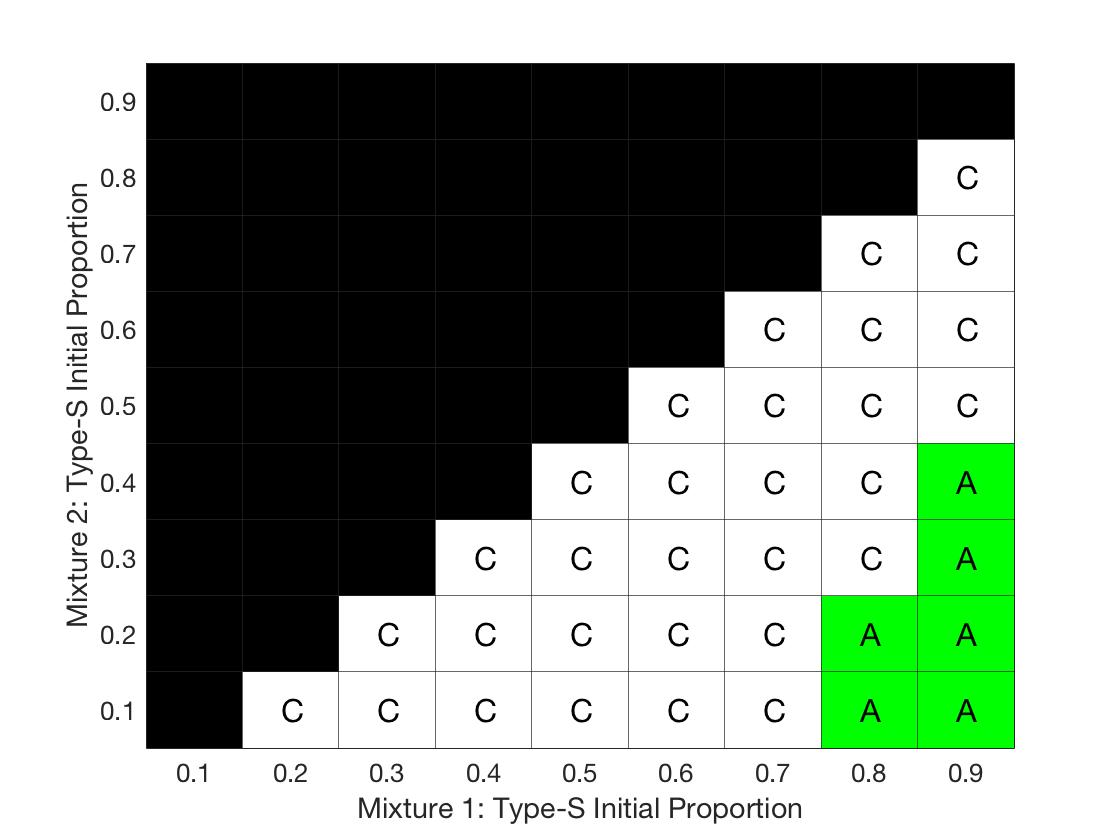}
\begin{minipage}{0.3\textwidth} \begin{center}(a)\end{center} \end{minipage}
\begin{minipage}{0.3\textwidth} \begin{center}(b)\end{center} \end{minipage}
\begin{minipage}{0.3\textwidth} \begin{center}(c)\end{center} \end{minipage}\\
\caption{(Parallel Calibration) Metric $E_1$ in Eq.~\eqref{eqn:E1} for the calibration of the Lotka-Volterra model to synthetic CA data for (a) competitive, (b) mutual, and (c) antagonistic interaction types, using CA intensity parameter $I=4$. Green indicates mixture combinations for which inferred interaction type in the LV model matches the interaction type used to generate the CA data.}
\label{fig:parallelCA_E1}
\end{figure}

Metric $E_1$ for the competitive, mutual, and antagonistic interaction types is illustrated in Figure \ref{fig:parallelCA_E1}. Recall, when calibrating with the parallel procedure using data generated from the LV model, the correct interaction type was inferred in all cases. Here, with calibration to the
CA data, the signs of the inferred interaction parameters in the LV model may not match those in the CA model, as expected given the results for the neutral case presented in Figure \ref{fig:parallelCA_E1neutral}. Although  all competitive CA simulations are represented by competitive interactions in the LV model fits, such a match is only observed for mutual and antagonistic simulations if the initial proportions of the two mixtures are at the far extremes (e.g., 1:9 and 9:1). Further, CA simulations with mutual interactions are often represented by antagonistic interactions when fit to the LV model. Similarly, CA simulations of antagonistic cases are often represented by competitive interactions in the LV fits.  

\begin{figure}[!bth]
\centerline{
\includegraphics[width=0.45\textwidth]{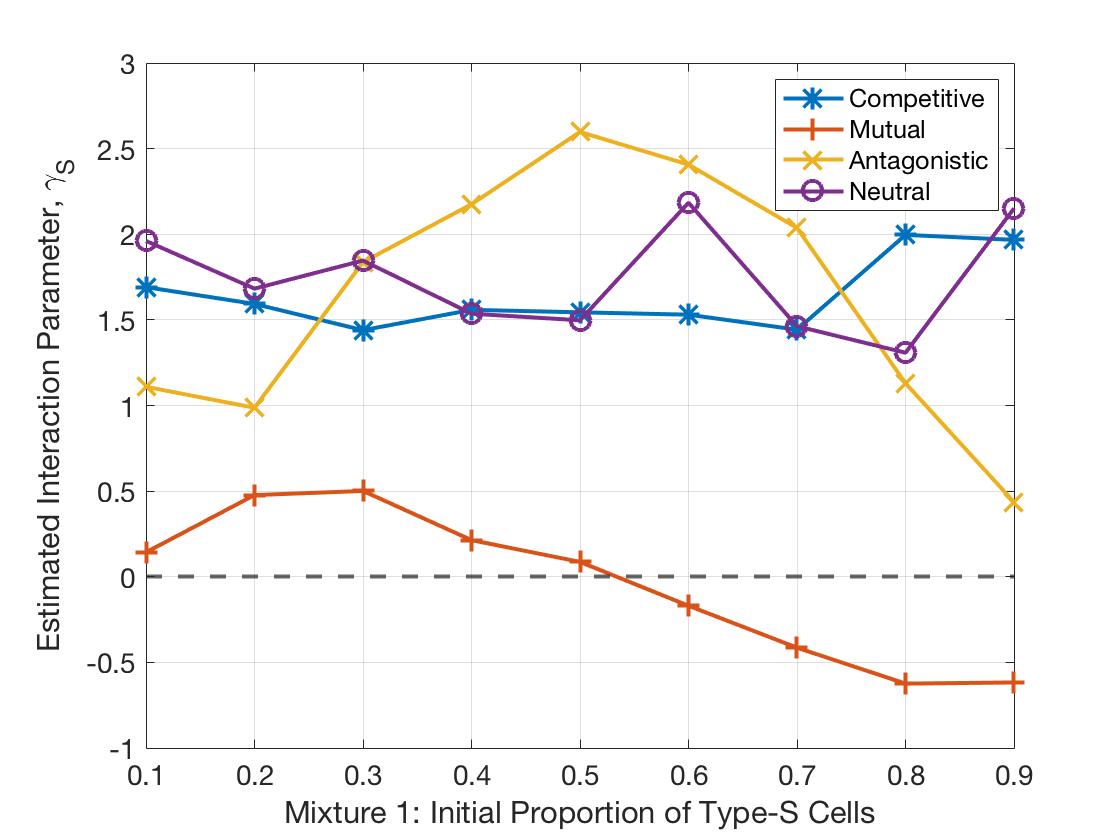}
\includegraphics[width=0.45\textwidth]{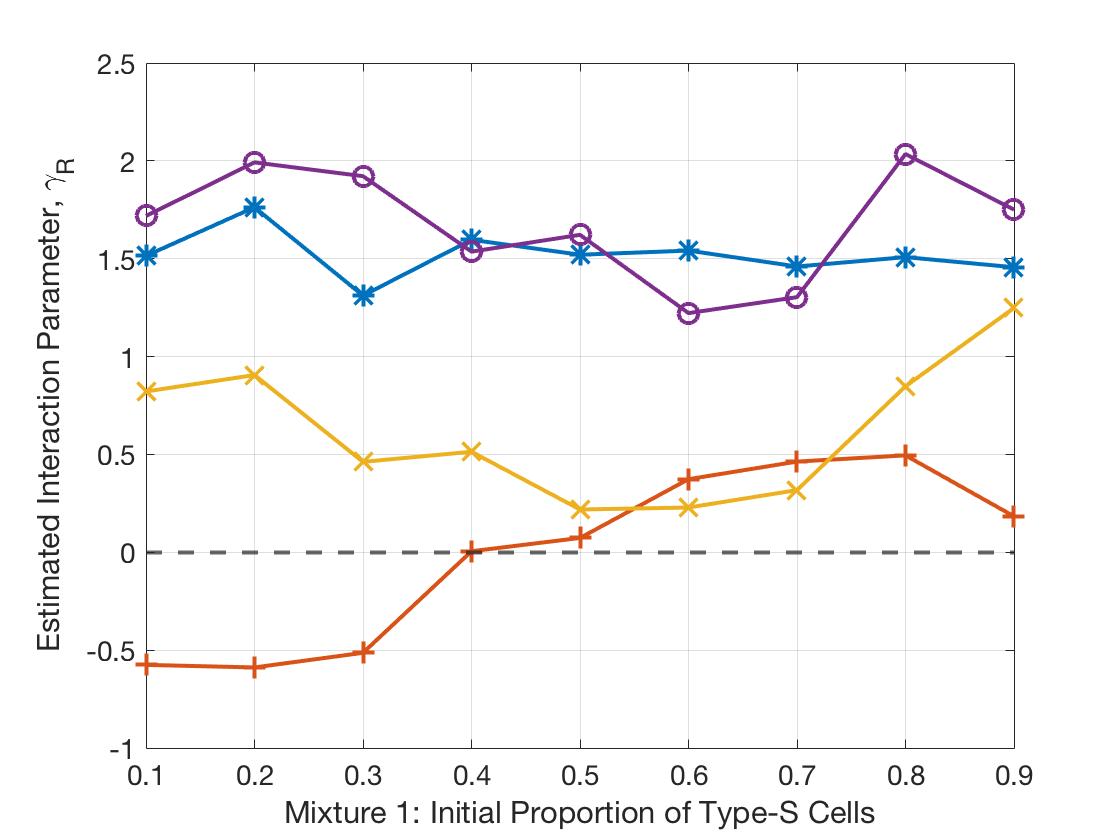}
}
\centerline{(a) \hspace{200pt} (b)}
\caption{(Parallel Calibration) Interaction parameter estimates for (a) $\gamma_S$ and (b) $\gamma_R$ according to first mixture Type-$S$:Type-$R$ ratio and interaction type, using CA intensity parameter $I=4$. Each data point represents the average interaction parameter over the eight remaining second mixture ratios.}
\label{fig:paramcomparison}
\end{figure}

To further quantify the bias toward a competitive interaction in the LV model, we present the estimated values of the interaction parameters for the four interaction types in Figure \ref{fig:paramcomparison}, demonstrating their dependence on both the interaction type imposed in the CA and the first mixture ratio employed.  For each first mixture ratio choice, we average the parameter estimates for the eight remaining second mixture ratios for easier visualization. Consistent with Figure \ref{fig:parallelCA_E1neutral}, the parameter estimates for $\gamma_S$ and $\gamma_R$ in the neutral case tend to positive values, such that the LV model infers a competitive interaction; in fact, the estimated parameter values for the neutral and competitive cases are similar in magnitude.  Additionally, it is rare for both interaction parameters in LV model fits to CA simulations with mutual interactions to be less than zero simultaneously, as was expected in our earlier idealistic calibration to spatially-averaged data. Thus, LV model fits to mutual CA simulations often infer antagonistic interactions. In general, applying a spatially-averaged model to fit spatially-resolved data can lead to a misinterpretation of the type of interaction present, as one is unable to distinguish between the effects of the interaction mechanism and the effects of the baseline competition for space and resources; both effects are absorbed into the values of the interaction parameters $\gamma_S$ and $\gamma_R$, and cannot be separately identified. 

\begin{figure}[!bth]
    \centering
    \begin{minipage}{0.3\textwidth} \hspace{30pt}\textbf{Intensity $I=3$} \end{minipage}t
    \begin{minipage}{0.3\textwidth} \hspace{35pt}\textbf{Intensity $I=4$} \end{minipage}
    \begin{minipage}{0.3\textwidth} \hspace{33pt}\textbf{Intensity $I=5$} \end{minipage}\\ 
    \rotatebox{0}{(a)}
    \begin{minipage}{0.3\textwidth}  \includegraphics[width=1.9in]{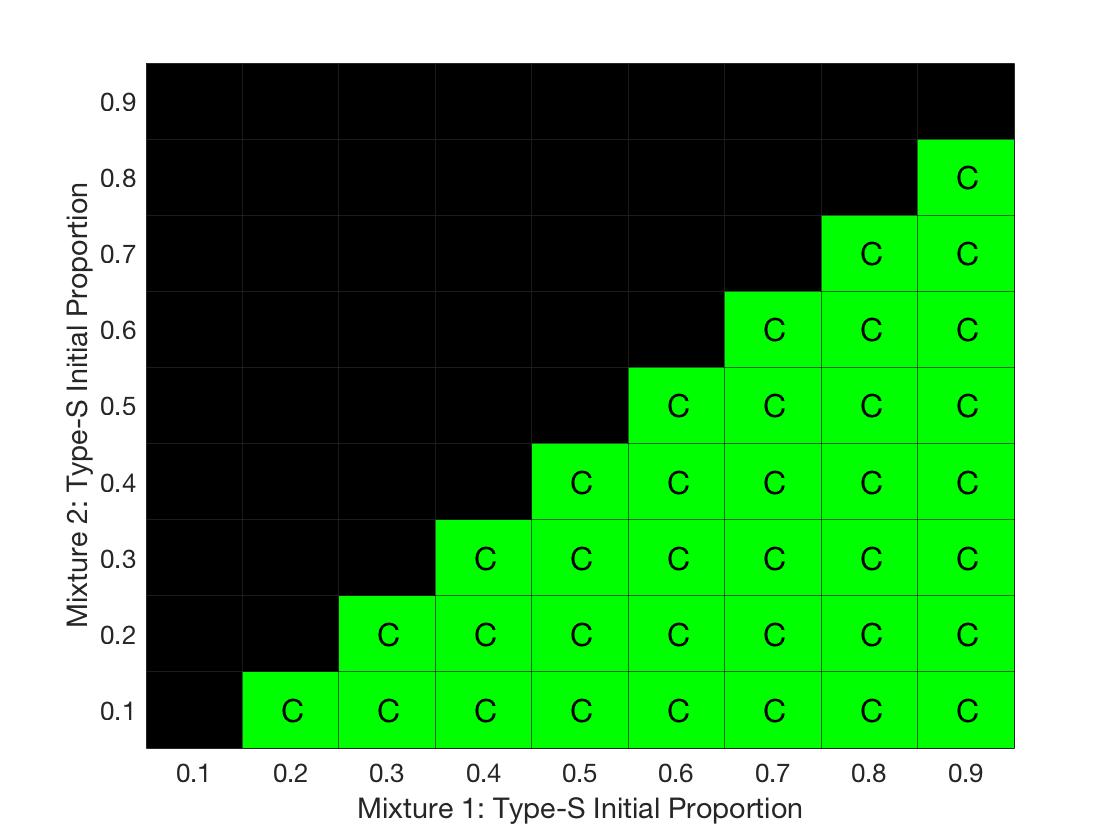}\end{minipage}
    \begin{minipage}{0.3\textwidth}
    \includegraphics[width=1.9in]{Fig/TwoMixtures/Int4CompetitiveMean_InteractionSignHeatMap.jpg}
    \end{minipage}
    \begin{minipage}{0.3\textwidth} \includegraphics[width=1.9in]{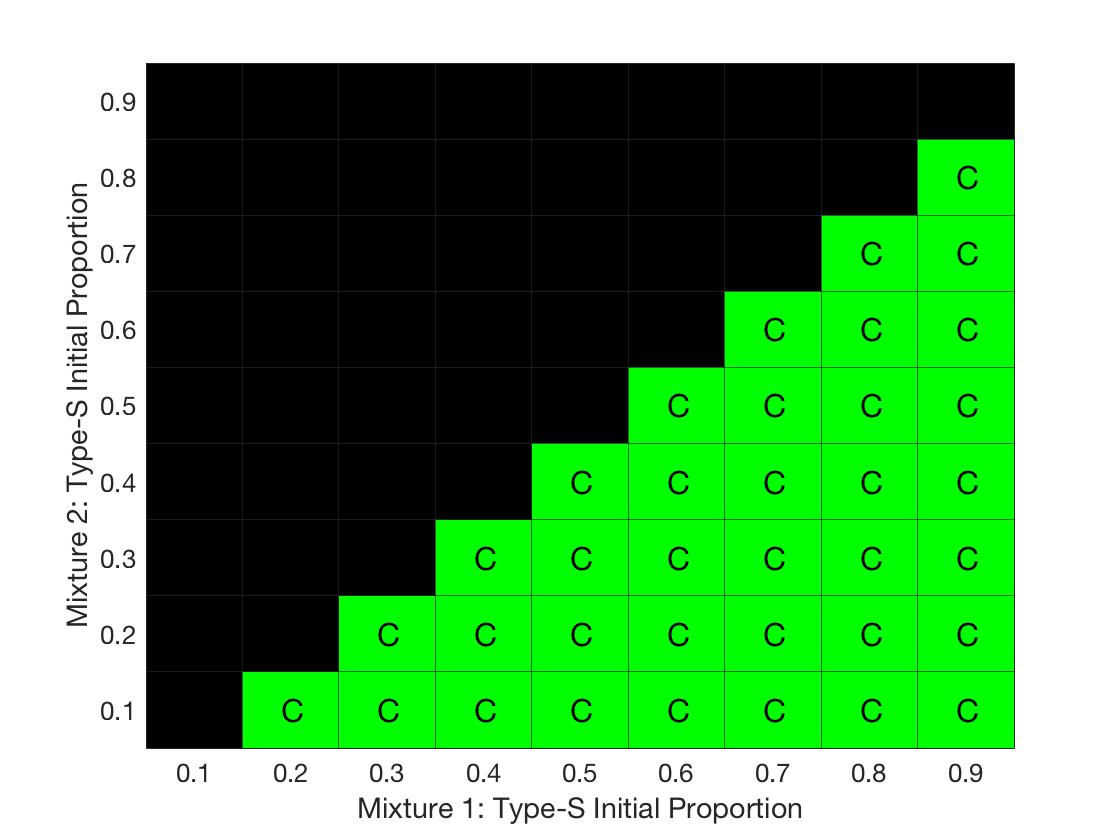}\end{minipage}\\
    \rotatebox{0}{(b)}
    \begin{minipage}{0.3\textwidth} \includegraphics[width=1.9in]{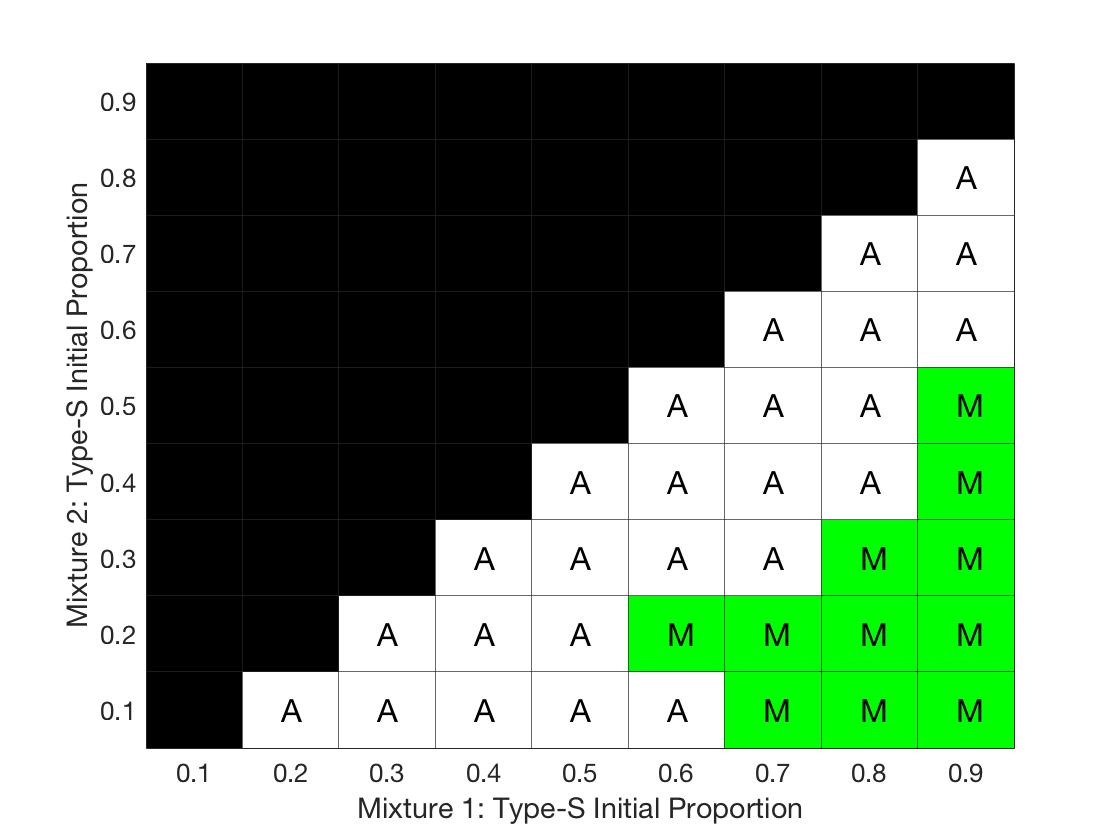}\end{minipage}
    \begin{minipage}{0.3\textwidth}
    \includegraphics[width=1.9in]{Fig/TwoMixtures/Int4MutualMean_InteractionSignHeatMap.jpg}
    \end{minipage}
    \begin{minipage}{0.3\textwidth} \includegraphics[width=1.9in]{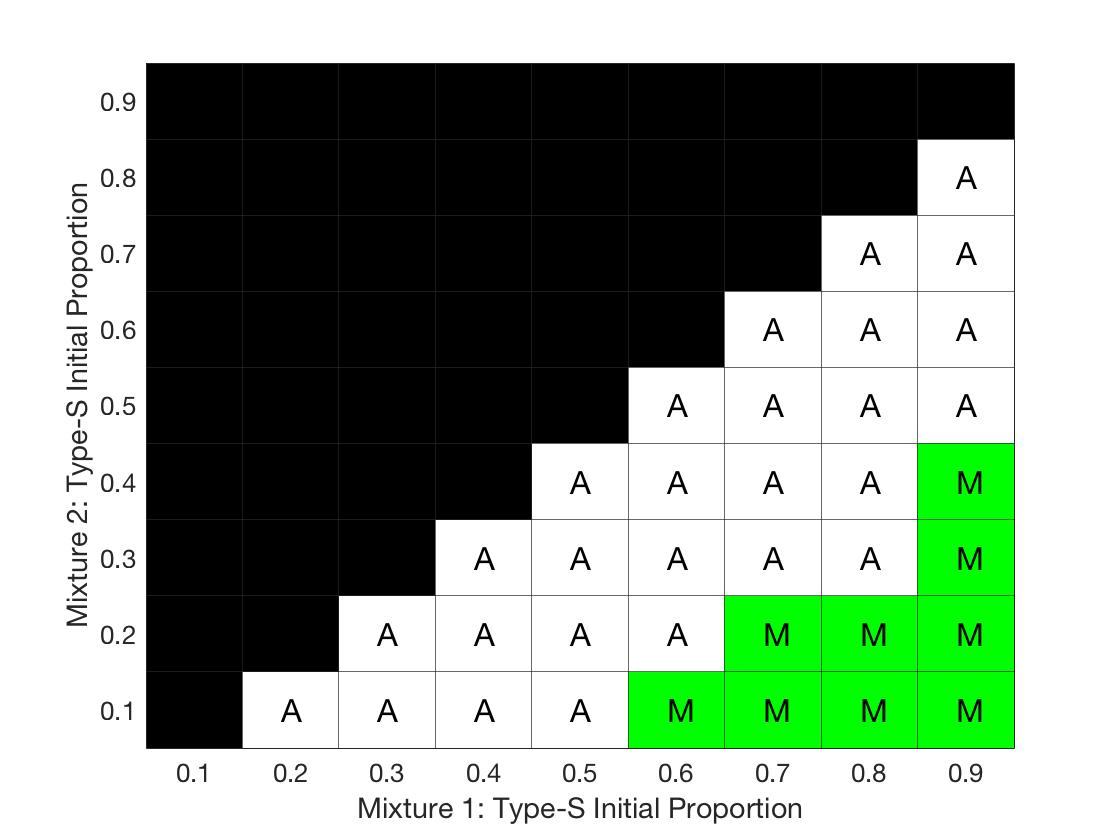}\end{minipage}\\
    \rotatebox{0}{(c)}
    \begin{minipage}{0.3\textwidth} \includegraphics[width=1.9in]{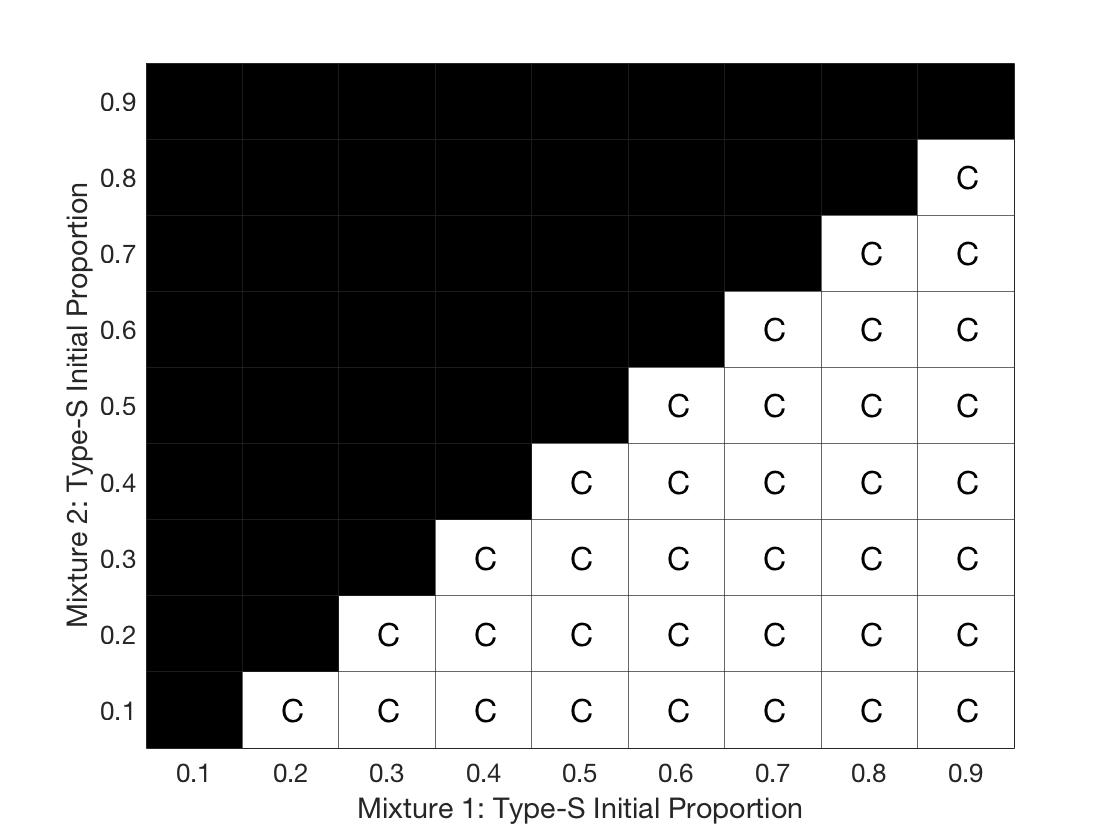}\end{minipage}
    \begin{minipage}{0.3\textwidth}
    \includegraphics[width=1.9in]{Fig/TwoMixtures/Int4RantagCMean_InteractionSignHeatMap.jpg}
    \end{minipage}
    \begin{minipage}{0.3\textwidth} \includegraphics[width=1.9in]{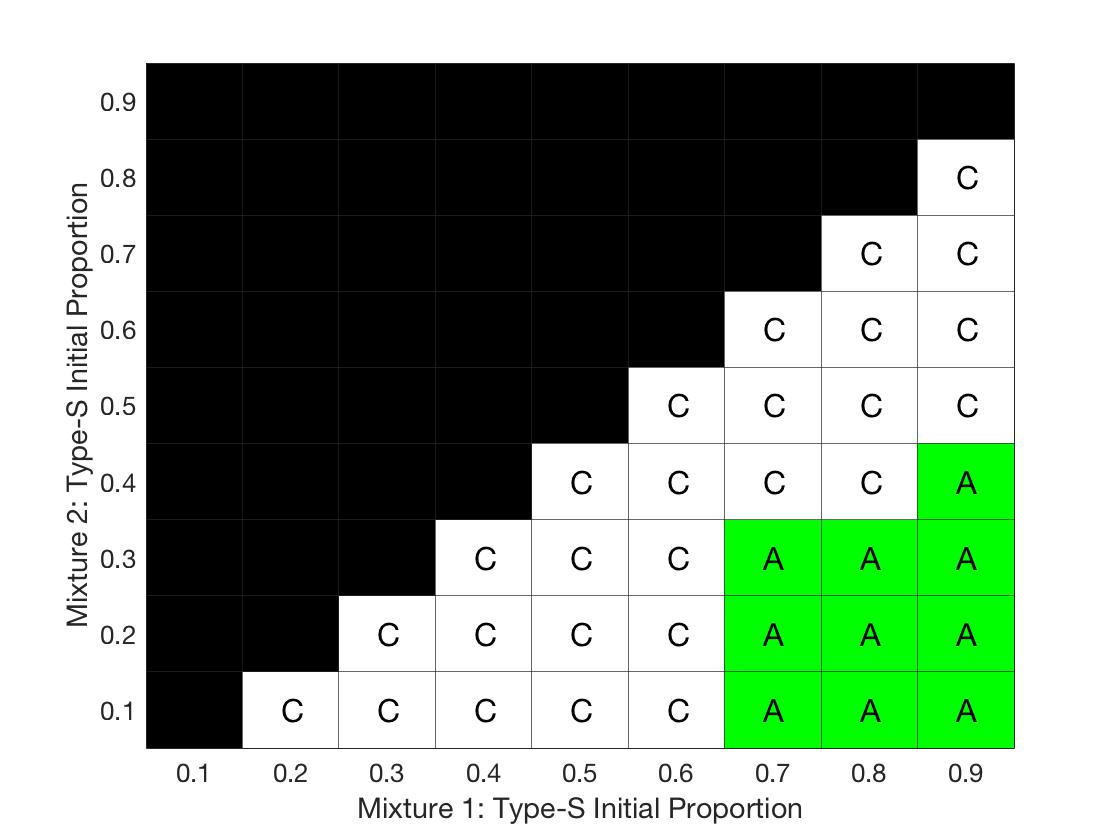}\end{minipage}\\
        \caption{(Parallel Calibration) Comparing LV model interaction inference across different CA interaction intensity levels using parallel calibration (left to right: $I=3$, $I=4$, and $I=5$) for (a) competitive, (b) mutual, and (c) antagonistic. Green indicates mixture combinations for which inferred interaction type in the LV model matches the interaction type used to generate the CA data.}
    \label{fig:comparesign_intlevel}
\end{figure}

Based on this result, we hypothesize that interaction inference in the LV model may depend upon the level of intensity of the interaction mechanism imposed between the two cell lines in the CA model.  Figure \ref{fig:comparesign_intlevel} illustrates the dependence of metric $E_1$ on the value of the intensity parameter, $I$, used to generate the CA data. For $I=3$ in the antagonistic case, we observe that there is no combination of mixtures for which the interaction inferred from the LV model matches the interaction type of the CA simulations.  As the intensity level increases, we observe that using two extreme mixture combinations enables correct inference of the interaction type used to generate the CA data. We conclude that 
the interaction inferred by the LV model will only match that used in the CA simulation if the intensity level in the CA model is sufficiently strong. 
 
\begin{figure}
\centering 
\includegraphics[width=0.32\textwidth]{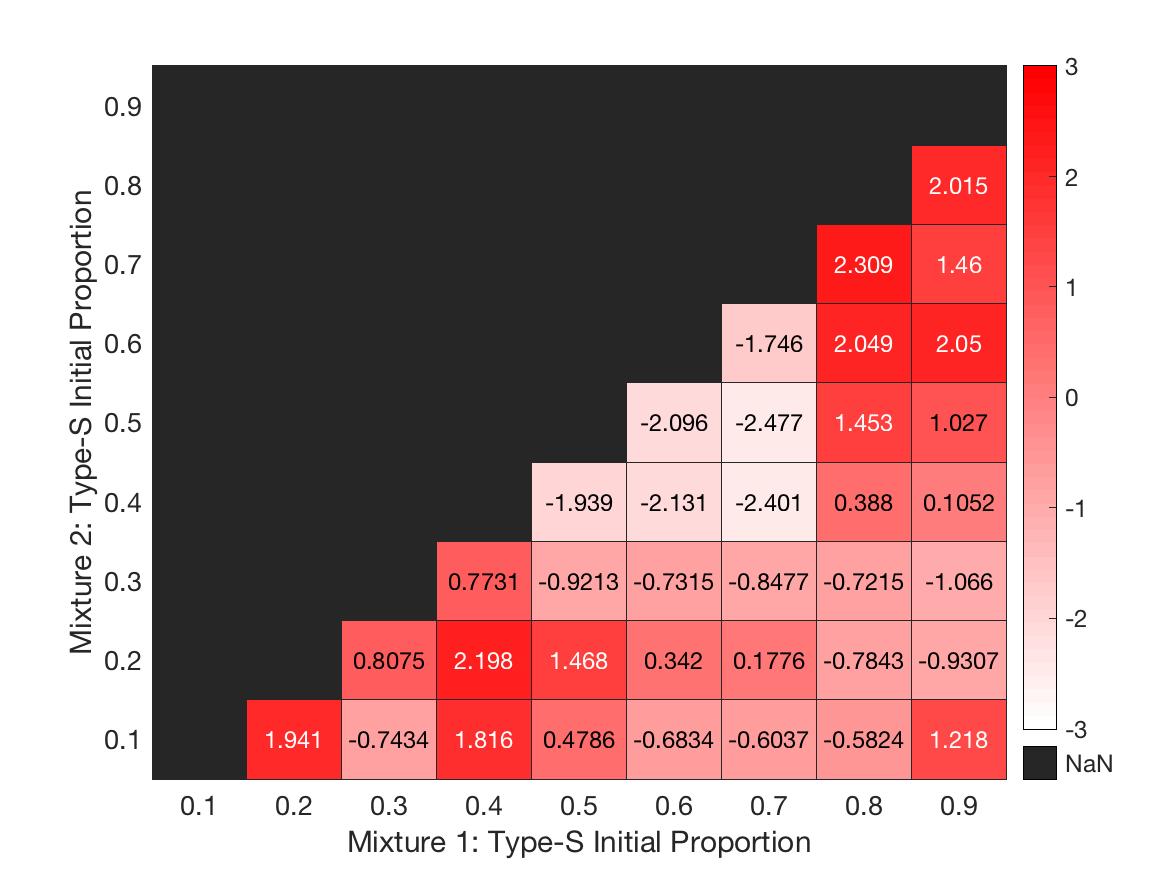}
\includegraphics[width=0.32\textwidth]{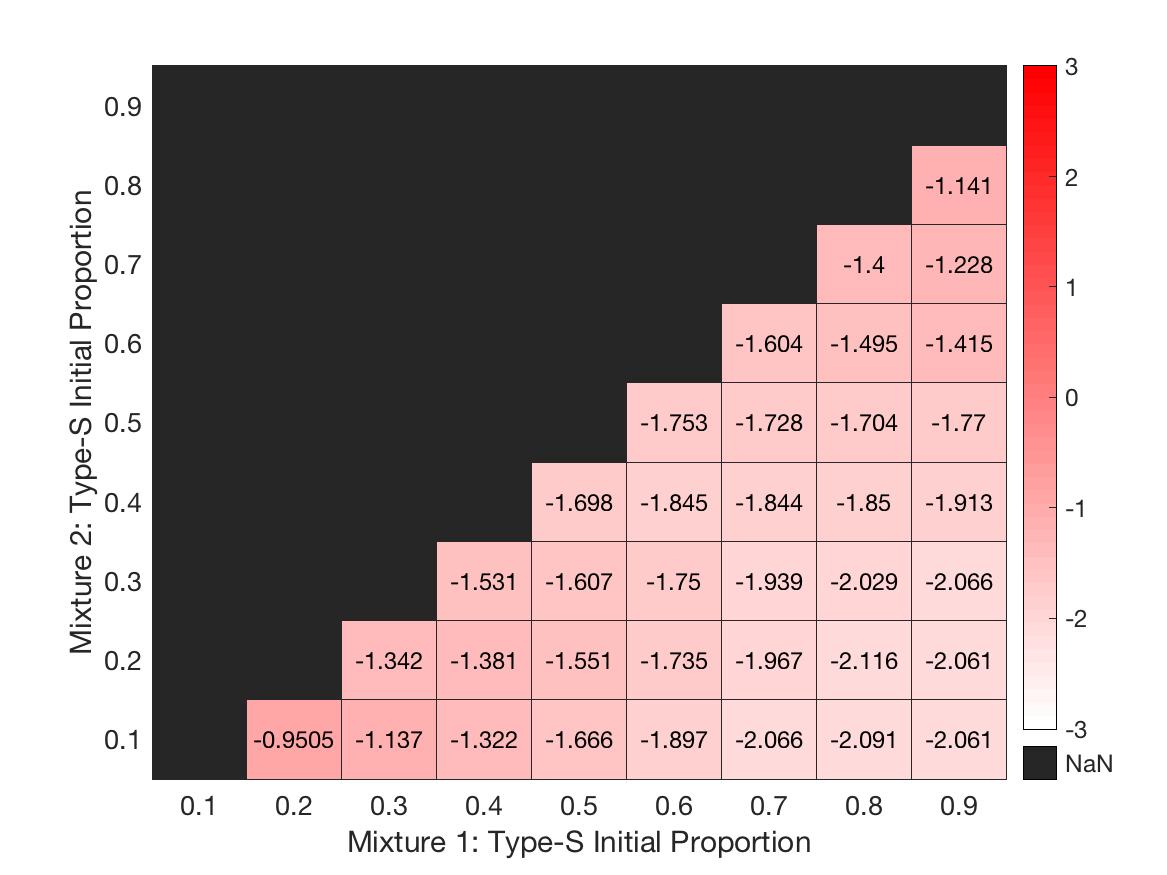}
\includegraphics[width=0.32\textwidth]{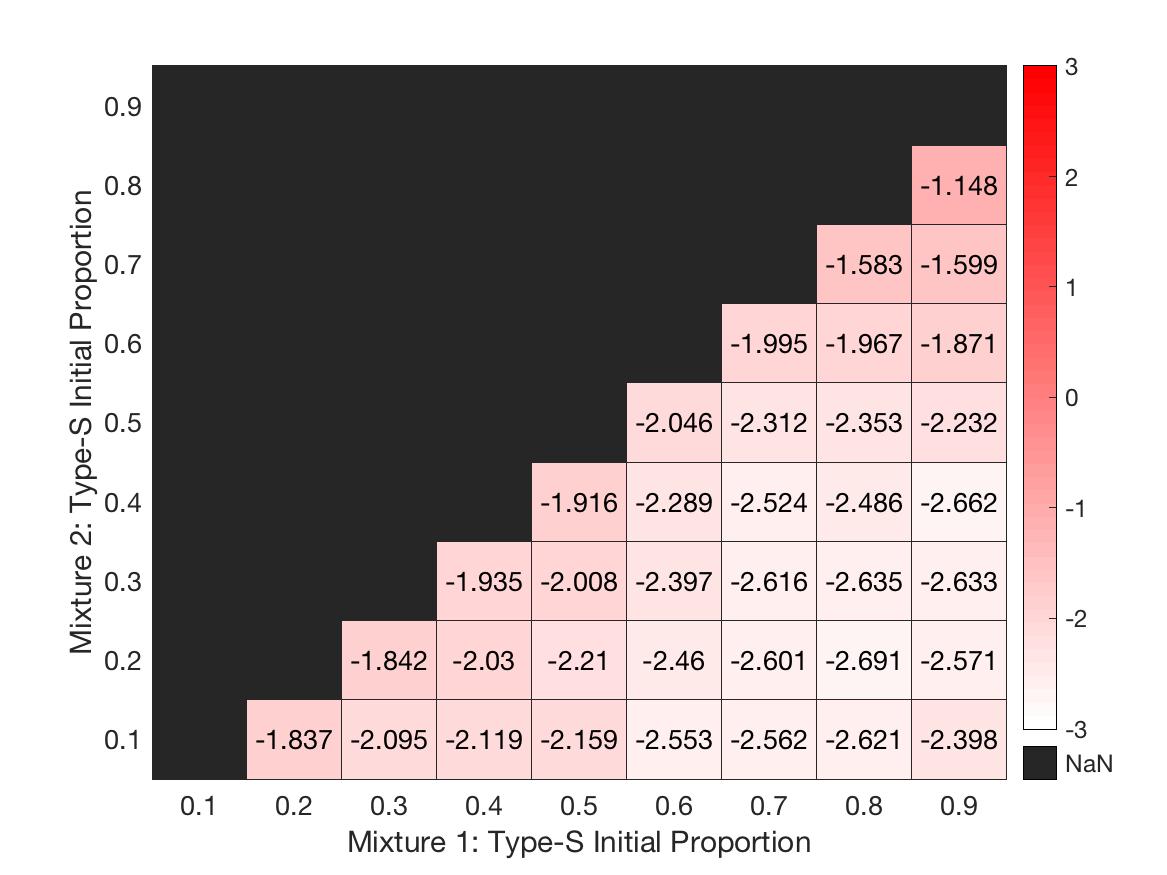}
\begin{minipage}{0.3\textwidth} \begin{center}(a)\end{center} \end{minipage}
\begin{minipage}{0.3\textwidth} \begin{center}(b)\end{center} \end{minipage}
\begin{minipage}{0.3\textwidth} \begin{center}(c)\end{center} \end{minipage}\\
\caption{(Parallel Calibration) Metric $E_2$ in Eq.~\eqref{eqn:E2} for the calibration of the Lotka-Volterra model to synthetic CA data for (a) competitive, (b) mutual, and (c) antagonistic interaction types, using CA intensity parameter $I=4$. }
\label{fig:parallelCA_E2}
\end{figure}

\begin{figure}
    \centering
    \includegraphics[width=0.47\textwidth]{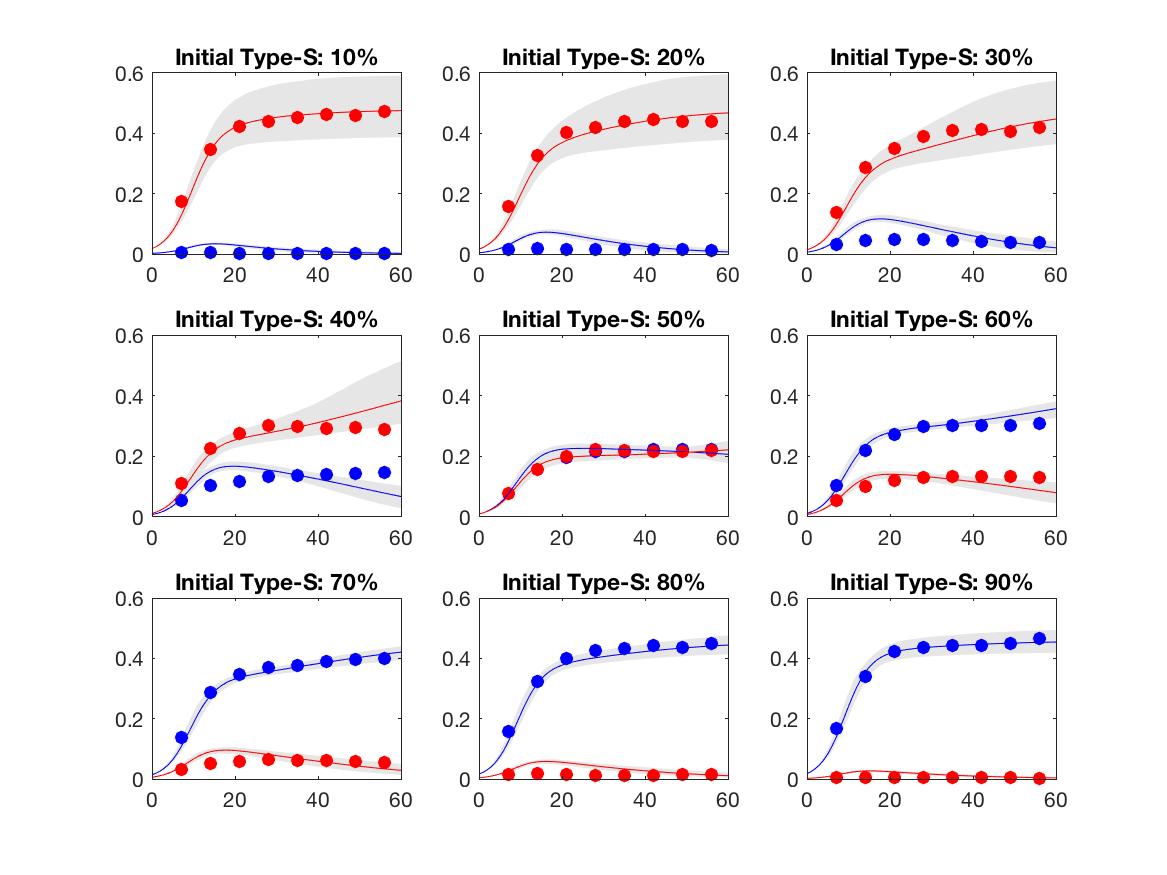} \hspace{7pt} \includegraphics[width=0.47\textwidth]{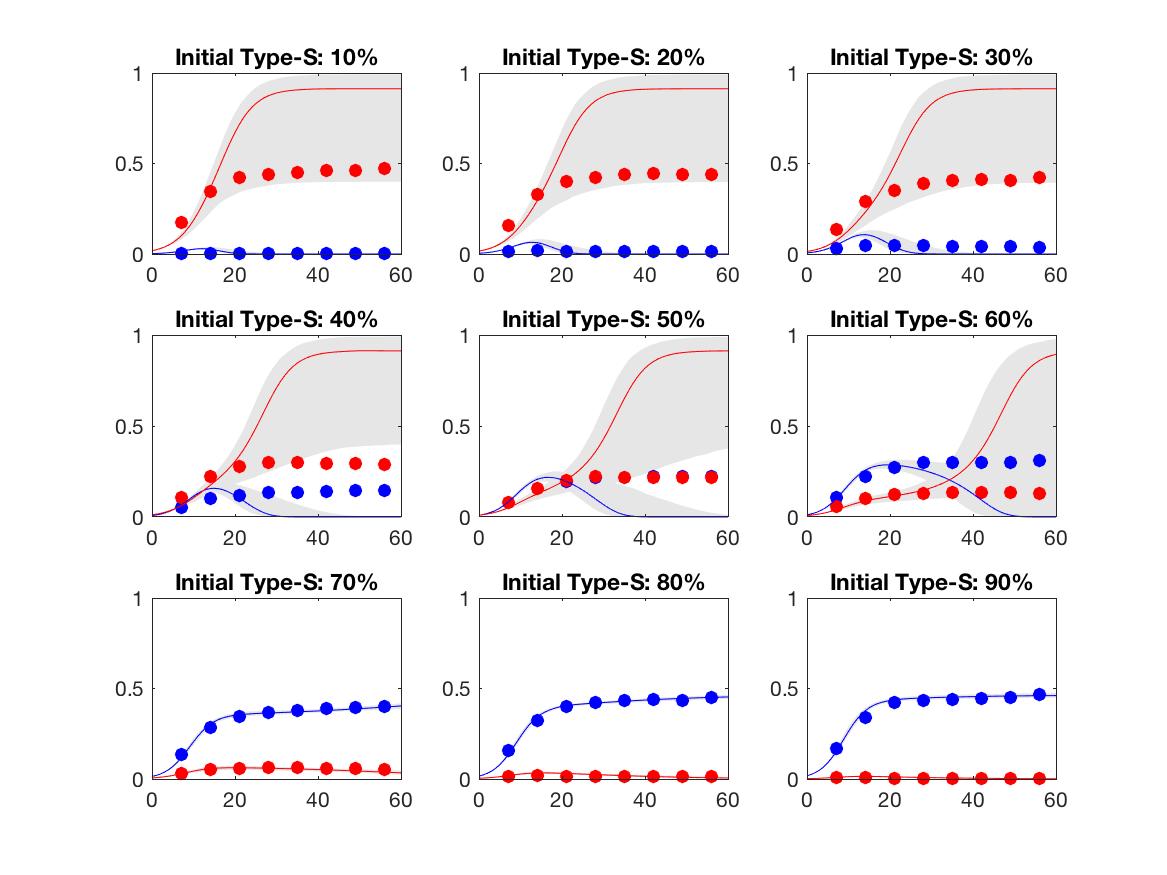} \\
    (a) \hspace{200pt} (b)
    \caption{(Parallel Calibration) Competitive model fits and credible intervals using (a) parameter values obtained from the 5:5 and 7:3 experiments, and (b) parameter values obtained from the 7:3 and 8:2 experiments, with synthetic data generated from the CA model with intensity parameter $I=4$. }
    \label{fig:modelfits_parallelCA}
\end{figure}

Figure \ref{fig:parallelCA_E2} details the $E_2$ error metric for the competitive, mutual, and antagonistic interaction types. In general, the errors in the model fits are much larger---particularly in the competitive and antagonistic cases---as compared to the idealistic calibration in Figure \ref{fig:errors_parallelLV}. For demonstration, we plot the model fits and associated credible intervals for two of the competitive scenarios in Figure \ref{fig:modelfits_parallelCA}.  The first set, using the 5:5 and 7:3 mixtures, is the best case scenario in terms of the $E_2$ metric, with $E_2 = -2.477.$ The second set, using the 7:3 and 8:2 mixtures, is the worst case scenario, with $E_2 = 2.309.$ In both cases, the model fits are reasonable for initial ratios close to one of the combinations used for calibration, but for other values of the initial ratio the fits (and uncertainties) are poor. Again, we attribute this phenomenon to model discrepancy: i.e., our use of a spatially-averaged model to fit spatially-resolved data. Comparing this figure to Figure \ref{fig:modelfits_ind}, where the data was generated from the LV model, reveals a major discrepancy in the behavior of the data. Generation of competitive interaction data from the LV model results in asymptotic coexistence between the two cell lines. However, the data shown in Figure \ref{fig:modelfits_parallelCA} tends toward elimination of one cell line over time; the additional competition for space and resources inhibits coexistence. The inability of the LV model to account for this added competition is an issue of model discrepancy, leading to larger model errors and uncertainties when calibrating to the synthetic CA data. 

\section{Discussion}\label{sec:Disc}

In this work, we investigated whether the type of interaction between two cell lines in a tumor spheroid (competitive, mutualistic, or antagonistic) can be inferred using the Lotka-Volterra model. We verified the structural identifiability of the system, and considered the practical identifiability of the model for three different experimental designs and calibration procedures. The first procedure tested whether the parameters of the Lotka-Volterra model could be uniquely inferred using a single data set containing information about both cell line volumes over time. The second design was a sequential calibration procedure, using two homogeneous cell-line experiments to calibrate the growth rate and capacity parameters and a mixture data set to calibrate the interaction parameters. The final scheme was a parallel calibration of all parameters using two mixture data sets. 

By testing the three experimental design procedures on data generated from the Lotka-Volterra model, we determined that both the sequential and parallel designs yield identifiable parameter sets, while the posterior densities resulting from the individual calibration scheme were not well-informed by the data. Though the sequential procedure produced the narrowest posteriors, the parallel procedure was shown to be more robust to noise and recovered point estimates closest to the true parameters. Thus, we favored the parallel procedure since it could be employed with one fewer data set, an important consideration for applications with budgetary restrictions on experiments. 

To assess the robustness of our parallel calibration procedure on a more realistic data set, we developed a cellular automaton model that tracks two cell types exhibiting these distinct types of interactions with parameters estimated from a prostate cancer cell line. The imposed interaction type dictated the speed at which cells divide and their corresponding oxygen consumption, when surrounded by a sufficient number of cells of the opposite type. We used this model to generate \textit{in silico} data, to which we calibrated the Lotka-Volterra model using the proposed parallel calibration design. The use of the spatially-averaged Lotka-Volterra model to fit data generated from the spatially-resolved cellular automaton model uncovered a model discrepancy issue; namely, the effects of both the primary interaction mechanism between the two cell lines and the baseline competition for space and resources were both absorbed into the Lotka-Volterra interaction parameters $\gamma_S$ and $\gamma_R$, with the result that the inferred interaction type by the Lotka-Volterra model would only match that used to generate the CA data if the intensity level of the interaction---as specified in the CA data generation---was strong enough to overcome the underlying competitive nature of the spatially-resolved data. With a strong enough intensity level, this match generally occurs for all interaction types when two asymmetric initial proportions are used for calibration, but is not guaranteed for other mixture combinations. Thus, researchers are cautioned to treat their interpretation of the Lotka-Volterra interaction parameters with care, as interaction types inferred from spatially-averaged data may not accurately reflect the interaction type of a spatially-resolved experiment. 

This work serves as a preliminary investigation into the identifiability of the Lotka-Volterra model using \textit{in silico} data, with special emphasis given to distinguishing the interaction type between the two cell lines. A number of possible directions are available from here. First, it remains to be seen whether there exists a correlation between the initial spatial distribution of the two cell lines and the type of interactions that result. A natural follow-up would be to expand the Lotka-Volterra into PDE form to account for spatial variation, and to determine whether there exists a relationship between spatial distribution of cell types and the model's ability to infer the interaction mechanism. This investigation could be performed using both the current version of the cellular automaton model described in Section \ref{sec:CAresults} and \ref{sec:app:CA}, which mimics two-dimensional growth such as that observed in $\textit{in vitro}$ laboratory settings, but could also be tested using a scaled-up version of the CA model in three-dimensions, a closer representation of \textit{in vivo} tumor composition. Eventually, we plan to verify these results using experimental data for two different cell lines.

Additionally, the analysis in this study is a first step towards the incorporation of radiotherapy treatment, and potentially other types of treatment. Specifically, we are interested in studying two cell types with differing intrinsic characteristics and different levels of radiosensitivity (e.g., Type-$S$ for ``sensitive" and Type-$R$ for ``resistant"), to determine whether these inherent differences in the two cell lines have an impact on the inferred interaction type when the Lotka-Volterra model is fit to experimental data.  \\




\bibliographystyle{cas-model2-names}

\bibliography{main_arxiv.bib}


\section*{Appendix}

\section*{A. Cellular Automaton Model}
\label{sec:app:CA}

We adapt our CA model from the models in \cite{Paczkowski2021,ChoSpringer,ChoJCM}, by allowing for different types of interactions between the two cell types. In the CA model, the interaction types affect the proliferation and oxygen consumption rates. 
Let $\tau$ denote the discrete time step length in the model simulation, and let $\tau_{cycle}(\mathbf{x},t)$ denote the cell cycle counter for a proliferating cell at site $\mathbf{x}$ at time $t$. When $\tau_{cycle}(\mathbf{x},t)\leq 0$, the cell at site $\mathbf{x}$ divides. During each time step, the reduction in the cycle counter for a given cell and its oxygen consumption depend on the composition of its neighborhood. For a proliferating Type-$S$ cell or Type-$R$ cell at site $\mathbf{x}$, its cell cycle counter reduces by $\Delta \tau_S(I)$ or $\Delta \tau_R(I)$, respectively, and it consumes oxygen at rate $\kappa_S(I)$ or $\kappa_R(I)$, respectively (mol cm$^{-3}$ s$^{-1}$). The values of $\Delta\tau_y(I)$ and $\kappa_y(I)$, for $y\in (S,R)$, depend on the local neighborhood of each cell, on the interaction type that we are modeling, and on the   
intensity parameter $I$, which quantifies the intensity of the interaction between the cell populations. Note that $I\geq 1$, where $I=1$ yields the neutral interaction case in the CA model, and higher values of $I$ correspond to stronger interactions between the cell types. In Section \ref{sec:CAresults} of the main text, we discuss the sensitivity of the Lotka-Volterra interaction parameter calibration to the CA interaction parameter $I$.

We consider four different interaction types between cell populations: neutral, competitive, mutualistic, antagonistic (with the Type-$R$ population antagonizing the Type-$S$ population, and the Type-$S$ population promoting the Type-$R$ population). In the competitive case, cells surrounded by a large number of cells of the other type consume less oxygen, and, as a result, divide more slowly. In the mutual case, cells surrounded by a large number of cells of the other type consume more oxygen and divide more quickly. In the antagonistic case, Type-$S$ cells surrounded by large numbers of Type-$R$ cells consume less oxygen and divide more slowly, while Type-$R$ cells surrounded by large numbers of Type-$S$ cells consume more oxygen and divide more quickly. In order to summarize the specific details of these effects, we first  define relevant terminology.
Let $N(\mathbf{x},t)$ denote the total number of neighbors in the two-dimensional Moore neighborhood of site $ \mathbf{x}$ at time $t$, with $0 \leq N(\mathbf{x},t)\leq 8$. Let $N_S(\mathbf{x},t)$ and $N_R(\mathbf{x},t)$ be the number of Type-$S$ cells and Type-$R$ cells, respectively, in the neighborhood of site $\mathbf{x}$ at time $t$, with $0 \leq N_y(\mathbf{x},t) \leq N(\mathbf{x},t)\leq 8$ for $y\in(S,R)$. We denote by $T_1$ and $T_2$ two threshold values for the neighborhood size, with $0\leq T_1<T_2\leq 8$, which are used to determine the values of $\Delta\tau_y(I)$ and $\kappa_y(I)$.

Table \ref{tbl:CArules_tau} summarizes the value for the cell cycle counter reduction for a Type-$S$ cell, $\Delta \tau_S(I)$, dependent upon  the total number of neighbors in its Moore neighborhood, the number of Type-$R$ neighbors, and the interaction type used in the simulation. We note that there are baseline levels of cell cycle counter reduction, dependent on the total number of neighbors, to simulate contact inhibition of proliferation. For interaction types with $I>1$, additional changes to the cell cycle counter reduction depend on the number of neighbors of the other type.  Similarly to the cell cycle counter reduction, Table \ref{tbl:CArules_kappa} displays the value for the corresponding oxygen consumption rate, $\kappa_S(I)$. Almost all values in Tables \ref{tbl:CArules_tau} and \ref{tbl:CArules_kappa} are identical for a Type-$R$ cell, when considering the number of Type-$S$ neighbors in its neighborhood, except the two values highlighted in blue, for the antagonistic case. In this case,  $\Delta \tau_R(I) = \frac{\tau I}{2}$ and $\kappa_R(I)=\frac{\kappa_P I}{2}$ when $T_1 < N(\mathbf{x},t) \leq T_2$  and $N_S(\mathbf{x},t)> T_1$, and $\Delta \tau_R(I) = \frac{\tau I}{4}$ and $\kappa_R(I)=\frac{\kappa_P I}{4}$ when $N(\mathbf{x},t) > T_2$  and $N_S(\mathbf{x},t)> T_2$.

 \begin{table}[htbp]
    \centering
        \renewcommand{\arraystretch}{1.5}
    \begin{tabular}{|c||c|c|c|c|} 
  \hline 
        &  Neutral & Competitive & Mutual & Antagonistic  \\ \hline    \hline
      $N(\mathbf{x},t) \leq T_1$    & $\tau$ & $\tau$ & $\tau$ & $\tau$  \\ \hline 
   
     $T_1 < N(\mathbf{x},t) \leq T_2$  and $N_R(\mathbf{x},t)\leq T_1$ & $\frac{\tau}{2}$ & $\frac{\tau}{2}$ & $\frac{\tau}{2}$ & $\frac{\tau}{2}$  \\ \hline 
      $T_1 < N(\mathbf{x},t) \leq T_2$  and $N_R(\mathbf{x},t)> T_1$ & $\frac{\tau}{2}$ & $\frac{\tau}{2I}$ & $\frac{\tau I}{2}$ & \textcolor{blue}{$\frac{\tau}{2I}$}  \\ \hline 
      $N(\mathbf{x},t) > T_2$  and $N_R(\mathbf{x},t)\leq T_2$ & $\frac{\tau}{4}$ & $\frac{\tau}{4}$ & $\frac{\tau}{4}$ & $\frac{\tau}{4}$  \\ \hline 
      $N(\mathbf{x},t) > T_2$  and $N_R(\mathbf{x},t)> T_2$ & $\frac{\tau}{4}$ & $\frac{\tau}{4I}$ & $\frac{\tau I}{4}$ & \textcolor{blue}{$\frac{\tau}{4I}$}  \\ \hline 
    \end{tabular}
    \caption{The value for the cell cycle counter reduction, $\Delta \tau_S(I)$, for each neighborhood scenario and interaction type. Alternatively, when considering a resistant cell at site $\text{x}$ and the number of sensitive neighbors, the only values that differ for $\Delta \tau_R(I)$ are those highlighted in blue.}
    \label{tbl:CArules_tau}
\end{table}

 \begin{table}[htbp]
    \centering
        \renewcommand{\arraystretch}{1.5}
    \begin{tabular}{|c||c|c|c|c|} 
  \hline 
        &  Neutral & Competitive & Mutual & Antagonistic  \\ \hline    \hline
      $N(\mathbf{x},t) \leq T_1$    & $\kappa_P$ & $\kappa_P$ & $\kappa_P$ & $\kappa_P$  \\ \hline 
   
     $T_1 < N(\mathbf{x},t) \leq T_2$  and $N_R(\mathbf{x},t)\leq T_1$ & $\frac{\kappa_P}{2}$ & $\frac{\kappa_P}{2}$ & $\frac{\kappa_P}{2}$ & $\frac{\kappa_P}{2}$  \\ \hline 
      $T_1 < N(\mathbf{x},t) \leq T_2$  and $N_R(\mathbf{x},t)> T_1$ & $\frac{\kappa_P}{2}$ & $\frac{\kappa_P}{2I}$ & $\frac{\kappa_P I}{2}$ & \textcolor{blue}{$\frac{\kappa_P}{2I}$}  \\ \hline 
      $N(\mathbf{x},t) > T_2$  and $N_R(\mathbf{x},t)\leq T_2$ & $\frac{\kappa_P}{4}$ & $\frac{\kappa_P}{4}$ & $\frac{\kappa_P}{4}$ & $\frac{\kappa_P}{4}$  \\ \hline 
      $N(\mathbf{x},t) > T_2$  and $N_R(\mathbf{x},t)> T_2$ & $\frac{\kappa_P}{4}$ & $\frac{\kappa_P}{4I}$ & $\frac{\kappa_P I}{4}$ & \textcolor{blue}{$\frac{\kappa_P}{4I}$}  \\ \hline 
    \end{tabular}
    \caption{The value for the oxygen consumption rate, $\kappa_S(I)$, for each neighborhood scenario and interaction type. Alternatively, when considering a resistant cell at site $\text{x}$ and the number of sensitive neighbors, the only values that differ for $\kappa_R(I)$ are those highlighted in blue.}
    \label{tbl:CArules_kappa}
\end{table}

As part of the CA model, we incorporate oxygen, the single growth-rate-limiting nutrient, using a reaction-diffusion equation. In particular, the change in oxygen concentration $c(\textbf{x},t)$ (mol cm$^{-3}$) at location $\textbf{x}$ at time $t$ is described by:
\begin{equation*}
    \frac{\partial c(\textbf{x},t)}{\partial t} = D\nabla^2c(\textbf{x},t) - \Gamma(\textbf{x},t),
\end{equation*}
where $D$ is the oxygen diffusion coefficient (cm$^2$ s$^{-1}$), and $\Gamma(\textbf{x},t)$ is the oxygen consumption rate (mol cm$^{-3}$ s$^{-1}$). In the neutral interaction case, $\Gamma(\textbf{x},t)$ is defined as follows:
\begin{align*}
    \Gamma(\textbf{x},t)= \begin{cases} \kappa_S(\textbf{x},t) & \text{ if }\textbf{x}\text{ is occupied by a proliferating Type-$S$ cell}\\
    \kappa_R(\textbf{x},t) & \text{ if }\textbf{x}\text{ is occupied by a proliferating Type-$R$ cell}\\
    \kappa_Q& \text{ if }\textbf{x}\text{ is occupied by a quiescent cell}\\
    0& \text{ otherwise.}\end{cases}
\end{align*}
Recall $\kappa_S(\textbf{x},t)$ and $\kappa_R(\textbf{x},t)$ are summarized in Table \ref{tbl:CArules_kappa}. The parameter $\kappa_Q>0$ denotes the constant rate, in mol cm$^{-3}$ s$^{-1}$, at which quiescent cells consume oxygen. We note that the concentration $c(\textbf{x},t)$ will not become negative because cells exposed to oxygen below the threshold $c_N$ will become necrotic, and necrotic cells do not consume oxygen. We also use the following initial and boundary conditions to simulate oxygen diffusion from the boundaries of a square Petri dish into the culture medium:
\begin{eqnarray*}
    &&c(x,y,0)=c_{\infty},\\ \ \\
    &&c(0,y,t)=c(L,y,t) = c(x,0,t) = c(x,L,t) = c_{\infty},
\end{eqnarray*}
where $L$ is the domain length, and $c_{\infty}$ is the background O$_2$ concentration.

Local oxygen levels determine the cell state at each lattice site. Sites can contain proliferating cells, quiescent cells, or necrotic cells, or they can be empty. The threshold oxygen levels $c_Q$ and $c_N$ determine the state at each site, as follows:
\begin{itemize}
    \item If $c_Q<c(\textbf{x},t)\leq c_{\infty}$, then the cell at site $\textbf{x}$ is proliferating.
    \item If $c_N<c(\textbf{x},t)\leq c_Q$, then the cell at site $\textbf{x}$ is quiescent.
    \item If $0 \leq c(\textbf{x},t)\leq c_N$, then the cell at site $\textbf{x}$ is necrotic.
\end{itemize}

Table \ref{table:CA_pars} summarizes the parameter values that we used to simulate the CA model throughout this work.\\

\begin{table}[!htb]
\begin{center}
\begin{tabular}[c]{|c|c|c|c|}
        \hline
        \textbf{Param.} &\textbf{Description} & \textbf{Value} & \textbf{Units} \\

\hline 
 $l$ & Cell size & 0.0018 & cm \\
\hline
 $L$ & Domain length & 0.36 & cm \\
\hline
 $\bar{\tau}_{cycle}$& Mean (SD) cell cycle time   & 18.3 (1.4) & h \\
\hline 
 $c_{\infty}$& Background O$_2$ concentration & $2.8\times 10^{-7}$ & mol cm$^{-3}$  \\
\hline 
 $D$ & O$_2$ diffusion constant & $1.8\times 10^{-5}$ & cm$^2$s$^{-1}$\\
\hline 
 $c_Q$ & O$_2$ concentration threshold for proliferating cells & 1.82$\times10^{-7}$ & mol cm$^{-3}$ \\
\hline 
$c_N$ & O$_2$ concentration threshold for quiescent cells & 1.68$\times10^{-7}$ & mol cm$^{-3}$ \\
\hline 
 $\kappa_P$ & O$_2$ consumption rate of proliferating cells & 1.0$\times10^{-8}$ &mol cm$^{-3}$s$^{-1}$\\
\hline 
 $\kappa_Q$ & O$_2$ consumption rate of quiescent cells & 5.0$\times10^{-9}$ & mol cm$^{-3}$s$^{-1}$  \\
\hline 
 $p_{NR}$ & Rate of lysis of necrotic cells & 0.01  & hr$^{-1}$ \\
 \hline
 $I$ & Interaction intensity level& Varies & -- \\
 \hline
 $T_1$ & Neighborhood threshold 1 for cell cycle reduction & 4 & cells \\
  \hline
 $T_2$ & Neighborhood threshold 2 for cell cycle reduction & 7 & cells \\
\hline 
\end{tabular}
\caption{A summary of the parameters used in the CA model and their default values. Parameter values are estimated using experimental data from the prostate cancer cell line, PC3, in \cite{Paczkowski2021}.}
\label{table:CA_pars}
\end{center}
\end{table}

We generate the \textit{in silico} data from the CA model with different initial proportions of Type-$S$ cells, varying this initial proportion across all 0.1 increments between 0 and 1. We generate data exhibiting each of the four interaction types: neutral, competitive, mutual, and antagonistic. In all cases, we initialize the cells in a circular region of radius 9 cells, in the center of a domain of size 200 cells $\times$ 200 cells. The type of cell at each initial site is randomly chosen, according to the specified initial distribution. All cells are initially proliferating cells, each with an initial cell cycle counter $\tau_{cycle}(\mathbf{x},0)$ chosen as a random integer between 0 and the mean cell cycle time, $\bar{\tau}_{cycle}$.  \\

\section*{B. Sequential Calibration to Synthetic CA Data}
\label{sec:app:CAseqresults}

\begin{figure}[!htb]
\centerline{
    \rotatebox{90}{$\qquad\,\,\,$ \rotatebox{-90}{(a)}}
    \includegraphics[width=0.78\textwidth]{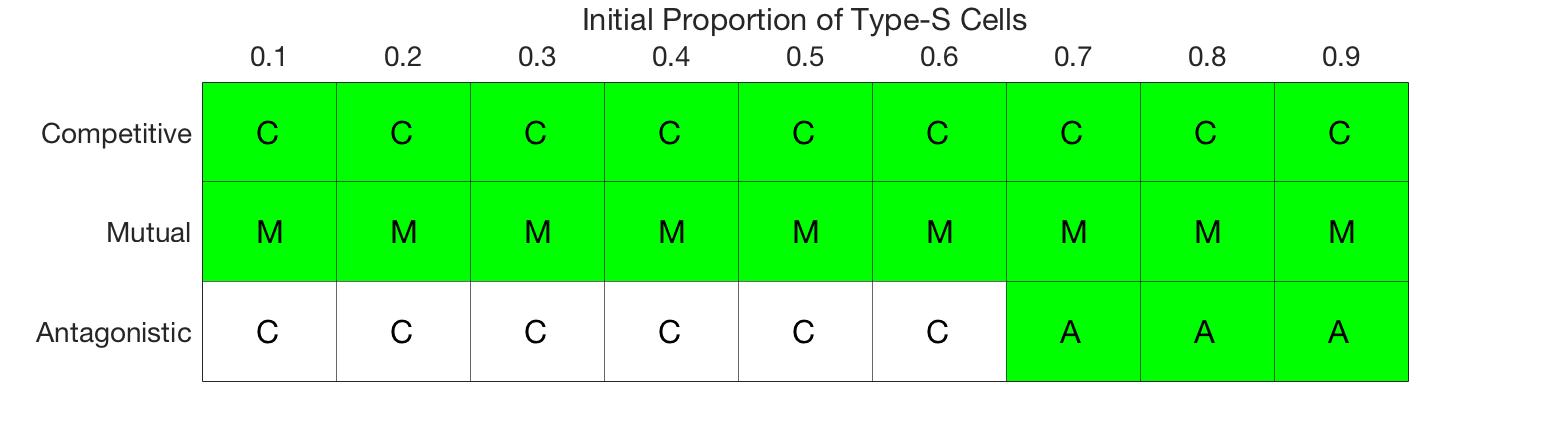}}
\centerline{
    \rotatebox{90}{$\qquad\,\,\,$ \rotatebox{-90}{(b)}}
    \includegraphics[width=0.78\textwidth]{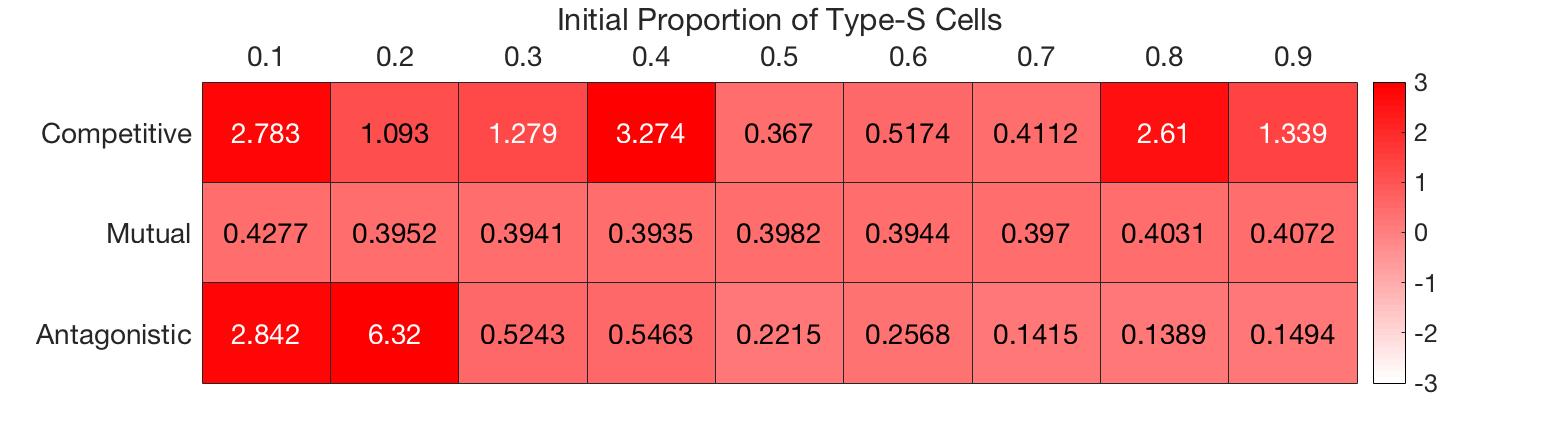}}
\centerline{
    \rotatebox{90}{$\qquad\qquad$ \rotatebox{-90}{(c)}}
    \includegraphics[width=0.78\textwidth]{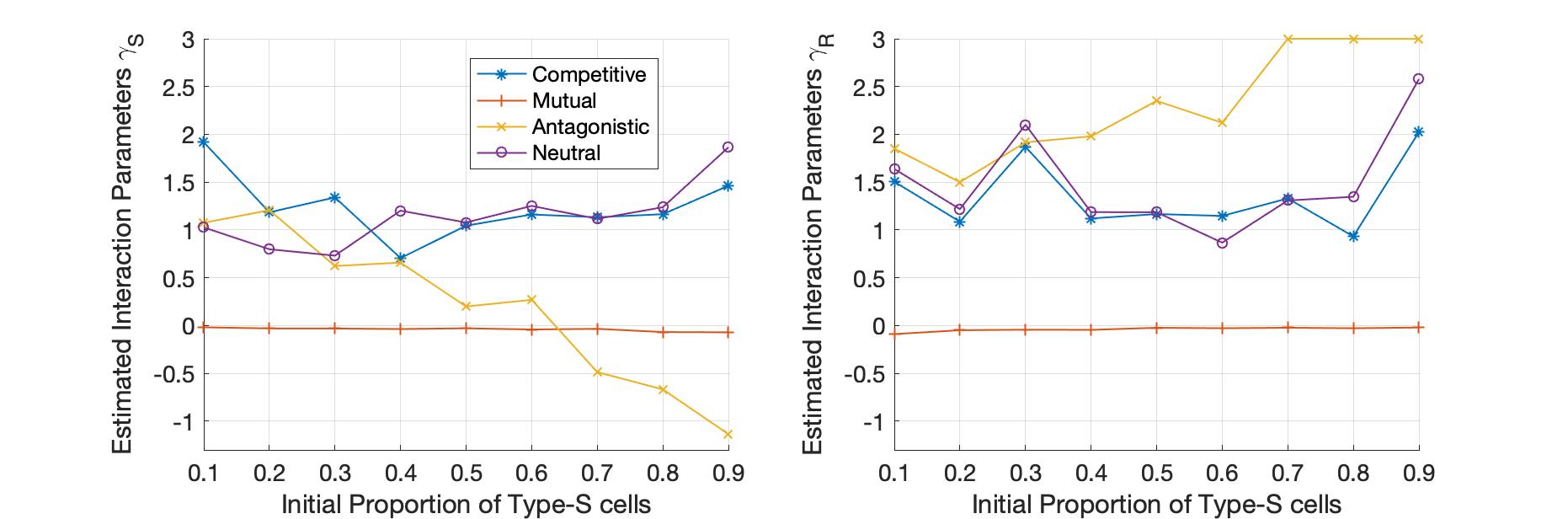}
}
\caption{ (Sequential Calibration) 
(a) Metric $E_1$ in Eq.~\eqref{eqn:E1} for the sequential calibration of the LV model to synthetic CA data. Green indicates mixtures for which the inferred interaction type in the LV model matches the interaction type used to generate the CA data. (b) Metric $E_2$ in Eq.~\eqref{eqn:E2}, measuring the log sum-of-squares error across all nine model fits. (c) The values of fitted parameters $\hat \gamma_S$ and $\hat \gamma_R$ for each interaction type across different initial proportions of Type-$S$ cells. 
}
\label{fig:purethen1mixed}
\end{figure}

Here, we present the sequential calibration results when fitting the LV model to synthetic CA data, for comparison to the results from the parallel procedure detailed in Section \ref{sec:CAresults}. Metric $E_1$, shown in Figure \ref{fig:purethen1mixed}(a), reveals that for antagonistic CA simulations, there is inconsistency in the type of interaction inferred by the LV model. Specifically, for Type-$S$ initial proportions from 0.1 to 0.6, the LV model predicts a competitive interaction, while for larger initial proportions of Type-$S$, an antagonistic relationship is inferred. Metric $E_2$ is shown in Figure \ref{fig:purethen1mixed}(b); in general, the errors are larger than those computed for the parallel calibration scheme in Figure \ref{fig:errors_parallelLV}, further motivating our choice to use the parallel procedure in Section \ref{sec:CAresults}. The parameter estimates for each of the interaction types and initial ratios are reported in Figure \ref{fig:purethen1mixed}(c). We note that the interaction parameters associated with  mutualistic interactions appear to vanish to zero, but they are actually slightly negative in value. As discussed in Section \ref{sec:CAresults}, the tendency of the interaction parameters to skew toward positive values is a result of the baseline competition for space and resources in the spatially-resolved CA data. As such, the LV model often predicts a competitive relationship for antagonistic CA simulations. In particular, if the Type-$R$ cells antagonize the Type-$S$ cells and a large number of Type-$R$ cells are mixed with a small number of Type-$S$ cells, then the type of interaction tends to be misclassified. Since the initial fraction of the Type-$S$ cells is small, it is likely that the Type-$R$ cells rapidly become dominant and the Type-$S$ cells are rapidly eliminated. Thus, when the antagonized population Type-$S$ comprise only 10\% of the initial tumor volume, the interaction parameters are estimated to have the same signs as for the competitive interaction case. However, as the initial proportion of Type-$S$ cells increases, the antagonizing effect of the Type-$R$ cells is discernible, since the Type-$R$ ultimately dominates despite its small initial portion. Specifically, we find that the interaction parameter $\gamma_S$ is negative when the initial proportion of Type-$S$ cells exceeds 0.6.

\end{document}